\documentclass[10pt,journal,compsoc]{IEEEtran}
\usepackage{hyperref}
\usepackage{amssymb}
\usepackage{mathtools}
\usepackage{amsmath}                   
\usepackage{color}
\usepackage{mathptmx}
\usepackage{amssymb}
\usepackage{enumitem}
\usepackage{multicol}
\usepackage{anyfontsize}
\usepackage{amsmath}
\usepackage{algorithmic}
\usepackage{algorithm}
\usepackage{esvect}
\usepackage{wrapfig}
\usepackage{multicol}
\usepackage[dvipsnames]{xcolor}

\DeclareMathAlphabet{\mathcal}{OMS}{cmsy}{m}{n}

\DeclareMathOperator*{\argmin}{arg\,min}

\ifCLASSOPTIONcompsoc
  \usepackage[nocompress]{cite}
\else
  \usepackage{cite}
\fi
\ifCLASSINFOpdf
\else
\fi
\begin{document}
%
\title{Principal Geodesic Analysis of Merge Trees\\ (and Persistence Diagrams)}
%
%
%
%

\author{Mathieu Pont, Jules Vidal and Julien Tierny
\IEEEcompsocitemizethanks{\IEEEcompsocthanksitem Mathieu Pont, Jules Vidal and 
Julien Tierny are with the CNRS and Sorbonne Universit\'e.
E-mails: \{firstname.lastname\}@sorbonne-universite.fr}
\thanks{Manuscript received April 19, 2005; revised August 26, 2015.}}

\IEEEtitleabstractindextext{%
\begin{abstract}
This paper presents a computational framework for the Principal
Geodesic Analysis of merge trees (MT-PGA), a novel adaptation of the
celebrated Principal Component Analysis (PCA) framework \cite{pearson01} to the
Wasserstein metric space of merge trees \cite{pont_vis21}.
We
formulate MT-PGA computation as a constrained optimization problem,
aiming at
adjusting a basis of orthogonal
geodesic axes, while minimizing a fitting energy.
We introduce
an efficient, iterative algorithm which exploits
shared-memory parallelism, as well as
an analytic expression of the
fitting energy gradient, to ensure fast iterations.
Our approach also trivially extends to
extremum persistence diagrams.
Extensive experiments on public ensembles
demonstrate the efficiency of our approach -- with MT-PGA computations in the
orders of minutes
for the largest examples.
We show the utility of our contributions by extending to merge trees two
typical PCA applications.
First, we apply MT-PGA to \emph{data reduction} and reliably compress
merge trees by concisely representing them by
their first coordinates in the MT-PGA basis.
Second, we present a \emph{dimensionality reduction} framework exploiting the
first two directions of the MT-PGA basis to generate two-dimensional layouts of
the ensemble.
We augment these layouts with
persistence correlation
views, enabling global and local visual inspections of the feature variability
in the
ensemble.
In both applications, quantitative experiments
assess the
relevance of our framework.
Finally, we provide a
C++ implementation that can be used to
reproduce our results.
\end{abstract}

}

\maketitle

\IEEEdisplaynontitleabstractindextext

%
\IEEEpeerreviewmaketitle

\newcommand{\domain}{\mathcal{M}}
\newcommand{\range}{\mathbb{R}}
\newcommand{\sublevelset}[1]{#1^{-1}_{-\infty}}
\newcommand{\superlevelset}[1]{#1^{-1}_{+\infty}}
\newcommand{\Star}{St}
\newcommand{\Link}{Lk}
\newcommand{\simplex}{\sigma}
\newcommand{\face}{\tau}
\newcommand{\lowerlink}{\Link^{-}}
\newcommand{\upperlink}{\Link^{+}}
\newcommand{\Index}{\mathcal{I}}
\newcommand{\offset}{o}
\newcommand{\Natural}{\mathbb{N}}
\newcommand{\criticalSet}{\mathcal{C}}
\newcommand{\diagram}{\mathcal{D}}
\newcommand{\wasserstein}[1]{W^{\diagram}_#1}
\newcommand{\projection}{\Delta}
\newcommand{\hierarchy}{\mathcal{H}}
\newcommand{\decimation}{D}
\newcommand{\xDimD}{L_x^\decimation}
\newcommand{\yDimD}{L_y^\decimation}
\newcommand{\zDimD}{L_z^\decimation}
\newcommand{\xDim}{L_x}
\newcommand{\yDim}{L_y}
\newcommand{\zDim}{L_z}
\newcommand{\Grid}{\mathcal{G}}
\newcommand{\GridD}{\mathcal{G}^\decimation}
\newcommand{\x}{\phantom{x}}
\newcommand{\Mod}{\;\mathrm{mod}\;}
\newcommand{\NN}{\mathbb{N}}
\newcommand{\forwardIntegralLine}{\mathcal{L}^+}
\newcommand{\backwardIntegralLine}{\mathcal{L}^-}
\newcommand{\triangulationOp}{\phi}
\newcommand{\decimationOp}{\Pi}
\newcommand{\isovalue}{w}
\newcommand{\persistence}{\mathcal{P}}
\newcommand{\pointMetric}{d}
\newcommand{\diagramSet}{\mathcal{S}_\mathcal{D}}
\newcommand{\diagramSpace}{\mathbb{D}}
\newcommand{\jointree}{\mathcal{T}^-}
\newcommand{\splittree}{\mathcal{T}^+}
\newcommand{\mergetree}{\mathcal{T}}
\newcommand{\mergetreeSet}{\mathcal{S}_\mathcal{T}}
\newcommand{\branchset}{\mathcal{S}_\mathcal{B}}
\newcommand{\branchspace}{\mathbb{B}}
\newcommand{\mergetreeSpace}{\mathbb{T}}
\newcommand{\editdistance}{D_E}
\newcommand{\wassersteinTree}{W^{\mergetree}_2}
\newcommand{\distanceSequence}{d_S}
\newcommand{\branchtree}{\mathcal{B}}
\newcommand{\branchtreeSet}{\mathcal{S}_\mathcal{B}}
\newcommand{\branchtreeSpace}{\mathbb{B}}
\newcommand{\forest}{\mathcal{F}}
\newcommand{\sequenceSpace}{\mathbb{S}}
\newcommand{\forestMatrix}{\mathbb{F}}
\newcommand{\treeMatrix}{\mathbb{T}}
\newcommand{\normalizedLocation}{\mathcal{N}}
\newcommand{\normalizedWasserstein}{W^{\normalizedLocation}_2}
\newcommand{\geodesictree}{\mathcal{G}}
\newcommand{\dummyVector}{\mathcal{V}}
\newcommand{\geodesictreeVec}{g}
\newcommand{\geodesicAxis}{\mathcal{A}}
\newcommand{\directionVector}{\mathcal{V}}
\newcommand{\geodesicdiagram}{\mathcal{G}^{\diagram}}
\newcommand{\reconstructionError}{E_{L_2}}
\newcommand{\pcaBasis}{B_{\mathbb{R}^d}}
\newcommand{\mtPgaBasis}{B_{\branchtreeSpace}}
\newcommand{\mtPgaError}{E_{\wassersteinTree}}
\newcommand{\frechetEnergy}{E_F}
\newcommand{\geodesicExtremity}{\mathcal{E}}
\newcommand{\vectorNotation}[1]{\protect\vv{#1}}
\newcommand{\axisNotation}[1]{\protect\overleftrightarrow{#1}}
\newcommand{\individualEnergy}{E}
\newcommand{\ensembleSize}{N}
\newcommand{\numberBranchinBarycenter}{N_1}
\newcommand{\numberGeodesicSamples}{N_2}
\newcommand{\planarGridX}{N_x}
\newcommand{\planarGridY}{N_y}
\newcommand{\regularGrid}{G}
\newcommand{\distanceMatrix}{\mathbb{D}}
\newcommand{\maxDimensions}{{d_{max}}}
\newcommand{\projectionOperator}{\mathcal{P}}
\newcommand{\reconstructed}[1]{\widehat{#1}}
\newcommand{\gt}{>}
\newcommand{\lt}{<}
\newcommand{\branch}{b}

\newcommand{\julien}[1]{\textcolor{red}{#1}}
\newcommand{\mathieu}[1]{\textcolor{green}{#1}}
\renewcommand{\mathieu}[1]{\textcolor{green}{#1}}
\newcommand{\jules}[1]{\textcolor{orange}{#1}}
\renewcommand{\jules}[1]{\textcolor{black}{#1}}
\newcommand{\note}[1]{\textcolor{magenta}{#1}}
\newcommand{\cutout}[1]{\textcolor{blue}{#1}}
\renewcommand{\cutout}[1]{}

\newcommand{\revision}[1]{\textcolor{black}{#1}}
\newcommand{\minorRevision}[1]{\textcolor{black}{#1}}

\renewcommand{\figureautorefname}{Fig.}
\renewcommand{\sectionautorefname}{Sec.}
\renewcommand{\subsectionautorefname}{Sec.}
\renewcommand{\equationautorefname}{Eq.}
\renewcommand{\tableautorefname}{Tab.}
\newcommand{\algorithmautorefname}{Alg.}
\newcommand{\lineautorefname}{Alg.}

\newcommand{\eqSpace}{-1.75ex}

\newcommand{\mycaption}[1]{
\caption{#1}
}


\IEEEraisesectionheading{\section{Introduction}\label{sec:introduction}}

\begin{figure*}
  \centering
  \includegraphics[width=\linewidth]{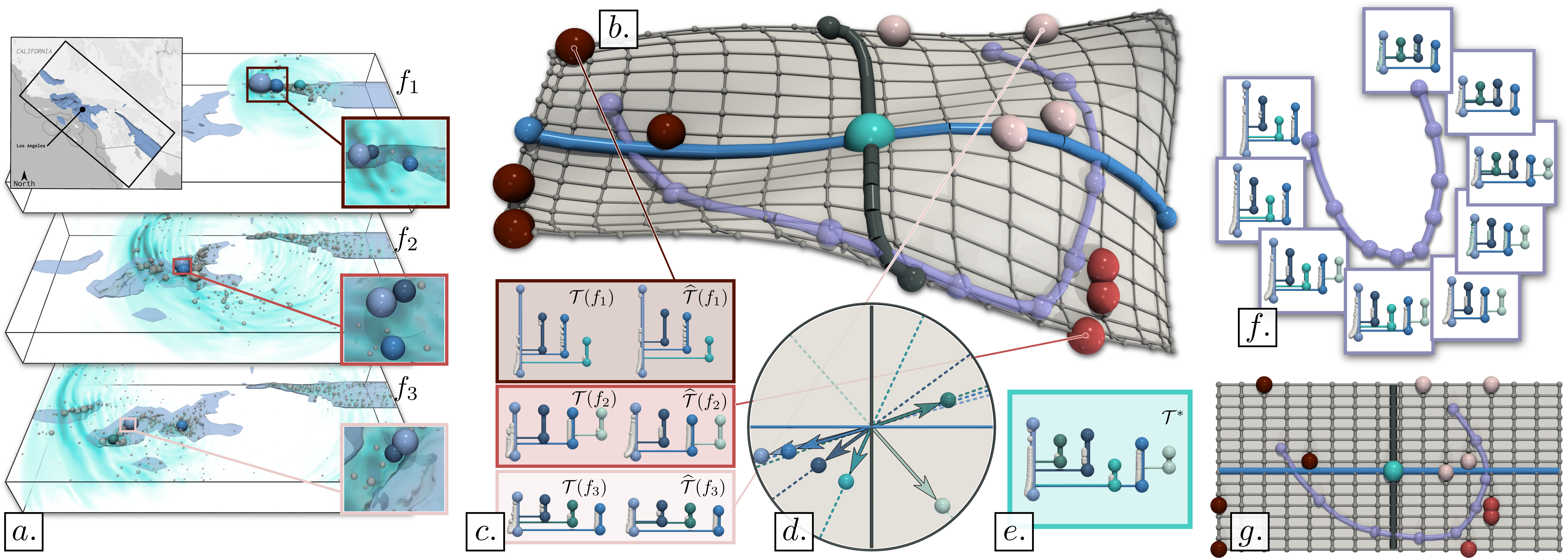}
  \caption{
  Visual analysis
  of the \emph{Earthquake} ensemble
  \revision{with Merge Tree Principal Geodesic Analysis (MT-PGA, \emph{(a)}: one member per ground-truth class).}
  Our framework computes
  a coordinate system
  \minorRevision{\emph{(b)}} for the Wasserstein
metric space of merge trees $\branchtreeSpace$ by adjusting geodesic axes (blue
and black \minorRevision{, \emph{(b)}}) to optimize a fitting energy.
This enables the adaptation to merge trees of typical applications of
Principal Component Analysis, such as
data
reduction, where the input trees are accurately reconstructed (\minorRevision{\emph{(c)}}, right),
by simply storing their MT-PGA coordinates,
or
dimensionality reduction.
MT-PGA enables the computation of \minorRevision{a} \emph{Principal Geodesic 
Surface} \minorRevision{\emph{(b)}},
which
complement\minorRevision{s} its planar layout \minorRevision{\emph{(g)}} by 
better conveying visually the curved
nature of $\branchtreeSpace$.
MT-PGA supports
the efficient reconstruction of user-defined locations,
for the
interactive exploration of $\branchtreeSpace$:
the reconstruction of the
purple curve \minorRevision{\emph{(f)}} enables the navigation from the trees of the
first cluster (\revision{dark red}, \minorRevision{\emph{(b)}}) to the second (\revision{orange}, \minorRevision{\emph{(b)}}) and third (\revision{pink},
\minorRevision{\emph{(b)}}) clusters.
MT-PGA also introduces
\emph{Persistence Correlation Views}
\revision{\emph{(d)}} which enable the visual identification of the features which are the most
responsible for the variability in the ensemble (high correlation,
near the disk boundary, \revision{\emph{(d)}}) as well as their direct inspection in the data
(matching colors \revision{\emph{(a)}}).
}
	\label{fig:teaser}
\end{figure*}


\IEEEPARstart{W}{ether}
they are acquired or simulated, modern datasets
are constantly gaining in 
detail and complexity, 
given
the continuous improvement of acquisition devices or computing resources.
This geometrical complexity is a difficulty for interactive data 
analysis and interpretation.
This observation
motivates the development of concise yet informative data representations, 
capable 
of encoding the main features of interest and visually representing them to the
users.
In that regard, 
Topological Data Analysis (TDA)
\cite{edelsbrunner09} has demonstrated 
its ability
to generically, robustly and efficiently reveal implicit structural
patterns hidden in complex datasets, in
particular in
support of analysis and visualization
tasks \cite{heine16}.
Examples of successful applications include
turbulent combustion \cite{laney_vis06, bremer_tvcg11, gyulassy_ev14},
 material sciences \cite{gyulassy_vis07, gyulassy_vis15, favelier16},
 nuclear energy \cite{beiNuclear16},
fluid dynamics \cite{kasten_tvcg11},
bioimaging \cite{carr04, topoAngler, beiBrain18},
quantum chemistry \cite{chemistry_vis14, harshChemistry, Malgorzata19} or
astrophysics \cite{sousbie11, shivashankar2016felix}.
Among the feature representations studied in TDA
(see \autoref{sec_relatedWork}), the merge tree \cite{carr00} 
(\autoref{fig_mergeTree}),
is a popular instance
in the visualization community \cite{carr04, bremer_tvcg11,
topoAngler}.

In many applications, 
on top of the increasing geometrical data
complexity, an additional challenge emerges, related to
\emph{ensemble datasets}. These describe a phenomenon
not only with a single dataset, but with a collection of datasets, called
\emph{ensemble members}, in order to characterize the
 variability of the phenomenon under study. 
 In principle,
a topological 
representation (like the merge tree) can
be computed for each ensemble member.
While this strategy has several
practical advantages (direct representations of the features of interest,
reduced memory footprint), it shifts the analysis problem from
an ensemble of datasets to an ensemble of merge trees. Then, a major challenge
consists in designing statistical tools for such an ensemble of topological
descriptors, to support its interactive analysis and interpretation. In this
direction, a series of recent works
focused on the notion of \emph{average
topological descriptor} \cite{Turner2014,
lacombe2018, vidal_vis19, YanWMGW20, pont_vis21}, with applications to ensemble
summarization and clustering. However, while such averages 
synthesize a topological descriptor which is well \emph{representative} of the
ensemble, they do not describe the topological variability of the ensemble.

This paper addresses this issue
and goes beyond simple averages by
adapting the celebrated framework of Principal Component Analysis (PCA)
\cite{pearson01}
to ensembles of merge trees\revision{. For that, we introduce the novel notion of \emph{``Merge-Tree Principal Geodesic Analysis''} (MT-PGA), which captures the most informative geodesics (i.e. analogs of straight lines
on the abstract space of merge trees) given the input ensemble, hence facilitating variability analysis and visualization}.
In particular, we formalize the computation of an orthogonal basis of principal
geodesics in the Wasserstein metric space of merge trees \cite{pont_vis21} as
a constrained
optimization problem (\autoref{sec_basis}), inspired by previous work on the
optimal transport of histograms \cite{SeguyC15, CazellesSBCP18}, which we extend
and specialize to merge trees. We introduce an efficient iterative algorithm
(\autoref{sec_algorithm}), which
exploits an analytic expression of the energy gradient
to ensure
fast
iterations. Moreover, we
document
accelerations with shared-memory parallelism. Extensive experiments
(\autoref{sec_results}) indicate that our algorithm
produces
bases of acceptable reconstruction quality within minutes, for
real-life
ensembles extracted from public benchmarks. We illustrate the utility of our
contribution in two applications. First, we show that the principal geodesic
bases computed by our algorithm
can result in
an important compression
of ensembles of merge trees, while still enabling a successful post-processing
for typical
visualization
tasks such as feature tracking or
ensemble
clustering.
Second, we present an extended application of our work to dimensionality
reduction, for the visual inspection of the ensemble variability via
two-dimensional
embeddings, where we show that the views generated by our approach preserve well
the intrinsic metric between merge trees, as well as the global structure of the
input ensembles.
Since our
framework is based on the Wasserstein distance between merge
trees \cite{pont_vis21}, which generalizes the
Wasserstein
distance between persistence diagrams \cite{Turner2014}, it trivially extends to
persistence diagrams by simply adjusting a parameter.

\subsection{Related work}
\label{sec_relatedWork}


The literature related to our work can be classified 
\revision{in}
three 
groups, reviewed in the following:
\emph{(i)} uncertainty visualization,
\emph{(ii)} ensemble visualization, and
\emph{(iii)} topological methods for ensembles.

\noindent
\textbf{Uncertainty visualization:}
Variability in data can be 
\revision{represented} in several ways.
\emph{Uncertain} datasets \cite{uncertainty_survey1,unertainty_survey2,
uncertainty1,
uncertainty2, uncertainty3, uncertainty4} capture variability by modeling each
point of the
domain as a random variable, whose variability is explicitly modeled by
a given probability density function (PDF).
Early techniques focused on
estimating the entropy of the random variables \cite{uncertainty_entropy},
their correlations \cite{uncertainty_correlation} or their gradient variations
\cite{uncertainty_gradient}.
The
positional uncertainty of level sets has been studied for several
interpolations
and PDF
models \cite{uncertainty_isosurface1,uncertainty_isosurface2,
    uncertainty_isosurface3, uncertainty_isosurface4, uncertainty_isocontour1,
uncertainty_nonparam,uncertainty_isocontour2, uncertainty_interp,athwale19}.
The positional uncertainty of critical points has been studied for
Gaussian \cite{liebmann1,otto1,otto2,petz} or uniform distributions
\cite{gunther,bhatia,szymczak}.
A general limitation of existing methods for uncertain data is their
dependence on the PDF model
for which they have been
specifically
designed.
This
reduces their usability for ensemble data, where the PDF estimated from
the ensemble members
can follow an arbitrary, unknown model. Also, most
techniques
do not consider multimodal PDFs,
which is however necessary when
multiple
trends are present in the
ensemble.

\noindent
\textbf{Ensemble visualization:}
Ensemble datasets encode data variability by directly modeling
empirical observations (i.e. the \emph{members} of the
ensemble).
Current approaches to ensemble visualization
typically compute some geometrical objects describing the features of interest
(level sets, streamlines, etc), for each member of the ensemble. Then, an
aggregation phase estimates a \emph{representative} object for the resulting
ensemble of geometrical objects. For instance, spaghetti plots
\cite{spaghettiPlot}
\minorRevision{can encode}
level-set variability,
especially for weather data \cite{Potter2009,Sanyal2010}. More specifically,
box-plots \cite{whitaker2013} describe the variability of contours and curves
\cite{Mirzargar2014}. For flow ensembles, Hummel et al. \cite{Hummel2013}
introduce a Lagrangian framework for classification purposes. Clustering
techniques have been introduced, to identify the main trends, and
their variability, in ensembles of streamlines \cite{Ferstl2016} and
isocontours \cite{Ferstl2016b}.

Regarding topological features,
Favelier et al.
\cite{favelier_vis18} and Athawale et al. \cite{athawale_tvcg19} introduced
approaches for visualizing the variability of critical points and
gradient separatrices.
Lohfink et al.
\cite{LohfinkWLWG20} introduced an approach for the consistent planar layout of
multiple contour trees, to support effective visual comparisons between the
contour
trees of distinct members.
Although the above techniques addressed the visualization of ensembles
of
topological objects, they did not focus explicitly on their
statistical analysis.

Principal Component Analysis (PCA) \cite{pearson01} is a classical approach for
the
analysis of variability in vectorized data (i.e. point clouds in Euclidean
spaces). 
Extensions have been
investigated for metric spaces \cite{gorban08, gorban09, FletcherLPJ04},
including  
transport based
distances \cite{cuturi2013}
between histograms
\cite{SeguyC15, CazellesSBCP18}. However, these methods are not directly
applicable to merge trees.
\revision{They focus on fundamentally different objects (histograms). Thus, 
their distances, geodesics and barycenters are defined differently (in 
particular in an entropic form \cite{cuturi2013, cuturi2014}) and the 
algorithms for their computations are drastically different (based on Sinkhorn 
matrix scaling \cite{Sinkhorn}).}
Our global strategy (also based, at a high level, on an
alternation of fitting and constraint enforcement, \autoref{sec_basis})
can be interpreted as an extension
of
this line of work, but we
revisit it completely,
to specialize it to merge trees.

\noindent
\textbf{Topological methods:} Concepts and algorithms from
computational
topology \cite{edelsbrunner09} have been investigated, adapted and extended by
the visualization community 
\cite{heine16, surveyComparison2021}.
Popular topological representations in data visualization include the
persistence diagram
\cite{edelsbrunner02, edelsbrunner09, dipha}
(\autoref{sec_background_persistenceDiagrams}),
the Reeb graph \cite{biasotti08, pascucci07, tierny_vis09, parsa12,
DoraiswamyN13, gueunet_egpgv19} and its variants
the
merge (\autoref{fig_mergeTree})
and contour trees \cite{tarasov98, carr00,
MaadasamyDN12, AcharyaN15, CarrWSA16, gueunet_tpds19},
or the
Morse-Smale complex \cite{forman98, EdelsbrunnerHZ01,
EdelsbrunnerHNP03,
BremerEHP03, Defl15, robins_pami11, ShivashankarN12, gyulassy_vis18}.

As detailed in \autoref{sec_basis}, the development of a computational
framework for Principal Geodesic Analysis (PGA) over an ensemble of topological
objects requires several key, low-level, geometrical ingredients, namely
\emph{(i)} a distance metric (to measure distances between objects),
\emph{(ii)} a geodesic estimation routine (to model the axes of the PGA
basis) and \emph{(iii)} a barycenter estimation routine (to compute the origin
of the PGA  basis). We review now the previous work related to these
aspects.

Distance
metrics
have been studied for most of the above topological objects. Inspired by the
literature in
optimal transport \cite{Kantorovich, monge81}, the Wasserstein distance between
persistence diagrams \cite{edelsbrunner09}
(\autoref{sec_background_persistenceDiagrams})
has
been extensively studied. It is
based on a bipartite assignment problem,
for which exact
\cite{Munkres1957} and approximate \cite{Bertsekas81, Kerber2016}
implementations are publicly available \cite{ttk17}.
However, the persistence diagram can lack specificity in its data
characterization and more advanced topological descriptors, such as
merge trees (\autoref{sec_background_mergeTrees}), are often reported to
differentiate datasets better
\cite{morozov14, BeketayevYMWH14,
SridharamurthyM20, pont_vis21}.
Several similarity
measures have been introduced for Reeb graphs \cite{HilagaSKK01} and their
variants \cite{SaikiaSW14_branch_decomposition_comparison}.
However, since these measures are not  metrics (the
preservation of the triangle inequality is not specifically enforced), they
\revision{are not}
conducive to the computation of geodesics.
Stable distance metrics between Reeb graphs \cite{bauer14} and merge trees
\cite{morozov14} have been studied from a theoretical point of view
\cite{bollen21}
but their
computation, following an exponential time complexity, is not tractable for
practical datasets in general.
Distances with polynomial
time computation algorithms have also been investigated.
Beketayev et al.
\cite{BeketayevYMWH14}
focus on
the \emph{branch decomposition tree}
(BDT, \autoref{sec_background_mergeTrees}),
and
 estimate their distances
by iteratively reducing a target mismatch term over a significantly large
search space.
Sridharamurthy et al.
\cite{SridharamurthyM20} specialize efficient algorithms for computing
constrained edit distances between trees \cite{zhang96} to the
special
case of
merge trees, resulting in a distance
which is computable for
real-life datasets and
with acceptable practical stability.
Based on this edit distance, a generalization of the $L_2$-Wasserstein distance
between persistence diagrams \cite{Turner2014} has been introduced for merge
trees \cite{pont_vis21},
thereby enabling the efficient
computation of distances, geodesics and barycenters of
merge trees.


Regarding the estimation of a \emph{representative} object from a set of
topological
representations, multiple approaches emerged recently.
Several
methods
\cite{Turner2014,
lacombe2018, vidal_vis19} have been introduced for the
estimation of barycenters
of persistence diagrams (or
vectorized variants
\cite{Adams2015, Bubenik15}).
A recent
work \cite{YanWMGW20}
introduced a framework for computing
a \emph{1-center} of a set of
merge trees
(i.e.
minimizing its maximum distance to the set),
for
an interleaving distance \cite{intrinsicMTdistance}.
However,
this
approach requires pre-existing, reliable correspondence labels between the
nodes of
the input trees, which is not practical for real-life datasets
(heuristics need to be considered).
Also,
the resulting representative merge tree is not a barycenter (it
does not minimize a Fr\'echet energy, i.e. a sum of distances to the set).
Thus, it cannot be used directly for PGA. In contrast, the barycentric
framework of Pont et al. \cite{pont_vis21} automatically minimizes a Fr\'echet
energy explicitly. Thus, the resulting barycenter merge tree can be
directly used for PGA.
Another line of approaches
aimed at directly applying the classical  PCA (or variants from the
matrix sketching literature \cite{woodRuff2014}) to \emph{vectorizations} of
topological descriptors
\cite{robins16, AnirudhVRT16, LI2021},
i.e. by first converting each
topological descriptor into a (high-dimensional) Euclidean vector and
then
leveraging traditional tools from linear algebra (e.g.
classical PCA) on these vectors.
However, vectorizations are in general subject to a number of limitations.
They
are prone to approximation errors
(due to quantization and linearization artifacts),
they can be difficult to revert
(which challenges their usage for visualization applications) and their
stability is not always established.
%
In contrast,
our approach \emph{directly} manipulates merge trees in the Wasserstein metric
space, and not
linear approximations
in a Euclidean space.
This results in a faithful 
formulation of merge tree PGA, which ensures improved accuracy and 
interpretability (cf.
\minorRevision{experiments},
\autoref{sec_quality}).

\subsection{Contributions}

This paper makes the following new contributions:
\begin{enumerate}
 \item{\emph{An approach to Principal Geodesic Analysis of Merge
Trees (MT-PGA):} We formulate
the definition
of an orthogonal basis
of 
principal
geodesics
in the Wasserstein metric space of merge trees
\cite{pont_vis21} as a constrained optimization problem. Our formulation
(\autoref{sec_basis})
extends
previous work on 
histograms \cite{SeguyC15, CazellesSBCP18}
and specializes it to
merge trees.}
 \item{\emph{An optimization algorithm for MT-PGA:}
 We introduce an efficient optimization algorithm
(\autoref{sec_algorithm}) for MT-PGA. Each iteration alternates between
\emph{(i)} a minimization of the
fitting energy of
the MT-PGA basis
and \emph{(ii)} a
constraint enforcement.
\revision{Our} algorithm
exploits an analytic expression of the
fitting energy gradient,
to
ensure fast iterations. We document accelerations based on
shared-memory parallelism and report running times in the orders of minutes for
real-life ensembles.}
 \item{\emph{An application to data reduction:} We present an application to
data reduction (\autoref{sec_dataReduction}), where the merge trees of the input
ensemble are significantly compressed, by solely storing the MT-PGA basis
\revision{and}
the coordinates of the input merge trees in the
basis\revision{. We illustrate the utility of this reduction with applications to feature tracking and ensemble clustering}.
}
 \item{\emph{An application to dimensionality reduction:} We present an
application to dimensionality reduction (\autoref{sec_dimensionalityReduction}),
by embedding each merge
tree as a point in a planar view, based on its first two coordinates in the
MT-PGA basis.
\revision{We also contribute derived visualizations -- the
\emph{Principal Geodesic Surface} and the \emph{Persistence Correlation
View} --  which enable the visual inspection of the individual features which are the most responsible for the variability in the ensemble.}
}
 \item\emph{Implementation:} We provide a 
 C++ implementation of our
algorithms that can be used for reproduction purposes:
\revision{\href{https://github.com/MatPont/MT-PGA}{https://github.com/MatPont/MT-PGA}}
\end{enumerate}


\section{Preliminaries}
\label{sec_preliminaries}


\noindent
This section presents the theoretical background of our work. It contains
definitions adapted from the Topology ToolKit \cite{ttk17}.
\revision{It is structured as follows. First, we formalize the input data (\autoref{sec_inputData}).
Second, we introduce a first topological data representation -- the persistence diagram (\autoref{sec_background_persistenceDiagrams}) -- which is closely related to the main representation studied in this paper -- the merge tree (\autoref{sec_background_mergeTrees}).
Finally, we describe geometrical tools (e.g. metrics, geodesics, barycenters) in the space of merge trees (\autoref{sec_metricSpace}), which have been recently generalized from the study of persistence diagrams \cite{pont_vis21}.
These geometrical tools will act as building blocks in our formulation of MT-PGA (\autoref{sec_basis}).
}
\minorRevision{Tables of abbreviations and notations are provided in Appendix A.}
We refer the
reader to textbooks \cite{edelsbrunner09} for 
an
introduction to
computational topology.


\subsection{Input data}
\label{sec_inputData}
The input data is an ensemble of $\ensembleSize$ piecewise linear (PL) scalar
fields
$f_i : \domain \rightarrow \range$, with $i \in \{1, \dots,  \ensembleSize\}$,
defined on a
PL $d$-manifold $\domain$, with
$d\leq3$ in our applications. The \emph{sub-level set} of $f_i$, noted
$\sublevelset{{f_i}}(\isovalue)=\{p \in \domain~|
~f_i(p) < \isovalue\}$, is defined as the pre-image of  $(-\infty, \isovalue)$
by
$f_i$.
The \emph{super-level set} of $f_i$ is defined symmetrically:
$\superlevelset{{f_i}}(\isovalue)=\{p \in \domain~|
~f_i(p) > \isovalue\}$.
As $\isovalue$ continuously increases, the topology of
$\sublevelset{{f_i}}(\isovalue)$ changes at specific vertices of $\domain$,
called the \emph{critical points} of $f_i$ \cite{banchoff70}.
\revision{Critical points are classified by their \emph{index}
$\Index_i$: $0$ for minima, $1$ for $1$-saddles, $d-1$ for $(d-1)$-saddles and
$d$ for maxima.
In practice, $f_i$ is enforced to contain only isolated, non-degenerate
critical points \cite{edelsbrunner90, EdelsbrunnerHZ01}.
}

\begin{figure}
\centering
\includegraphics[width=\linewidth]{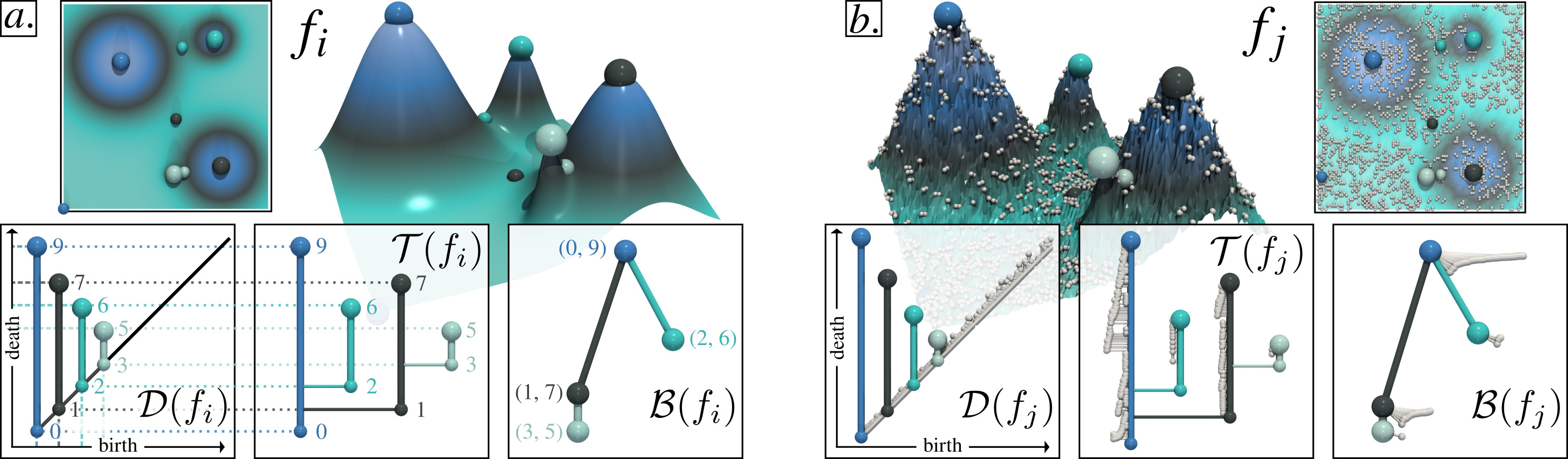}
\mycaption{
Critical points (spheres, larger radius: maxima),
persistence diagram (left \revision{inset}), merge tree (center \revision{inset}) and branch
decomposition tree (right \revision{inset}) of a clean \revision{\emph{(a)}} and noisy \revision{\emph{(b)}}
scalar field.
In both cases, four main hills are clearly
represented with salient features in the persistence diagram and the
merge tree.
Branches with low persistence (less than $10\%$ of the function range) are
shown with small white arcs.
}
\label{fig_mergeTree}
\end{figure}

\subsection{Persistence diagrams}
\label{sec_background_persistenceDiagrams}
The persistence diagram is a visual summary of the topological features (i.e.
connected components, independent cycles, voids) of
$\sublevelset{{f_i}}(\isovalue)$. 
\revision{As shown in \autoref{fig_mergeTree}, it is closely related to the \emph{merge tree}, which is the main topological data representation studied in this paper. We first describe the persistence diagram however, as the metric used in our work to measure distances between merge trees (\autoref{sec_metricSpace}) is a generalization of an established metric between persistence diagrams.}
\revision{Specifically, in the domain, each} topological feature of
$\sublevelset{{f_i}}(\isovalue)$ can be associated with a unique pair of
critical points $(c, c')$, corresponding to its \emph{birth} and \emph{death}.
The Elder rule \cite{edelsbrunner09} states that critical points can be
arranged
in pairs
according to this observation,
such that each
critical point appears in only one pair $(c, c')$, with $f_i(c) < f_i(c')$ and
$\Index_i(c) = \Index_i(c') - 1$.
For instance, if two connected components of $\sublevelset{{f_i}}(\isovalue)$
meet at a critical point $c'$, the \emph{younger} component
(created
last, in $c$) \emph{dies}, in favor of the \emph{older} one (created
first).
The
persistence diagram $\diagram(f_i)$ embeds each pair to a 
single
point in 2D at coordinates $\big(f_i(c), f_i(c')\big)$.
The \emph{persistence} of a pair
is given by
its height $f_i(c') - f_i(c)$.
The
persistence diagram provides 
a visual 
overview of the features
of a dataset
(\autoref{fig_mergeTree}), where salient features stand out
from the diagonal while pairs corresponding to noise are located
near
the diagonal.



\subsection{Merge trees}
\label{sec_background_mergeTrees}
\revision{In the following, we introduce the main topological data representation studied in this paper: the \emph{merge tree}. We also describe a specific representation of the merge tree called the \emph{branch decomposition tree},
which can be interpreted as a generalization of the extremum persistence diagram, and
which plays a central role in the computation of distances between merge trees (\autoref{sec_metricSpace}).}

The \emph{join} tree, noted $\jointree(f_i)$, is a visual summary of the
connected components of $\sublevelset{{f_i}}(\isovalue)$ \cite{carr00}. It is
a 1-dimensional simplicial complex defined as the quotient space
$\jointree(f_i) = \domain / \sim$ by the equivalence relation $\sim$
which states that
$p_1$ and $p_2$ are equivalent
if
$f_i(p_1) = f_i(p_2)$ and if $p_1$ and $p_2$ belong to the same connected
component
of $\sublevelset{{f_i}}\big(f_i(p_1)\big)$.

The \emph{split} tree (\autoref{fig_mergeTree}), noted $\splittree(f_i)$, is
defined symmetrically and describes the connected components of the super-level
set $\superlevelset{{f_i}}(\isovalue)$. Each of these two \emph{directed} trees
is called a
\emph{merge} tree (MT), noted generically $\mergetree(f_i)$ in the following.
Intuitively, these trees track the creation of connected components of
the sub
(or super) level sets at their leaves, and merge events at their
interior nodes.
To mitigate a phenomenon called \emph{saddle swap}, these trees are often
post-processed \cite{SridharamurthyM20, pont_vis21}, by merging adjacent
saddles in the tree if their relative difference in scalar value is smaller
than a threshold $\epsilon_1 \in [0, 1]$.
MTs
are often
visualized according to a persistence-driven \emph{branch decomposition}
\cite{pascucci_mr04}, to make the persistence pairs captured by the tree stand
out. In this context, a \emph{persistent branch} is a monotone path on the tree
connecting the nodes corresponding to the creation and destruction (according
to the Elder rule, \autoref{sec_background_persistenceDiagrams})
of a connected component of sub (or super) level set. Then, the branch
decomposition provides a planar layout of the MT,
where each
persistent
branch is
represented as a vertical segment \revision{(center insets in \autoref{fig_mergeTree})}.

The \emph{branch decomposition tree} (BDT),
noted
$\branchtree(f_i)$, is a directed tree
whose nodes are the persistent branches
captured
by the branch decomposition
and whose arcs denote adjacency relations between them in the
MT.
In \autoref{fig_mergeTree}, the BDTs (right
insets) can be interpreted as the dual of the branch decompositions
(center insets, with matching colors): each vertical segment in the
branch decomposition (center) corresponds to a node in the BDT (right) and each
horizontal segment (center, denoting an adjacency relation between branches)
corresponds to an arc in the BDT.
The BDT can be interpreted as a generalization of the extremum persistence
diagram:
like $\diagram(f_i)$, $\branchtree(f_i)$
describes the
population of (extremum) persistence pairs present in the data. However,
unlike the persistence diagram, it additionally captures adjacency relations
between them \revision{(\autoref{fig_mergeTree})}.
Note that, the birth and death
of each persistent branch $b_i \in \branchtree(f_i)$, noted $(x_i, y_i)$, span
by construction an interval included in that of its parent $b_i' \in
\branchtree(f_i)$: $[x_i, y_i] \subseteq [x_i', y_i']$. This \emph{nesting
property} of BDTs \cite{pont_vis21} is a direct consequence of the Elder rule
(\autoref{sec_background_persistenceDiagrams})\revision{, and it plays an important role in the computation of geodesics
between
MTs
(\autoref{sec_metricSpace})}.

\cutout{As discussed in \autoref{sec_relatedWork}, several metrics have been
introduced
for measuring distances between merge trees. We focus in the remainder on the
edit distance introduced by Sridharamurthy et al. \cite{SridharamurthyM20}, as
its balance between computability and
acceptable
stability in practice seems particularly
promising. The edit distance between two merge trees $\mergetree(f_i)$ and
$\mergetree(f_j)$, noted $\editdistance\big(\mergetree(f_i),
\mergetree(f_j)\big)$, is defined as follows.
Let $N_i$ be a subset of the nodes of $\mergetree(f_i)$ and $\overline{N_i} $
its complement. Let $\phi'$ be a \emph{partial} assignment
between $N_i$ and a subset $N_j$ of the nodes of
$\mergetree(f_j)$ (with complement $\overline{N_j}$).
Then $\editdistance\big(\mergetree(f_i),
\mergetree(f_j)\big)$ is given by:
\begin{eqnarray}
\label{eq_edit_distance}
 \editdistance\big(\mergetree(f_i), \mergetree(f_j)\big) =
  & \underset{(\phi', \overline{N_i}, \overline{N_j}) \in \Phi'}\min&  \Big(
    \label{eq_edit_distance_mapping}
    \sum_{n_i \in N_i} \gamma\big(n_i \rightarrow \phi'(n_i)\big)\\
    \label{eq_edit_distance_destroying}
    & + &  \sum_{n_i \in \overline{N_i}} \gamma(n_i \rightarrow \emptyset)\\
    \label{eq_edit_distance_creating}
    & + & \sum_{n_j \in \overline{N_j}} \gamma(\emptyset \rightarrow n_j)
  \Big)
  \label{eq_editDistance}
\end{eqnarray}
%
%
%
where $\Phi'$ is the space of \emph{constrained} partial assignments (i.e.
$\phi'$ maps disjoint subtrees of $\mergetree(f_i)$ to disjoint subtrees of
$\mergetree(f_j)$) and where $\gamma$ refers to the cost
for:
\emph{(i)} mapping a node $n_i \in \mergetree(f_i)$ to a node $\phi'(n_i) = n_j
\in \mergetree(f_j)$ (line \ref{eq_edit_distance_mapping}),
\emph{(ii)} deleting a node
$n_i \in \mergetree(f_i)$ (line \ref{eq_edit_distance_destroying}) and
\emph{(iii)} creating
a node $n_j \in \mergetree(f_j)$ (line \ref{eq_edit_distance_creating}),
$\emptyset$
being the empty tree.}
%

%

\cutout{Zhang \cite{zhang96} introduced a polynomial time algorithm for
computing a
constrained sequence of edit operations with minimal edit distance
(\autoref{eq_edit_distance_mapping}), and showed that the resulting distance is
indeed a
metric if each cost $\gamma$ for the above three edit operations is itself
a metric (non-negativity, identity, symmetry, triangle inequality).
Sridharamurthy et al. \cite{SridharamurthyM20} exploited this property to
introduce their metric,
by defining a distance-based cost model, inspired by the Wasserstein distance
between persistence diagrams (\autoref{sec_jackground_persistenceDiagrams}),
where $p_i$ and $p_j$ stand for the persistence pairs \emph{containing} the
nodes $n_i \in \mergetree(f_i)$ and $n_j \in \mergetree(f_j)$:
%
\begin{eqnarray}
  \gamma(n_i \rightarrow n_j) & = & \min\big(\pointMetric_\infty(p_i, p_j),
    \gamma(n_i \rightarrow \emptyset) + \gamma(\emptyset \rightarrow n_j)\big)\\
  \gamma(n_i \rightarrow \emptyset) & = & \pointMetric_\infty\big(p_i,
\projection(p_i)\big)\\
  \gamma(\emptyset \rightarrow n_j) & = & \pointMetric_\infty\big(p_j,
\projection(p_j)\big)
\end{eqnarray}
In our work, as described next, we introduce an alternative
edit distance
which adheres even further to the Wasserstein distance between persistence
diagrams,
to ease geodesic computation for merge
trees.}



\begin{figure}
\includegraphics[width=\linewidth]{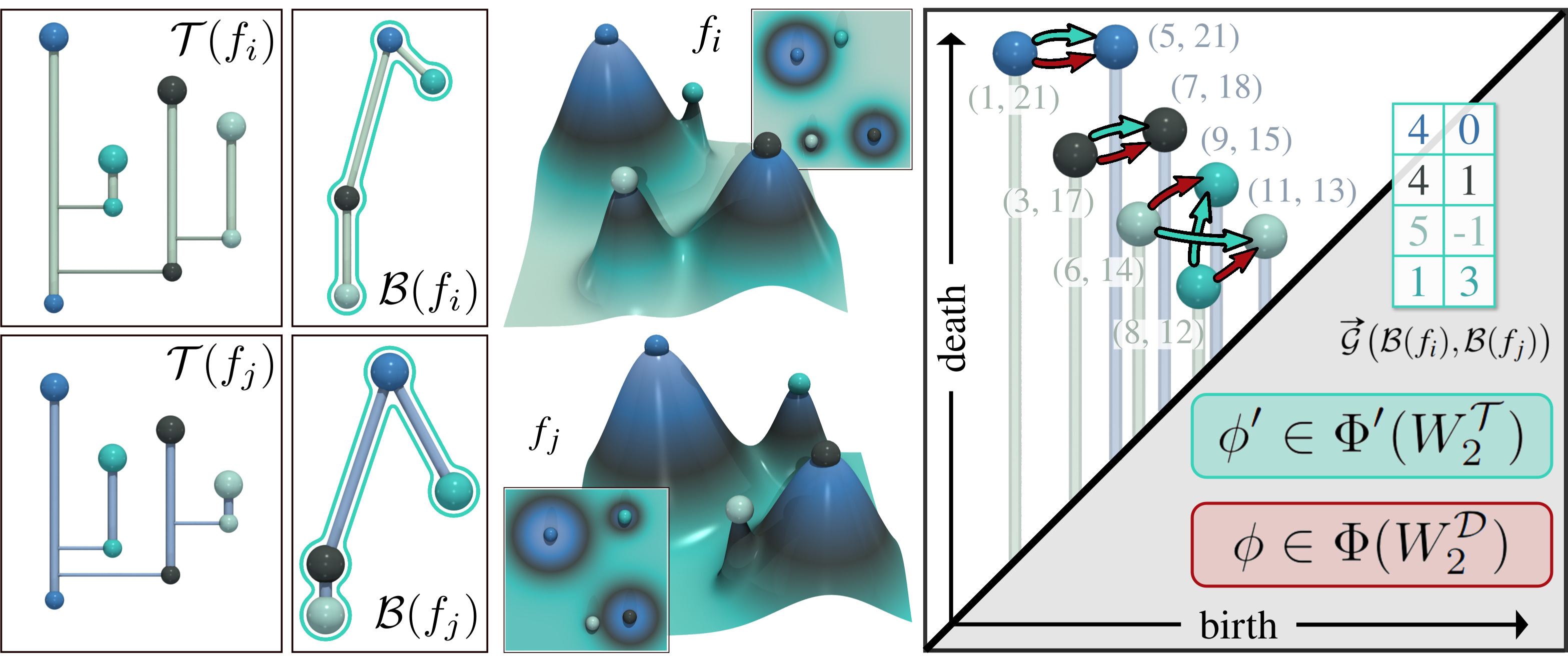}
\mycaption{The Wasserstein distance $\wassersteinTree$ between the BDTs
$\branchtree(f_i)$ (white) and $\branchtree(f_j)$ (blue) is computed by
\revision{assignment optimization}
(\autoref{eq_wasserstein}) in the 2D 
birth/death plane (cyan
arrows, right), given a search space of assignments  $\Phi'$ describing partial
rooted isomorphisms between $\branchtree(f_i)$ and $\branchtree(f_j)$
(\minorRevision{an example is shown with a} cyan halo
on the BDTs). For $\epsilon_1 = 1$, the structure of the BDTs is 
ignored and $\wassersteinTree$ becomes equivalent to the
Wasserstein
distance between persistence diagrams $\wasserstein{2}$ (red arrows), which
\minorRevision{yields}
a small distance
\minorRevision{here}
despite a significant structural
change in the data (two hills have been swapped from $f_i$ to $f_j$).}
\vspace{1.5ex}
\label{fig_wassersteinMetric}
\end{figure}

\subsection{Wasserstein metric space}
\label{sec_metricSpace}
\revision{In the following, we introduce an established metric between persistence diagrams (the Wasserstein distance) which has been recently generalized to merge trees \cite{pont_vis21}. This generalization allows our PGA framework to support both persistence diagrams and merge trees
(as discussed next).
After introducing this metric, we detail further derived concepts, such as the notions of geodesics and Wasserstein barycenters, which will act as building blocks in our formulation of MT-PGA (\autoref{sec_basis}).
}

To measure the distance between two diagrams $\diagram(f_i)$ and
$\diagram(f_j)$, a typical pre-processing 
step 
consists in augmenting
each
diagram with the diagonal projection of the off-diagonal points of the other
diagram:
\begin{eqnarray}
  \nonumber
 \diagram'(f_i) = \diagram(f_i) \cup \{
\projection(p_j) ~ | ~ p_j \in \diagram(f_j)
\}\\
\nonumber
  \diagram'(f_j) = \diagram(f_j) \cup \{
\projection(p_i) ~ | ~ p_i \in \diagram(f_i)
\},
\end{eqnarray}

\noindent
where $\projection(p_i) =
(\frac{x_i+y_i}{2},\frac{x_i+y_i}{2})$ stands for the diagonal projection of
the off-diagonal point $p_i = (x_i, y_i) \in \diagram(f_i)$. Intuitively, this
augmentation phase inserts dummy features in the diagram (with zero
persistence, along the diagonal), hence preserving the topological information
of the diagrams. This augmentation guarantees that the two
diagrams now have the same number of points ($| \diagram'(f_i)| = |
\diagram'(f_j)|$), which facilitates the evaluation of their distance, as
described next.

Given two points $p_i = (x_i, y_i) \in
\diagram'(f_i)$ and $p_j = (x_j, y_j) \in \diagram'(f_j)$, the ground
distance $\pointMetric_q$ ($q > 0$) in the 2D birth/death space is given by:
\begin{eqnarray}
  \nonumber
\label{eq_pointMetric}
\pointMetric_{q}(p_i, p_j) = (|x_j - x_i|^q + |y_j - y_i|^q)^{1/q}
= \| p_i - p_j
\|_q.
\end{eqnarray}

\noindent
By convention,
$\pointMetric_q(p_i, p_j)$ is set to zero between diagonal points
($x_i = y_i$ and $x_j = y_j$).
 Then, the
$L^q$-Wasserstein
distance
$\wasserstein{q}$
is:
\begin{eqnarray}
\label{eq_wasserstein}
\wasserstein{q}\big(\diagram'(f_i), \diagram'(f_j)\big)
=
\label{eq_wasserstein_mapping}
&
\hspace{-.15cm}
\underset{\phi \in \Phi}\min
 \hspace{-.15cm}
&
\Big(
\sum_{p_i \in \diagram'(f_i)}
\pointMetric_q\big(p_i,
\phi(p_i)\big)^q\Big)^{1/q},
\end{eqnarray}

\noindent
where $\Phi$ is the set of all possible assignments $\phi$ mapping a point $p_i
\in \diagram'(f_i)$ to a point $p_j \in \diagram'(f_j)$ (possibly its diagonal
projection, indicating the destruction of the corresponding feature).

A variant of this metric, noted
$\wassersteinTree\big(\branchtree(f_i), \branchtree(f_j)\big)$, has been
introduced recently for BDTs \cite{pont_vis21}.
Its
expression is identical to \autoref{eq_wasserstein} (for $q = 2$), at the
notable exception of the search space of possible assignments, noted $\Phi'
\subseteq \Phi$, constrained to describe (rooted) partial isomorphisms \cite{pont_vis21} between
$\branchtree(f_i)$ and $\branchtree(f_j)$
\revision{(cyan halo on the BDTs of \autoref{fig_wassersteinMetric})}.
Intuitively,
$\wassersteinTree$ can be understood as a variant of $\wasserstein{2}$, which
takes into account the structures of the BDTs when evaluating candidate
assignments in the optimization of \autoref{eq_wasserstein}.
Since $\Phi'
\subseteq \Phi$, we have:
$\wassersteinTree\big(\branchtree(f_i), \branchtree(f_j)\big) \geq
\wasserstein{2}\big(\diagram'(f_i), \diagram'(f_j)\big)$. In other words,
$\wassersteinTree$ is more discriminative than $\wasserstein{2}$. In
the special case where the parameter $\epsilon_1$
(\autoref{sec_background_mergeTrees}) equals $1$ (all the saddles in the
MTs are merged), we have: $\wassersteinTree\big(\branchtree(f_i),
\branchtree(f_j)\big) =
\wasserstein{2}\big(\diagram'(f_i), \diagram'(f_j)\big)$. In other words,
$\wassersteinTree$ generalizes $\wasserstein{2}$ and $\epsilon_1$ acts as a
control parameter, which adjusts the importance of the structure of the BDTs in
the metric.
\revision{Then, our PGA framework will be able to support both persistence diagrams and merge trees, thanks to this metric generalization.}
Unless specified otherwise, we assume in the following that
$\epsilon_1$ is set to the default recommended value:
$\epsilon_1 = 0.05$ \cite{pont_vis21}.
In the remainder, we note $\branchtreeSpace$ the metric space induced
by the Wasserstein metric between BDTs.

\revision{Once a metric is established, we can now introduce the derived notions of geodesics and Wasserstein barycenters, which act as building blocks in our formulation of MT-PGA (\autoref{sec_basis}).}

A \emph{geodesic} (i.e.
length minimizing path on $\branchtreeSpace$) between two BDTs
$\branchtree(f_i)$ and
$\branchtree(f_j)$, noted $\vectorNotation{\geodesictree}\big(\branchtree(f_i),
\branchtree(f_j)\big)$, is simply given by linearly interpolating the
optimal assignment $\phi' \in \Phi'$ \cite{pont_vis21}.
Technically, it
is
a vector in $\mathbb{R}^{2 \times
|\branchtree(f_i)|}$ \minorRevision{(\autoref{fig_wassersteinMetric}, right, cyan array)}, obtained by concatenating the $|\branchtree(f_i)|$
\minorRevision{2D}
vectors
\minorRevision{(cyan arrows)}
representing the optimal assignment $\phi' \in \Phi'$ in the 2D birth/death
space.
\autoref{fig_wassersteinMetric} (right) illustrates two geodesics:
one with respect to $\wasserstein{2}$ ($\epsilon_1 = 1$, red arrows), the
other with
respect to $\wassersteinTree$ ($\epsilon_1 = 0$, cyan arrows).
Note that an arbitrary vector in $\mathbb{R}^{2\times
|\branchtree(f_i)|}$ represents an arbitrary
\revision{displacement in $\branchtreeSpace$,}
which is
consequently
not
necessarily optimal
%
with regard to \autoref{eq_wasserstein}\revision{, and which therefore, does not necessarily represent a geodesic}.
To
guarantee the invertibility of the interpolated BDTs into valid MTs,
a local normalization, turning $\branchtree(f_i)$ into
$\normalizedLocation\big(\branchtree(f_i)\big)$,  needs to be introduced in a
pre-processing step \cite{pont_vis21}, where the persistence of each branch $b_i
\in \branchtree(f_i)$ is normalized with regard to that of its parent $b'_i \in
\branchtree(f_i)$, displacing $b_i$ in the 2D birth/\minorRevision{death} space from $(x_i, y_i)$ to
$\normalizedLocation(b_i) = \big(\normalizedLocation_x(b_i),
\normalizedLocation_y(b_i)\big)$:
\begin{eqnarray}
\begin{array}{c}
\normalizedLocation_x(b_i) = (x_i - x'_i) / (y'_i - x'_i)\\
\vspace{-1.5ex}\\
\normalizedLocation_y(b_i) = (y_i - x'_i) / (y'_i - x'_i).
\end{array}
\label{eq_localNormalization}
\end{eqnarray}

\noindent
After this pre-normalization, any interpolated BDT is reverted into a
valid MT by recursively reverting \autoref{eq_localNormalization},
hence constructively
enforcing
the \emph{nesting property} of BDTs ($[x_i, y_i] \subseteq
[x'_i, y'_i]$, \autoref{sec_background_mergeTrees}) thereby guaranteeing the
validity of the interpolated MT. Pont et al. \cite{pont_vis21} introduce two
parameters to tune the effect of this
pre-normalization \revision{($\epsilon_2$ balances the normalized persistence of small branches, selected via the threshold $\epsilon_3$). We set these to their} default recommended values
($\epsilon_2 = 0.95$, $\epsilon_3 = 0.9$). In the remainder, we consider that
all the input BDTs are normalized this way.

Once distances and geodesics for BDTs are available, the
notion of \emph{barycenter} can be introduced.
Given a set $\branchtreeSet = \{\branchtree(f_1),
\dots, \branchtree(f_\ensembleSize)\}$,
let $\frechetEnergy(\branchtree)$ be the Fr\'echet energy of
a candidate BDT $\branchtree \in \branchtreeSpace$:
\begin{eqnarray}
\label{eq_frechet_energy}
\frechetEnergy(\branchtree) =
 \sum_{i = 1}^{\ensembleSize}
        \wassersteinTree\big(\branchtree,
          \branchtree(f_i)\big)^2.
\end{eqnarray}

\noindent
Then a BDT $\branchtree^*  \in \branchtreeSpace$
which minimizes
$\frechetEnergy$ is called a \emph{Wasserstein
barycenter} of the set $\branchtreeSet$ (or its Fr\'echet mean under the metric
$\wassersteinTree$). \autoref{eq_frechet_energy} can be optimized with an
iterative algorithm \cite{pont_vis21}, alternating assignment and update
phases. After optimization, a barycenter MT can be obtained
by recursively reverting
\autoref{eq_localNormalization}.

%
%
%


\section{Formulation}
\label{sec_basis}


\noindent
This section describes
our novel extension of the Principal Component
Analysis (PCA) framework to the Wasserstein metric space of merge trees, 
leading to 
\revision{the new
notion of}
merge tree Principal Geodesic Analysis (MT-PGA). 
\revision{This section is structured as follows. First, we describe a geometric interpretation of PCA (\autoref{sec_geometricPCA}), which our approach extends.
Next, we describe how to generalize each low-level geometrical tool used in PCA (\autoref{sec_geometricPCA}) to the
Wasserstein
metric space of merge trees
$\branchtreeSpace$
(\autoref{sec_defMtPga}). Finally, once these tools are available, we formalize our notion of MT-PGA  as a constrained optimization problem (\autoref{sec_detailedFormulation}), for which we provide an algorithm in \autoref{sec_algorithm} (an overview of the algorithm is given in \autoref{sec_overview}).}
\minorRevision{Tables of abbreviations and notations are provided in Appendix A.}


\begin{figure}
\centering
\includegraphics[width=\linewidth]{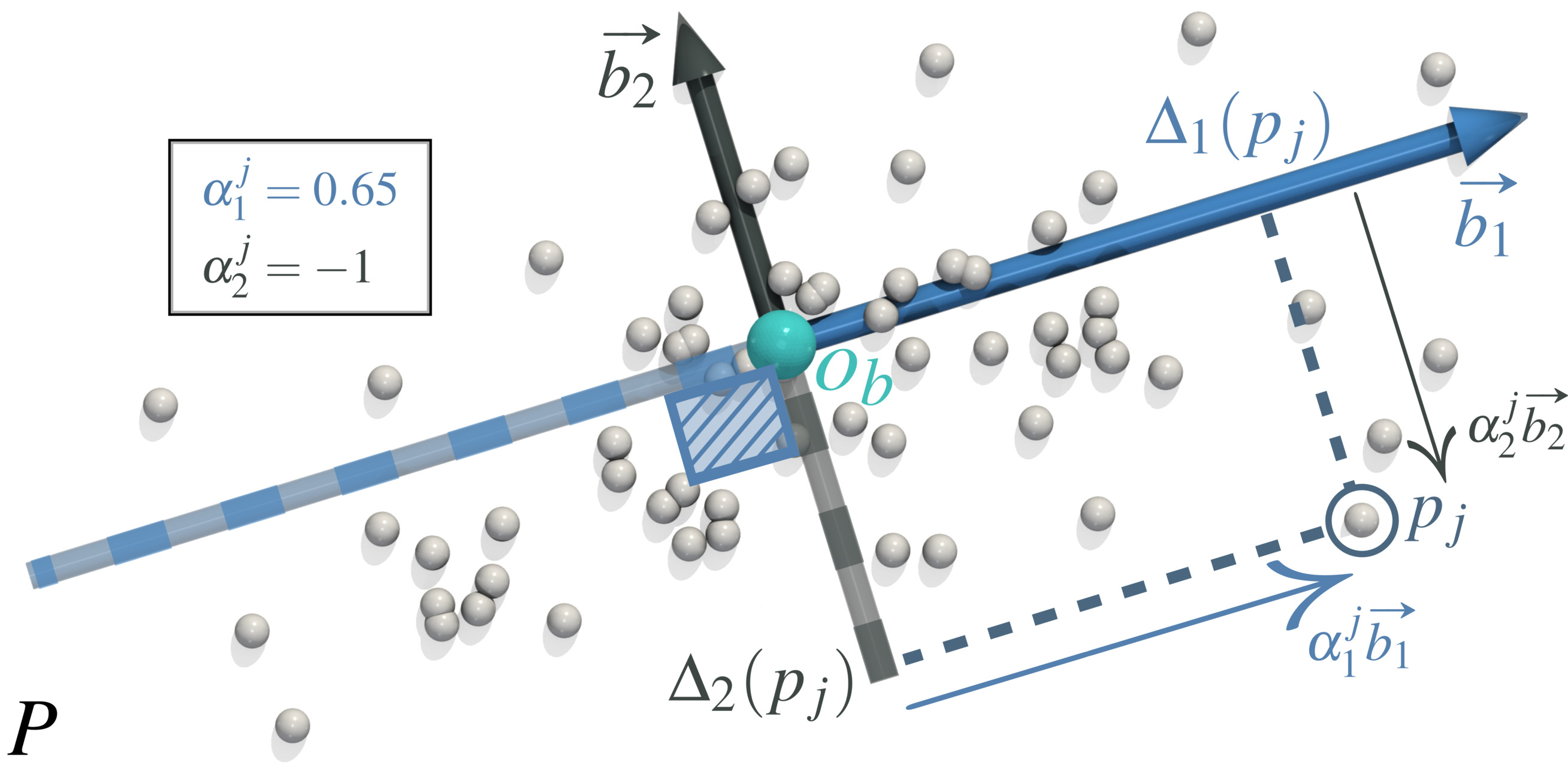}
\mycaption{The classical Principal Component Analysis (PCA) of a point
cloud $P$  in
$\mathbb{R}^d$ (white spheres) can be computed
by iteratively optimizing the fitting to $P$ of orthogonal directions (blue
axis first, then black axis).
}
\label{fig_classicalPCA}
\end{figure}

\begin{figure*}
\includegraphics[width=\linewidth]{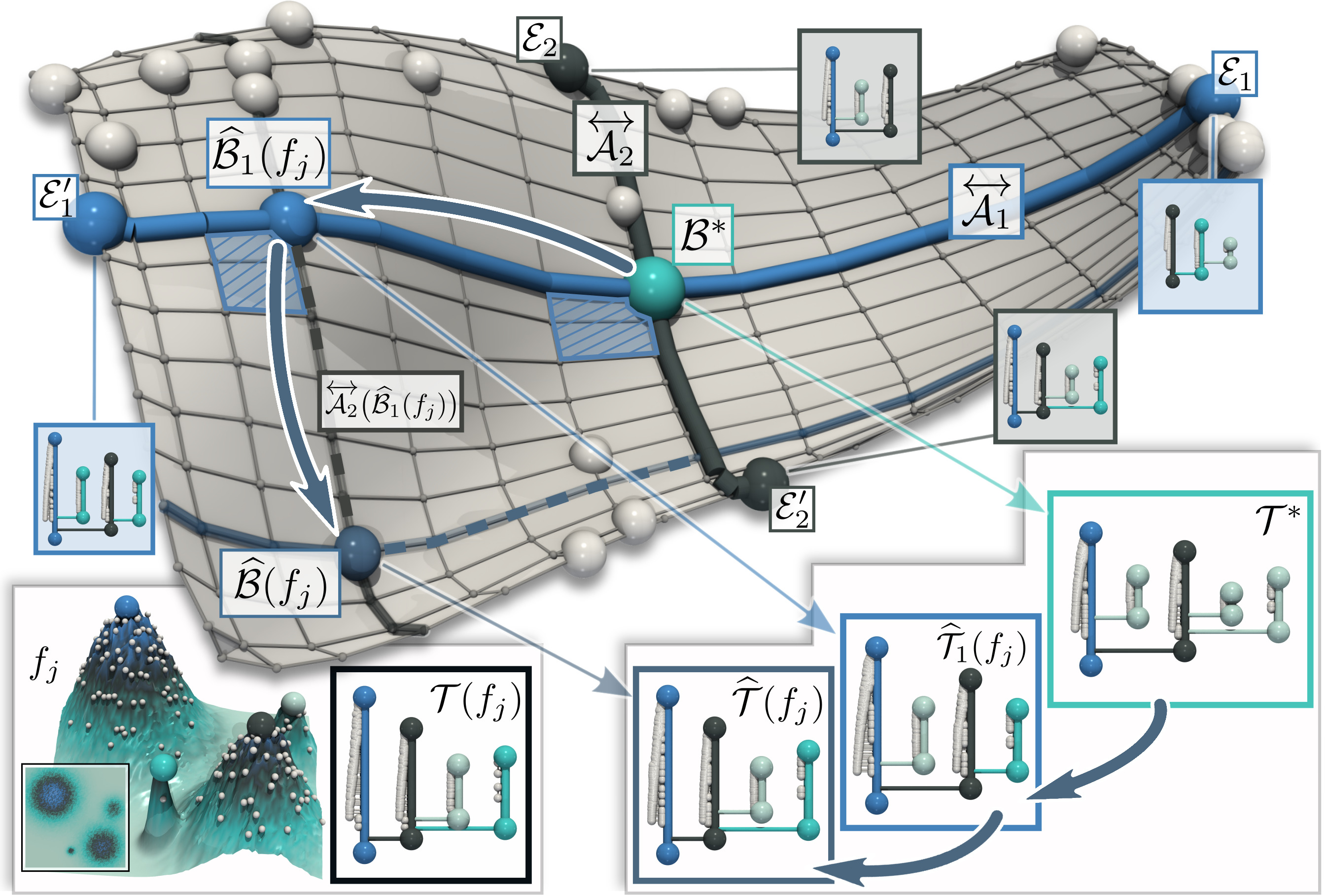}
\mycaption{
\revision{Approach overview:}
Our merge tree Principal Geodesic Analysis (MT-PGA)
defines a basis $\mtPgaBasis$ on the Wasserstein metric space of BDTs
$\branchtreeSpace$ (grey surface).
It is computed by iteratively optimizing
a fitting energy (\autoref{eq_mtPgaEnergy}) to an input ensemble of BDTs (white
spheres) for orthogonal \emph{geodesic axes} (blue:
$\axisNotation{\geodesicAxis_{1}}$,
black:
$\axisNotation{\geodesicAxis_2}$). Each axis (e.g.
$\axisNotation{\geodesicAxis_{1}}$) is defined
as a pair of \emph{geodesics}
between its extremities ($\geodesicExtremity_1$ and $\geodesicExtremity_1'$) and
the
Fr\'echet mean
of the ensemble (cyan: $\branchtree^*$). After the
optimization of the MT-PGA basis, any input BDT $\branchtree(f_j)$ can be
reconstructed
\revision{into $\reconstructed{\branchtree}(f_j)$ (dark blue sphere)}
by successive geodesic
\revision{displacements}
(dark blue arrows,
\autoref{eq_reconstruction}) along \emph{translations} of the axes (e.g.
$\axisNotation{\geodesicAxis_2}\big(\widehat{\branchtree}_1(f_j)\big)$\revision{, black dashed curve}),
\revision{given}
the coordinates
$\alpha^j$
of $\branchtree(f_j)$ in
the basis.
While our
framework manipulates BDTs, each of them can be directly inverted
(transparent
arrows)
back to
MTs
(insets),
resulting in reconstructions
($\widehat{\mergetree}(f_j)$) that are
\revision{visually}
similar
to the input
($\mergetree(f_j)$).
}
\label{fig_mtpgaOverview}
\end{figure*}

\subsection{Geometric interpretation of PCA}
\label{sec_geometricPCA}

Principal Component Analysis (PCA) \cite{pearson01} can be described with
several, different, yet equivalent, formulations.
In the following, we focus on a
geometric interpretation
(\autoref{fig_classicalPCA})
which
simplifies the
transition to merge trees.
Given a set of points $P = \{p_1, p_2, \dots, p_\ensembleSize\}$ in a
Euclidean space $\mathbb{R}^d$, PCA
defines
an orthogonal basis $\pcaBasis = \{\vectorNotation{b_1}, \vectorNotation{b_2},
\dots, \vectorNotation{b_d}\}$
of
vectors
$\vectorNotation{b_i} \in \mathbb{R}^d$
($i \in \{1, \dots, d\}$),
with origin $o_b \in
\mathbb{R}^d$,
such that:
\begin{enumerate}[leftmargin=.35cm]
\item{the origin $o_b$
coincides with
the arithmetic mean of $P$,}
\item{the line $l_i$ (defined by
$o_b$ and
$\vectorNotation{b_i}$)  is orthogonal to all
previous lines $l_{i'}$ ($i' \in \{1, \dots, i-1\}$) and
%
minimizes
its average squared distance to $P$.}
\end{enumerate}

Let $\projection_i(p_j)$ be the orthogonal projection of the point $p_j$ on
$l_i$ (i.e. $\projection_i(p_j)$ is the closest
point to $p_j$ on $l_i$, \autoref{fig_classicalPCA}). It can be expressed as
a
displacement
from
$o_b$
on
$l_i$: $\projection_i(p_j) = o_b +
\alpha_i^j
\vectorNotation{b_i}$, with
$\alpha_i^j \in
[-1, 1]$. Then,
$\pcaBasis$
can be formulated as an orthogonal basis which minimizes the following
data
fitting energy
(with $d' = d$ in general):
\begin{eqnarray}
\label{eq_euclideanPCA}
\reconstructionError(\pcaBasis) = \sum_{j = 1}^\ensembleSize ||p_j - \big(o_b +
\sum_{i = 1}^{d'}
\alpha_i^j
\vectorNotation{b_i}\big)||_2^2.
\end{eqnarray}

In practice, $\pcaBasis$ can be optimized
iteratively, one dimension at a time (starting with $d' = 1$ and finishing
with $d' = d$), by finding at each iteration a vector $\vectorNotation{b_{d'}}$
which is
orthogonal to all the previous vectors in $\pcaBasis$ ($\vectorNotation{b_{d'}}
\cdot \vectorNotation{b_{d''}} =
0,
\forall d'' \in  \{1, \dots, d'-1\}$) and which minimizes $\reconstructionError$
(\autoref{eq_euclideanPCA}).
After all dimensions have been processed ($d' = d$), $\pcaBasis$ provides a new
coordinate system for $P$ (composed
of the vectors $\vectorNotation{b_i}$ and the coordinates \revision{$\alpha_i^j$} for
each point $p_j$), such that each direction of $\pcaBasis$
successively provides an optimal fit to $P$ (\autoref{eq_euclideanPCA}).
Note that, {by construction,
the variance of the projected data (i.e. the variance of
\julien{}$\projection_{d'}(p_j)$ along $l_{d'}$})
will be maximized for the first direction ($d'=1$)
and it will keep on decreasing
for increasing $d'$.
%
This
motivates early terminations of the optimization (for $d' < d$), as
the most \emph{informative} directions are identified in the first iterations of
the algorithm. 

\subsection{From PCA to MT-PGA}
\label{sec_defMtPga}
When the input data is not given as a point cloud in a Euclidean space
(\autoref{sec_geometricPCA}) but as an abstract set
equipped with a metric,
the above formulation of Principal Component Analysis (PCA)
needs to be \revision{extended. Such an extension can be done}
with the more general notion of Principal Geodesic
Analysis (PGA)\revision{, which needs itself to be specifically instantiated given the specific metric space under study} \cite{FletcherLPJ04, SeguyC15, CazellesSBCP18}.
This
\revision{instantiation}
consists in
\revision{redefining the}
low-level geometrical tools
used in PCA
\revision{, but within the considered metric space.}
\revision{For instance,}
\emph{geodesic\revision{s}} (length-minimizing path\revision{s}) \revision{between merge trees (\autoref{sec_metricSpace})} \revision{extend to $\branchtreeSpace$}
the notion
of
straight line\revision{s}, and the
Fr\'echet mean \revision{(\autoref{sec_metricSpace}) extends}
the
arithmetic mean.
\revision{In this section, we specifically formulate such an extension for
merge trees (MTs). In particular, we formalize the following low-level 
geometrical notions for the Wasserstein metric space of merge trees 
$\branchtreeSpace$ \minorRevision{(see Appendix B for 
detailed illustrations)}:
\renewcommand{\theenumi}{\roman{enumi}}%
\begin{enumerate}
 \item BDT geodesic dot product;
 \item Orthogonal BDT geodesics;
 \item Collinear BDT geodesics;
 \item BDT geodesic axis;
 \item BDT geodesic axis arc-length parametrization;
 \item BDT geodesic axis projection;
 \item BDT geodesic axis translation;
 \item BDT geodesic orthogonal basis.
\end{enumerate}
}

\revision{The above notions derive sequentially from one another, to eventually result in the concept of \emph{orthogonal bases of BDT geodesic axes}, which is precisely the main variable
of the constrained optimization problem for MT-PGA formulation (\autoref{sec_detailedFormulation}).}

\noindent
\revision{\textbf{\emph{(i)} BDT geodesic dot product.}}
A \emph{geodesic}
on
\revision{the Wasserstein metric space of merge trees}
$\branchtreeSpace$
is
the shortest path
$\vectorNotation{\geodesictree}\big(\geodesicExtremity,
\geodesicExtremity'\big)$ between its
two
\revision{extremity BDTs}
$\geodesicExtremity$ and $\geodesicExtremity'$.
As detailed in \autoref{sec_metricSpace}, it is
\revision{an optimal assignment with regard to \autoref{eq_wasserstein}, which can be represented
as}
a vector in
$\mathbb{R}^{2 \times |\geodesicExtremity|}$.
Then, given two geodesics
$\vectorNotation{\geodesictree}\big(\geodesicExtremity,
\geodesicExtremity'\big)
\in
\mathbb{R}^{2 \times |\geodesicExtremity|}$ and
$\vectorNotation{\geodesictree}\big(\geodesicExtremity,
\geodesicExtremity''\big) \in
\mathbb{R}^{2 \times |\geodesicExtremity|}$ sharing an extremity
$\geodesicExtremity$, their \emph{dot product}, noted
$\vectorNotation{\geodesictree}\big(\geodesicExtremity,
\geodesicExtremity'\big)
\cdot
\vectorNotation{\geodesictree}\big(\geodesicExtremity,
\geodesicExtremity''\big)$ can be
naturally introduced by considering their Cartesian dot product (i.e. sum of
component-wise products \revision{between two vectors in $\mathbb{R}^{2 \times |\geodesicExtremity|}$}).
Note
that these two vectors
\revision{must}
be consistently parametrized with regard
to their common extremity $\geodesicExtremity$: for both vectors, the $i^{th}$
entry represents a 2D vector in the birth/death space modeling the \revision{optimal} assignment \revision{(with regard to \autoref{eq_wasserstein})}
of the $i^{th}$ point of $\geodesicExtremity$ to points of
$\geodesicExtremity'$ and $\geodesicExtremity''$
(small arrows, \autoref{fig_wassersteinMetric}, right).

\noindent
\revision{\textbf{\emph{(ii)} Orthogonal BDT geodesics.}}
Two geodesics
$\vectorNotation{\geodesictree}\big(\geodesicExtremity,
\geodesicExtremity'\big)$ and
$\vectorNotation{\geodesictree}\big(\geodesicExtremity,
\geodesicExtremity''\big)$ are
\emph{orthogonal} if $\vectorNotation{\geodesictree}\big(\geodesicExtremity,
\geodesicExtremity'\big)
\cdot \vectorNotation{\geodesictree}\big(\geodesicExtremity,
\geodesicExtremity''\big) = 0$.

\noindent
\revision{\textbf{\emph{(iii)} Collinear BDT geodesics.}}
Two geodesics $\vectorNotation{\geodesictree}\big(\geodesicExtremity,
\geodesicExtremity'\big)$ and
$\vectorNotation{\geodesictree}\big(\geodesicExtremity,
\geodesicExtremity''\big)$
are \emph{collinear} if
$\vectorNotation{\geodesictree}\big(\geodesicExtremity,
\geodesicExtremity'\big) = \lambda
\vectorNotation{\geodesictree}\big(\geodesicExtremity,
\geodesicExtremity''\big)$, with $\lambda \in \mathbb{R}$. In particular, they
are
\emph{positively collinear} if $\lambda \gt 0$, and \emph{negatively
collinear} if $\lambda \lt 0$.

\noindent
\revision{\textbf{\emph{(iv)} BDT geodesic axis.}}
\revision{Now that the notion of collinear geodesics is available, we proceed to the introduction of the concept of \emph{geodesic axis}.}
Given an \emph{origin}
BDT $\geodesicExtremity_O$ and two extremities $\geodesicExtremity_i$ and
$\geodesicExtremity_i'$,
a \emph{geodesic axis}, noted
$\axisNotation{\geodesicAxis_i}$,
is defined as a pair
of negatively collinear geodesics
$ \vectorNotation{\geodesictree_i} =
\vectorNotation{\geodesictree}\big(\geodesicExtremity_O,
\geodesicExtremity_i\big)$,
and $ \vectorNotation{\geodesictree_i'} =
\vectorNotation{\geodesictree}\big(\geodesicExtremity_O,
\geodesicExtremity_i'\big)$.
\revision{It follows that a geodesic axis $\axisNotation{\geodesicAxis_i}$ is itself a geodesic between $\geodesicExtremity_i'$ and $\geodesicExtremity_i$, which is guaranteed to pass through a given origin $\geodesicExtremity_O$.}
\revision{An example of geodesic axis is given in \autoref{fig_mtpgaOverview} with the axis $\axisNotation{\geodesicAxis_1}$ (represented as a blue curve) between the BDTs $\geodesicExtremity_1'$ and $\geodesicExtremity_1$, with origin $\branchtree^*$.}
\revision{This notion will be instrumental in the definition of an orthogonal basis in $\branchtreeSpace$, to constrain multiple geodesics to indeed pass through the origin of the basis, while still allowing for an optimization of their extremities (\autoref{sec_algorithm}). Each geodesic axis}
$\axisNotation{\geodesicAxis_i}$
is associated to a \emph{direction
vector}:
\revision{$\vectorNotation{\directionVector_i} =
\vectorNotation{\geodesictree_i} - \vectorNotation{\geodesictree_i'} = \vectorNotation{\geodesictree}\big(\geodesicExtremity_i',
\geodesicExtremity_i\big)$}.

\noindent
\revision{\textbf{\emph{(v)} BDT geodesic axis arc-length parametrization.}}
By construction, any BDT
$\branchtree$ located on an axis $\axisNotation{\geodesicAxis_i}$
can be expressed with the following
arc-length parametrization $\vectorNotation{\geodesicAxis_i}(\branchtree)$
(with $\alpha_i \in [0, 1]$):
\begin{eqnarray}
\label{eq_arcLengthParametrization}
\branchtree =
\geodesicExtremity_O +
\vectorNotation{\geodesicAxis_i}(\branchtree) =
\geodesicExtremity_O + \alpha_i \times
\vectorNotation{\geodesictree_i}
+ (1 - \alpha_i) \times \vectorNotation{\geodesictree_i'}.
\end{eqnarray}
\revision{This arc-length parametrization is the analog for the metric space
$\branchtreeSpace$ of the Euclidean line parametrization used in PCA (\autoref{eq_euclideanPCA}).}

\noindent
\revision{\textbf{\emph{(vi)} BDT geodesic axis projection.}}
The \emph{projection} $\branchtree_{\axisNotation{\geodesicAxis_i}}$ of
an arbitrary BDT $\branchtree$
on the axis
$\axisNotation{\geodesicAxis_i}$
is
its closest BDT on $\axisNotation{\geodesicAxis_i}$.
\revision{It minimizes the
following projection energy:
\begin{eqnarray}
\label{eq_projectionEnergy}
\nonumber
\branchtree_{\axisNotation{\geodesicAxis_i}}
& = &
\argmin_{\branchtree' \in
\axisNotation{\geodesicAxis_i}}
\Big(\wassersteinTree\big(\branchtree, \branchtree'\big)\Big)\\
& = &
\argmin_{\branchtree' \in
\axisNotation{\geodesicAxis_i}}
\Big(\wassersteinTree\big(\branchtree, \geodesicExtremity_O +
\vectorNotation{\geodesicAxis_i}(\branchtree')\big)\Big).
\end{eqnarray}}

\revision{An example of projection is given in \autoref{fig_mtpgaOverview}, with the blue sphere noted $\reconstructed{\branchtree}_1(f_j)$, which is the projection on the axis $\axisNotation{\geodesicAxis_1}$ (blue curve) of an input BDT $\branchtree(f_j)$.}
\revision{The projection $\branchtree_{\axisNotation{\geodesicAxis_i}}$ of $\branchtree$ on $\geodesicAxis_i$ will act as the analog, for the metric space
$\branchtreeSpace$, of the projection $\projection_i(p_j)$ of  a point $p_j$ along a line $l_i$ in the Euclidean PCA (\autoref{sec_geometricPCA}).}

\noindent
\revision{\textbf{\emph{(vii)} BDT geodesic axis translation.}}
Let
$\axisNotation{\geodesicAxis_i}$ and
$\axisNotation{\geodesicAxis_j}$
be two axes
sharing the same origin
$\geodesicExtremity_O$.
The axis
$\axisNotation{\geodesicAxis_j}(\branchtree)$ is called the \emph{translation}
of
$\axisNotation{\geodesicAxis_j}$
%
along $\axisNotation{\geodesicAxis_i}$, with origin
 $\branchtree \in \axisNotation{\geodesicAxis_i}$ if:
\label{eq_axisTranslation}
$
\axisNotation{\geodesicAxis_j}(\branchtree) =
\big(
(\branchtree, \branchtree +
\vectorNotation{\geodesictree_j}),
(\branchtree, \branchtree +
\vectorNotation{\geodesictree'_j})
\big)$.
\revision{Intuitively, a translated axis $\axisNotation{\geodesicAxis_j}(\branchtree)$ is a BDT axis, which is parallel
to
the axis
$\axisNotation{\geodesicAxis_j}$, and which passes through a specific BDT $\branchtree$.
In \autoref{fig_mtpgaOverview}, an example is given with the axis $\axisNotation{\geodesicAxis_2}\big(\reconstructed{\branchtree}_{1}(f_j)\big)$ (black dashed curve), which is the translation of the axis $\axisNotation{\geodesicAxis_2}$ (black curve) at the
BDT
$\reconstructed{\branchtree}_{1}(f_j)$
(blue sphere).
This notion of translated axis is needed
when reconstructing a BDT given its coordinates in the MT-PGA basis, as detailed next.}
%

\noindent
\revision{\textbf{\emph{(viii)} BDT geodesic orthogonal basis.}}
\revision{We now introduce the notion of orthogonal basis in the metric space of merge trees $\branchtreeSpace$, which will be the variable at the core of the constrained optimization problem for MT-PGA formulation (\autoref{sec_detailedFormulation}).}
Let $\mtPgaBasis = \{\axisNotation{\geodesicAxis_1},
\axisNotation{\geodesicAxis_2}, \dots, \axisNotation{\geodesicAxis_{d'}}\}$ be
an
 \emph{orthogonal basis} of $d'$ 
geodesic axes
(i.e. with pairwise orthogonal
direction vectors $\{\vectorNotation{\directionVector_1},
\vectorNotation{\directionVector_2}, \dots,
\vectorNotation{\directionVector_{d'}}\}$),
with origin
$\geodesicExtremity_O$.
Let $\widehat{\branchtree}_1$ be the projection of
an arbitrary BDT
$\branchtree$ on
$\axisNotation{\geodesicAxis_1}$
($\widehat{\branchtree}_1 =
\branchtree_{\axisNotation{\geodesicAxis_1}}
=
\branchtree_{\axisNotation{\geodesicAxis_1}(\geodesicExtremity_O)}$,
\revision{e.g.} blue sphere on $\axisNotation{\geodesicAxis_1}$,
\autoref{fig_mtpgaOverview}\revision{, noted $\widehat{\branchtree}_1(f_j)$}).
Now, let $\widehat{\branchtree}_2$ be the projection of $\branchtree$ on the
\emph{translation} of $\axisNotation{\geodesicAxis_2}$ along
$\axisNotation{\geodesicAxis_1}$,
with origin
$\widehat{\branchtree}_1$:
$\widehat{\branchtree}_2$
is the BDT of the \emph{translated} axis
$\axisNotation{\geodesicAxis_2}(\widehat{\branchtree}_1)$ which is the closest
to $\branchtree$
(i.e.
$\widehat{\branchtree}_2 =
\branchtree_{\axisNotation{\geodesicAxis_2}(\widehat{\branchtree}_1)}$,
\revision{e.g.} dark
blue sphere on $\axisNotation{\geodesicAxis_2}(\widehat{\branchtree}_1)$ in
\autoref{fig_mtpgaOverview}\revision{, noted $\widehat{\branchtree}(f_j)$}).

By recursion, we
can then define the \emph{translated projection} of $\branchtree$  for the axis
$\axisNotation{\geodesicAxis_{d'}}$, noted
$\widehat{\branchtree}_{d'}$,
as
$\widehat{\branchtree}_{d'}
=
\branchtree_{\axisNotation{\geodesicAxis_{d'}}(\widehat{\branchtree}_{d'-1})}$
(with $\widehat{\branchtree}_{0} = \geodesicExtremity_O$).
\revision{Intuitively, the translated projection $\widehat{\branchtree}_{d'}$ of $\branchtree$ is the closest BDT to $\branchtree$ (dark blue sphere in \autoref{fig_mtpgaOverview}), which can be obtained by a sequence of $d'$
geodesic displacements
(dark blue arrows in \autoref{fig_mtpgaOverview}) along the translated axes of the orthogonal basis $\mtPgaBasis$. }

\revision{By using \autoref{eq_arcLengthParametrization}, }
$\widehat{\branchtree}_{d'}$ \revision{can be}
associated with a collection of
arc-length
parameterizations
$\{\vectorNotation{\geodesicAxis_1}(\widehat{\branchtree}_1),
\vectorNotation{\geodesicAxis_2}(\widehat{\branchtree}_2),
\dots,
\vectorNotation{\geodesicAxis_{d'}}(\widehat{\branchtree}_{d'})
\}$,
\revision{s.t.}:
\begin{eqnarray}
\revision{\reconstructed{\branchtree}_{d'}} = \geodesicExtremity_O +  \sum_{i =
1}^{d'} \vectorNotation{\geodesicAxis_i}(\widehat{\branchtree}_{i}).
\end{eqnarray}

\revision{
$\reconstructed{\branchtree}_{d'}$ can then be interpreted as the \emph{reconstruction} of $\branchtree$ (up to the dimension $d'$), given the basis $\mtPgaBasis$: it is obtained by starting from the origin $\geodesicExtremity_O$ of the basis, and successively minimizing the projection energy (\autoref{eq_projectionEnergy}) along translated axes of $\mtPgaBasis$, yielding a coordinate vector $\alpha \in [0, 1]^{d'}$ (\autoref{eq_arcLengthParametrization}) which can be interpreted as the coordinates of $\branchtree$ in $\mtPgaBasis$.}

%



\begin{algorithm}[t]
    \fontsize{5.625}{9}\selectfont
    \algsetup{linenosize=\tiny}
    \caption{\textcolor{black}{\footnotesize Merge Tree Principal Geodesic
Analysis (MT-PGA) Algorithm}}
\label{algo_mtPga}
\hspace*{\algorithmicindent} \textbf{Input}: Set of BDTs $\branchtreeSet =
\{\branchtree(f_1),
\dots, \branchtree(f_\ensembleSize)\}$.

\hspace*{\algorithmicindent} \textbf{Output1}: Basis origin
$\branchtree^* $;

\hspace*{\algorithmicindent} \textbf{Output2}: Basis
geodesic axes $\mtPgaBasis =
\{\axisNotation{\geodesicAxis_1}, \axisNotation{\geodesicAxis_2}, \dots,
\axisNotation{\geodesicAxis_{\maxDimensions}}\}$;

\hspace*{\algorithmicindent} \textbf{Output3}: Coordinates $\alpha^j
\in [0, 1]^{\maxDimensions}$ of the input BDTs in $\mtPgaBasis$ (with $j \in
\{1, 2,
\dots, \ensembleSize\}$).

\begin{algorithmic}[1]

\STATE $\branchtree^* \leftarrow $WassersteinBarycenter$(\branchtreeSet);$
\label{alg_line_barycenter}

\FOR{$d' \in \{1, 2, \dots,  \maxDimensions\}$}
\label{alg_dimensionLoop}
  \STATE $\axisNotation{\geodesicAxis_{d'}} \leftarrow
$InitializeGeodesicAxis$(\branchtreeSet,
  \branchtree^*,
  \mtPgaBasis);$
  \label{alg_lineInit}
  \WHILE{$\mtPgaError(\mtPgaBasis)$ decreases}
  \label{alg_whileLoop}
  \STATE // Optimize the current geodesic axis
$\axisNotation{\geodesicAxis_{d'}}$
(\autoref{sec_optimization})
  \STATE $\alpha_{d'}^{j \in \{1, 2, \dots \ensembleSize\}} \leftarrow
$ProjectTrees$(\branchtreeSet, \branchtree^*,
\mtPgaBasis,
\alpha^{j \in \{1, 2, \dots \ensembleSize\}});$
  \label{alg_line_projection}
  \STATE $\mtPgaError(\mtPgaBasis)
\leftarrow$EvaluateFittingEnergy$(\branchtreeSet, \branchtree^*,
\mtPgaBasis, \alpha^{j \in \{1, 2, \dots \ensembleSize\}});$
  \label{alg_evaluateEnergy}
  \STATE $\axisNotation{\geodesicAxis_{d'}} \leftarrow
$OptimizeFittingEnergy$(\branchtreeSet, \branchtree^*, \mtPgaBasis,
\alpha^{j \in \{1, 2, \dots \ensembleSize\}});$
\label{alg_line_optimize}
  \STATE // Enforce the constaints (\autoref{sec_constraints})
  \WHILE{$\axisNotation{\geodesicAxis_{d'}}$ evolves}
    \label{alg_line_whileConstraints}
    \STATE  $\axisNotation{\geodesicAxis_{d'}} \leftarrow
    $EnforceGeodesics$(\branchtree^*, \axisNotation{\geodesicAxis_{d'}});$
    \label{alg_line_firstConstraint}
    \STATE $\axisNotation{\geodesicAxis_{d'}} \leftarrow
    $EnforceNegativeCollinearity$(\axisNotation{\geodesicAxis_{d'}});$
    \label{alg_line_secondConstraint}
    \STATE $\axisNotation{\geodesicAxis_{d'}} \leftarrow
    $EnforceOrthogonality$(\mtPgaBasis, \axisNotation{\geodesicAxis_{d'}});$
    \label{alg_line_lastConstraint}
    \ENDWHILE
  \ENDWHILE
\ENDFOR
\label{alg_line_endFor}
\end{algorithmic}
\end{algorithm}

\subsection{\revision{MT-PGA formulation}}
\label{sec_detailedFormulation}
Now that the above low-level geometrical tools have been introduced for the
Wasserstein metric space $\branchtreeSpace$, we can formulate the MT-PGA by
direct analogy to the classical PCA (\autoref{sec_geometricPCA}).

Given a set
$\branchtreeSet = \{\branchtree(f_1),
\dots, \branchtree(f_\ensembleSize)\}$ of input BDTs, let $\branchtree^*$ be
their
Fr\'echet mean
(\autoref{eq_frechet_energy}).
Then, similarly to PCA, MT-PGA defines an
orthogonal
basis $\mtPgaBasis = \{
\axisNotation{\geodesicAxis_1}, \axisNotation{\geodesicAxis_2}, \dots,
\axisNotation{\geodesicAxis_{2\times|\branchtree^*|}}\}$ of geodesic
axes
$\axisNotation{\geodesicAxis_i}$ ($i \in \{1 ,\dots, 2\times|\branchtree^*|\}$)
s.t.:

\begin{enumerate}[leftmargin=.35cm]
\item{the origin $\geodesicExtremity_O$ of $\mtPgaBasis$ coincides with the
Fr\'echet mean
$\branchtree^*$ of
$\branchtreeSet$,}
\item{the axis $\axisNotation{\geodesicAxis_i}$
is orthogonal to all
axes $\axisNotation{\geodesicAxis_{i'}}$ ($i' \in
\{1, \dots, i-1\}$) and it 
minimizes its average squared Wasserstein
distance to $\branchtreeSet$.}
\end{enumerate}
Then,
$\mtPgaBasis$ can
be formulated as an orthogonal basis which minimizes the
following
(non-convex)
fitting energy
(with $d' = 2\times|\branchtree^*|$ in general):
\begin{eqnarray}
\label{eq_mtPgaEnergy}
\mtPgaError(\mtPgaBasis) = \sum_{j = 1}^\ensembleSize
\wassersteinTree\Big(\branchtree(f_j),
\branchtree^* +
\sum_{i =
1}^{d'}\vectorNotation{\geodesicAxis_i}\big(\widehat{\branchtree}_i(f_j)
\big)
\Big)^2
.
\end{eqnarray}

The above equation is a direct analogy to the classical PCA
(\autoref{eq_euclideanPCA}): the $L_2$ norm is replaced by the
Wasserstein distance $\wassersteinTree$, the arithmetic mean $o_b$ by
$\branchtree^*$ and the term $\alpha_i^j\vectorNotation{b_i}$ by
$\vectorNotation{\geodesicAxis_i}\big(\widehat{\branchtree}_i(f_j)
\big)$.

\section{Algorithm}
\label{sec_algorithm}
This section presents our novel 
\revision{algorithm},
based on the
constrained minimization of \autoref{eq_mtPgaEnergy}, for 
the 
estimation
of 
an
orthogonal basis
$\mtPgaBasis$ for the Principal Geodesic Analysis of merge trees.




\subsection{Overview}
\label{sec_overview}


\autoref{algo_mtPga} provides an overview of our approach. Our algorithm takes
an ensemble $\branchtreeSet$ of BDTs as an input and provides three outputs:
\emph{(i)} the basis origin $\branchtree^*$,
\emph{(ii)} its $\maxDimensions$ geodesic axes
$\{\axisNotation{\geodesicAxis_1},
\axisNotation{\geodesicAxis_2},
\dots, \axisNotation{\geodesicAxis_{\maxDimensions}}\}$ (where $1 \leq
\maxDimensions  \leq
2\times|\branchtree^*|$ is an input parameter) and \emph{(iii)} the
coordinates $\alpha^j \in [0, 1]^{\maxDimensions}$ of
the input BDTs
in
$\mtPgaBasis$ (\autoref{eq_arcLengthParametrization}, with $j \in \{1, 2,
\dots, \ensembleSize\}$).

First, the origin of the basis $\mtPgaBasis$  is computed as the
Wasserstein barycenter $\branchtree^*$ of $\branchtreeSet$ (\revision{line}
\ref{alg_line_barycenter}). This is done with an iterative algorithm
\cite{pont_vis21}, which initializes $\branchtree^*$ at the BDT in
$\branchtreeSet$ minimizing \autoref{eq_frechet_energy}, and which further
minimizes \autoref{eq_frechet_energy}
by iteratively alternating assignment and update phases
(see
\cite{pont_vis21}).
After this optimization has completed,
only the $\numberBranchinBarycenter$ most persistent branches are kept in
$\branchtree^*$, where $\numberBranchinBarycenter$
is an input parameter controlling the memory footprint of the basis (see
\autoref{sec_computationalAspects}).

Next, the MT-PGA basis is computed by adapting the iterative strategy
described in the case of PCA (\autoref{sec_geometricPCA}).
In particular, geodesic
axes are computed one dimension at a time (starting with $d' = 1$ and finishing
with $d' = \maxDimensions$, \emph{for} loop, \revision{lines} \autoref{alg_dimensionLoop} \revision{to \autoref{alg_line_endFor}}).

\revision{In particular, at each step $d'$:}

\begin{figure}
\includegraphics[width=\linewidth]{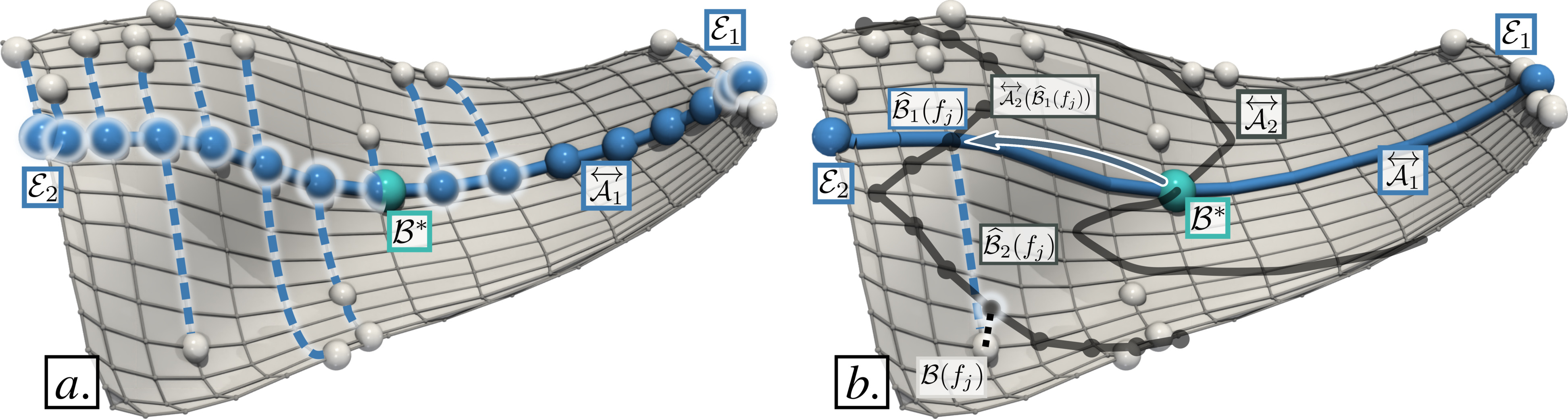}
\mycaption{Computing the \emph{translated projection} of each input BDT
$\branchtree(f_j)$ (white sphere).
For the first
axis \revision{\emph{(a)}}, $\numberGeodesicSamples$ samples evenly spaced on
$\axisNotation{\geodesicAxis_{1}}$ are considered (blue spheres). Then, for each
$\branchtree(f_j)$, the sample which minimizes its
$\wassersteinTree$
distance to
$\branchtree(f_j)$  is selected as
$\widehat{\branchtree}_1(f_j)$ (dashes).
Next \revision{\emph{(b)}},
the axis $\axisNotation{\geodesicAxis_{2}}$ is \emph{translated} to
$\widehat{\branchtree}_1(f_j)$ along $\axisNotation{\geodesicAxis_{1}}$ (arrow)
yielding the
translated axis
$\axisNotation{\geodesicAxis_{2}}\big(\widehat{\branchtree}_1(f_j)\big)$,
which is
also evenly sampled. The sample of
$\axisNotation{\geodesicAxis_{2}}\big(\widehat{\branchtree}_1(f_j)\big)$
minimizing
its
$\wassersteinTree$
distance to
$\branchtree(f_j)$ is selected as
$\widehat{\branchtree}_2(f_j)$.}
%
\label{fig_axisProjection}
\end{figure}

\begin{figure}[b]
\includegraphics[width=\linewidth]{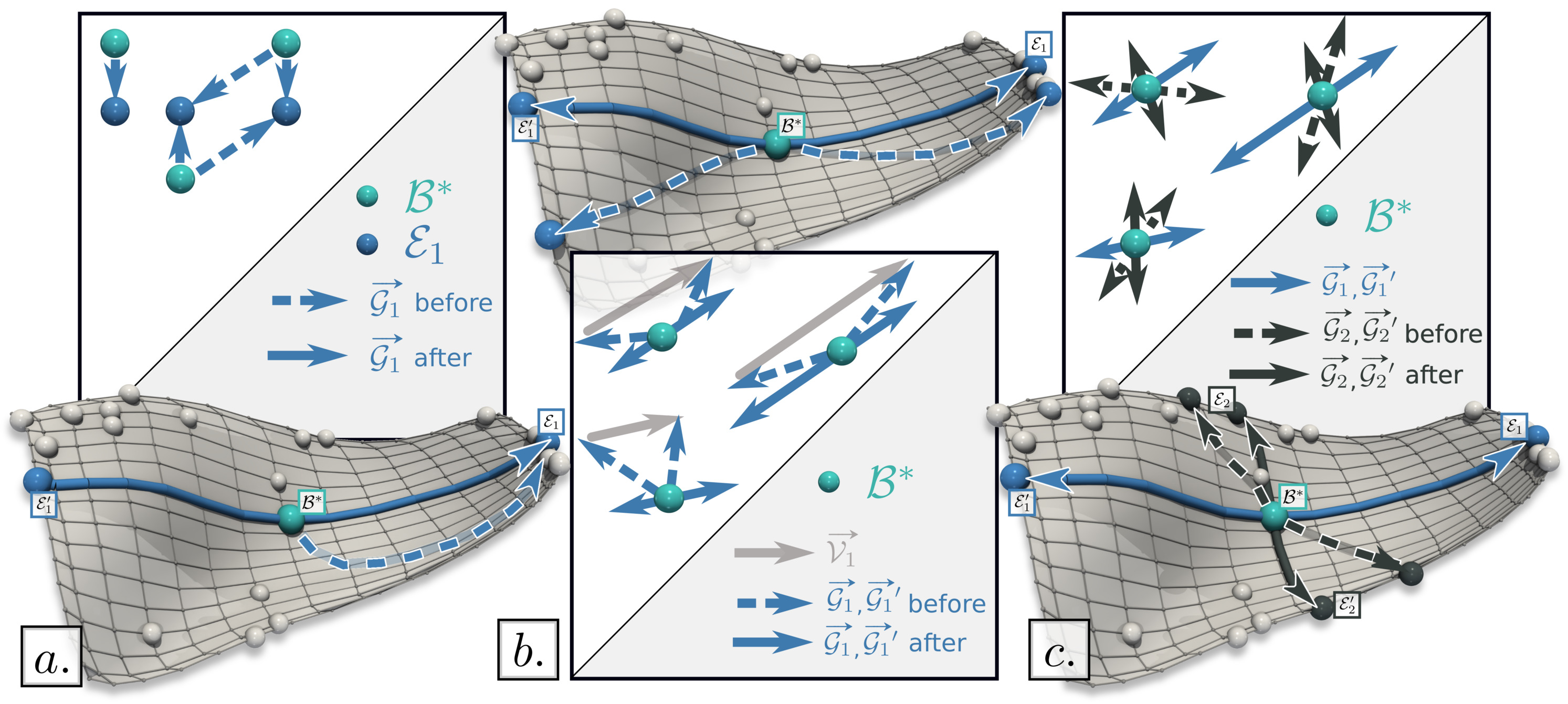}
  \caption{Constraint enforcements:
  \revision{\emph{(a)}} Geodesic enforcement guarantees that the vector
$\vectorNotation{\geodesictree_1}$ describes an assignment which minimizes
\autoref{eq_wasserstein};
  \revision{\emph{(b)}} Collinearity enforcement constructively aligns
$\vectorNotation{\geodesictree_1}$ and $\vectorNotation{\geodesictree'_1}$;
  \revision{\emph{(c)}} Orthogonality enforcement updates 
$\vectorNotation{\geodesictree_2}$ and
$\vectorNotation{\geodesictree'_2}$ via \emph{Gram-Schmidt} orthogonalization.}
%
  \label{fig_constraints}
\end{figure}

\begin{enumerate}
\item
The geodesics $\vectorNotation{\geodesictree_{d'}}$ and
$\vectorNotation{\geodesictree_{d'}'}$ defining the axis
$\axisNotation{\geodesicAxis_{d'}}$
are optimized in order to minimize
\autoref{eq_mtPgaEnergy} (\revision{line} \autoref{alg_line_optimize}).
This optimization \revision{(described in \autoref{sec_optimization} and illustrated in \autoref{fig_axisProjection})} involves the
translated projection $\widehat{\branchtree}_{d'}(f_j)$ of each
input
BDT $\branchtree(f_j)$
(\revision{line}
\autoref{alg_line_projection}),
providing an
estimation of its
coordinate $\alpha_{d'}^j$
(\autoref{eq_arcLengthParametrization}).
\item
\revision{The axis}
$\axisNotation{\geodesicAxis_{d'}}$
is updated into a valid solution
\revision{(as described in \autoref{sec_constraints} and illustrated in \autoref{fig_constraints})},
by enforcing
each constraint one after the other (\revision{lines} \autoref{alg_line_firstConstraint} to
\autoref{alg_line_lastConstraint}).
\end{enumerate}

This
alternated procedure optimization/constraint is iterated until
$\mtPgaError(\mtPgaBasis)$ does not evolve significantly anymore (see
\autoref{sec_optimization}).
Once
this is achieved, the current axis $\axisNotation{\geodesicAxis_{d'}}$ is
considered as
finalized and $d'$ is incremented 
until
$\axisNotation{\geodesicAxis_{\maxDimensions}}$ is finalized.


\subsection{Geodesic axis optimization}
\label{sec_optimization}

This section describes the optimization of a
\revision{single}
axis
$\axisNotation{\geodesicAxis_{d'}}$.






First, the  geodesic $\vectorNotation{\geodesictree_{d'}}$ of
$\axisNotation{\geodesicAxis_{d'}}$ is
initialized
by
considering as its extremity $\geodesicExtremity_{d'}$ the input BDT in
$\branchtreeSet$ which
maximizes its Wasserstein distance to all the previous geodesic axes
$\axisNotation{\geodesicAxis_{d''}}$, $\forall d'' \in \{1, \dots d' - 1\}$
(to $\branchtree^*$ when $d' = 1$).
The second geodesic
$\vectorNotation{\geodesictree_{d'}'}$ is initialized at
$-\vectorNotation{\geodesictree_{d'}}$.

Next, both $\vectorNotation{\geodesictree_{d'}}$ and
$\vectorNotation{\geodesictree_{d'}'}$ are optimized, to minimize
\autoref{eq_mtPgaEnergy}.
 This is achieved with an iterative algorithm
(\autoref{algo_mtPga}, \emph{while} loop,
\revision{line}
 \autoref{alg_whileLoop}) which, similarly to the minimization of the
 Fr\'echet energy \cite{pont_vis21}, alternates an \emph{(i)} \emph{Assignment}
 and an \emph{(ii)} \emph{Update} phase.
 \revision{Intuitively, the \emph{(i)} \emph{Assignment} phase will compute 
 geodesics  (dashed blue curves, \autoref{fig_axisProjection}a) between the 
input BDTs (white spheres, \autoref{fig_axisProjection}a) and 
 the axis to 
optimize (plain blue curve, \autoref{fig_axisProjection}a), while 
the \emph{(ii)} \emph{Update}  phase will optimize freely the axis under these optimal 
assignments. After the \emph{(ii)} \emph{Update}, the initial assignments 
may no longer
 be optimal  
 (since the axis has changed) and the \emph{Assignment/Update} sequence needs to be iterated again, 
 as detailed next.}

The \emph{Assignment} phase \emph{(i)}
computes a geodesic between each input BDT $\branchtree(j)$ and its
translated
projection
$\widehat{\branchtree}_{d'}(f_j)$ (\autoref{fig_axisProjection}).
%
In particular, these
translated
projections (\autoref{algo_mtPga},
\revision{line}
\autoref{alg_line_projection}) are
estimated in practice by sampling the translated axis
$\axisNotation{\geodesicAxis_{d'}}\big(\widehat{\branchtree}_{d'-1}(f_j)\big)$
along
$\numberGeodesicSamples$ evenly
spaced samples and by selecting, for each
$\branchtree(f_j)$, the
sample
$\widehat{\branchtree}_{d'}(f_j)$
\revision{which minimizes
\autoref{eq_projectionEnergy}.}

Next, the \emph{Update} phase \emph{(ii)} consists in optimizing
$\vectorNotation{\geodesictree_{d'}}$ and
$\vectorNotation{\geodesictree_{d'}'}$ to minimize
\autoref{eq_mtPgaEnergy} under the assignments computed by the
\emph{Assignment} phase \emph{(i)}.
As detailed in Appendix \minorRevision{C}, for fixed
assignments, the fitting energy $\mtPgaError$ is convex with the
variables $\vectorNotation{\geodesictree_{d'}}$ and
$\vectorNotation{\geodesictree_{d'}'}$. We provide in Appendix 
\minorRevision{C} the analytic
expression of the gradient of this energy and
we
derive the analytic
expressions of $\vectorNotation{\geodesictree_{d'}}$ and
$\vectorNotation{\geodesictree_{d'}'}$ which minimize \autoref{eq_mtPgaEnergy}.

This overall \emph{Assignment}/\emph{Update} sequence is then iterated
(\autoref{algo_mtPga}, \emph{while} loop,
\revision{line}
 \autoref{alg_whileLoop}).
 Each iteration
decreases $\mtPgaError$ constructively: while the \emph{Update}
phase \emph{(ii)} minimizes it under the current assignment, the next
\emph{Assignment} phase \emph{(i)} further improves (by construction) the
assignments, hence decreasing $\mtPgaError$ overall.
In our
implementation, the algorithm stops when the fitting energy has decreased by
less than $1\%$ between two consecutive iterations.

\subsection{Constraints}
\label{sec_constraints}

The previous section described an algorithm for optimizing the geodesic axis
$\axisNotation{\geodesicAxis_{d'}}$, in order to minimize the fitting
energy $\mtPgaError$ (\autoref{eq_mtPgaEnergy}). However, this algorithm did
not consider yet key constraints present in the 
definition of MT-PGA
(\autoref{sec_defMtPga}), such as axis orthogonality.
We describe now our strategy for extending the above algorithm
with a succession of
constraint enforcements.


\noindent
\textbf{\emph{(i)} Geodesic enforcement.}
After the axis $\axisNotation{\geodesicAxis_{d'}}$ has been optimized
(\autoref{sec_optimization}) to minimize the fitting energy $\mtPgaError$
(\autoref{eq_mtPgaEnergy}), its associated vectors
$\vectorNotation{\geodesictree_{d'}} \in \mathbb{R}^{2 \times|\branchtree^*|}$
and $\vectorNotation{\geodesictree'_{d'}} \in \mathbb{R}^{2
\times|\branchtree^*|}$ may
represent displacements
in
the 2D birth/death space which are no longer geodesics \revision{in $\branchtreeSpace$ (as illustrated in \autoref{fig_constraints}a with the dashed arrows).}
\revision{Concretely, this situation occurs if
the assignment described by the (optimized) vector
$\vectorNotation{\geodesictree_{d'}}$
between the barycenter $\branchtree^*$ and the
vector
extremity $\geodesicExtremity_{d'}$ is no longer optimal with
regard to
$\wassersteinTree$ (\autoref{eq_wasserstein}),
as
$\vectorNotation{\geodesictree_{d'}}$ has been  optimized freely when minimizing
$\mtPgaError$.}
This can be easily
corrected
by
re-computing the optimal assignment between $\branchtree^*$ and
$\geodesicExtremity_{d'}$,
yielding an updated vector
$\vectorNotation{\geodesictree_{d'}}$,
now describing
a valid geodesic
between
$\branchtree^*$ and
$\geodesicExtremity_{d'}$.
\revision{In \autoref{fig_constraints}a, this is illustrated by the switch from the dashed arrows
to the plain blue arrows.}
The same correction is applied to second vector of
the axis, $\vectorNotation{\geodesictree'_{d'}}$.


\noindent
\textbf{\emph{(ii)} Negative collinearity enforcement.} Up to now, the
two geodesics $\vectorNotation{\geodesictree_{d'}}$ and
$\vectorNotation{\geodesictree'_{d'}}$ of the axis
$\axisNotation{\geodesicAxis_{d'}}$ have been optimized independently. Then,
they may not be negatively collinear and thus, they may not describe a valid
geodesic axis, as defined in \autoref{sec_defMtPga}.
Let $\beta' =
||\vectorNotation{\geodesictree_{d'}}|| /
(||\vectorNotation{\geodesictree_{d'}}||
+ ||\vectorNotation{\geodesictree'_{d'}}|| )$ be the ratio describing the
relative norm of $\vectorNotation{\geodesictree_{d'}}$ with regard to
$\axisNotation{\geodesicAxis_{d'}}$.
Negative collinearity is now constructively enforced
by updating
$\vectorNotation{\geodesictree_{d'}}$ and $\vectorNotation{\geodesictree'_{d'}}$
(\autoref{fig_constraints}b)
such that $\vectorNotation{\geodesictree_{d'}} \leftarrow \beta' \times
\vectorNotation{\dummyVector_{d'}}$ and $\vectorNotation{\geodesictree'_{d'}}
\leftarrow \revision{-} (1 - \beta') \times
\vectorNotation{\dummyVector_{d'}}$, where $\vectorNotation{\dummyVector_{d'}}$
is the direction vector of $\axisNotation{\geodesicAxis_{d'}}$
(\autoref{sec_defMtPga}, $\vectorNotation{\dummyVector_{d'}} =
\vectorNotation{\geodesictree_{d'}} - \vectorNotation{\geodesictree'_{d'}}$).

%
%
%
%
%
%


\noindent
\textbf{\emph{(iii)} Orthogonality enforcement.} We now describe the
enforcement of the orthogonality of $\axisNotation{\geodesicAxis_{d'}}$ to all
previous axes in the basis. In particular,
given the direction vector
$\vectorNotation{\dummyVector_{d'}} = \vectorNotation{\geodesictree_{d'}} -
\vectorNotation{\geodesictree'_{d'}}$,
we want to
update $\vectorNotation{\geodesictree_{d'}}$ and
$\vectorNotation{\geodesictree'_{d'}}$,
such
that
$\vectorNotation{\dummyVector_{d'}}  \cdot \vectorNotation{\dummyVector_{d''}}
= 0$, for all $d'' \in \{1, 2, \dots, d' - 1\}$.
Let
$\projectionOperator_{\vectorNotation{\dummyVector_{d''}}}(
\vectorNotation{\dummyVector_{d'}}) = \big((\vectorNotation{\dummyVector_{d'}}
\cdot \vectorNotation{\dummyVector_{d''}})/(\vectorNotation{\dummyVector_{d''}}
\cdot \vectorNotation{\dummyVector_{d''}})\big) \times
\vectorNotation{\dummyVector_{d''}}$ be the projection of
$\vectorNotation{\dummyVector_{d'}}$ onto the direction spanned by
$\vectorNotation{\dummyVector_{d''}}$.
The orthogonality of
$\vectorNotation{\dummyVector_{d'}}$ to all
vectors
$\vectorNotation{\dummyVector_{d''}}$ is enforced
(\autoref{fig_constraints}c)
via \emph{Gram–Schmidt} orthogonalization \cite{cheney09},
by updating
$\vectorNotation{\dummyVector_{d'}}$ as follows:
$
\vectorNotation{\dummyVector_{d'}} \leftarrow
\vectorNotation{\dummyVector_{d'}} -
\sum_{d'' = 1}^{d'-1} \projectionOperator_{\vectorNotation{\dummyVector_{d''}}}(
\vectorNotation{\dummyVector_{d'}})$.


Note that
the above constraints may go against
each other:
e.g.
after orthogonality enforcement \emph{(iii)},
$\vectorNotation{\geodesictree_{d'}}$ may no longer represent a valid geodesic
\emph{(i)}. Thus, we iterate the above succession of constraint
enforcements, until $\axisNotation{\geodesicAxis_{d'}}$ no longer evolves
significantly (\autoref{algo_mtPga},
\revision{line}
\autoref{alg_line_whileConstraints}).
In practice, using a constant number of iterations (specifically $4$)
is
sufficient to obtain satisfactory results
%
(see
\autoref{sec_limitations} for
further details \revision{regarding convergence}).



%


\subsection{From BDTs to MTs}

\revision{Given an MT-PGA basis,}
for
an
arbitrary coordinate vector $\alpha \in [0, 1]^{\maxDimensions}$, the
resulting
BDT
$\widehat{\branchtree}$ can be reconstructed
(\autoref{sec_defMtPga})
by considering a BDT isomorphic to
$\branchtree^*$ such that:
\begin{eqnarray}
\label{eq_reconstruction}
\reconstructed{\branchtree} = \widehat{\branchtree}_{\maxDimensions} =
\branchtree^* + \sum_{i =
1}^{\maxDimensions}
\alpha_i \times \vectorNotation{\geodesictree_i}
+ (1 - \alpha_i) \times \vectorNotation{\geodesictree'_i}.
\end{eqnarray}

\noindent
Applying the above computation embeds each branch $\branch$ of
$\widehat{\branchtree}$ as a point $(x_{\branch}, y_{\branch})$ in the 2D
birth/death
space (\autoref{fig_wassersteinMetric}, right). Since we consider that the input
BDTs have been pre-normalized (\autoref{sec_metricSpace}), a valid MT
$\widehat{\mergetree}$ can in principle be reverted from
$\widehat{\branchtree}$, by recursively
reverting \autoref{eq_localNormalization}, assuming that $[x_{\branch},
y_{\branch}] \subseteq [0, 1]$ (to enforce the nesting property of BDTs,
\autoref{sec_background_mergeTrees}).
However, the latter assumption is not explicitly enforced in our
constrained optimization (Secs. \ref{sec_optimization} and
\ref{sec_constraints}).
Thus,
in
our entire framework (optimization included),
when
re-constructing a BDT (\autoref{eq_reconstruction}), for each branch
$\branch^* \in \branchtree^*$,
the corresponding
direction
\begin{wrapfigure}{r}{3cm}
  \begin{center}
\vspace{-0.45cm}
\includegraphics[width=\linewidth]{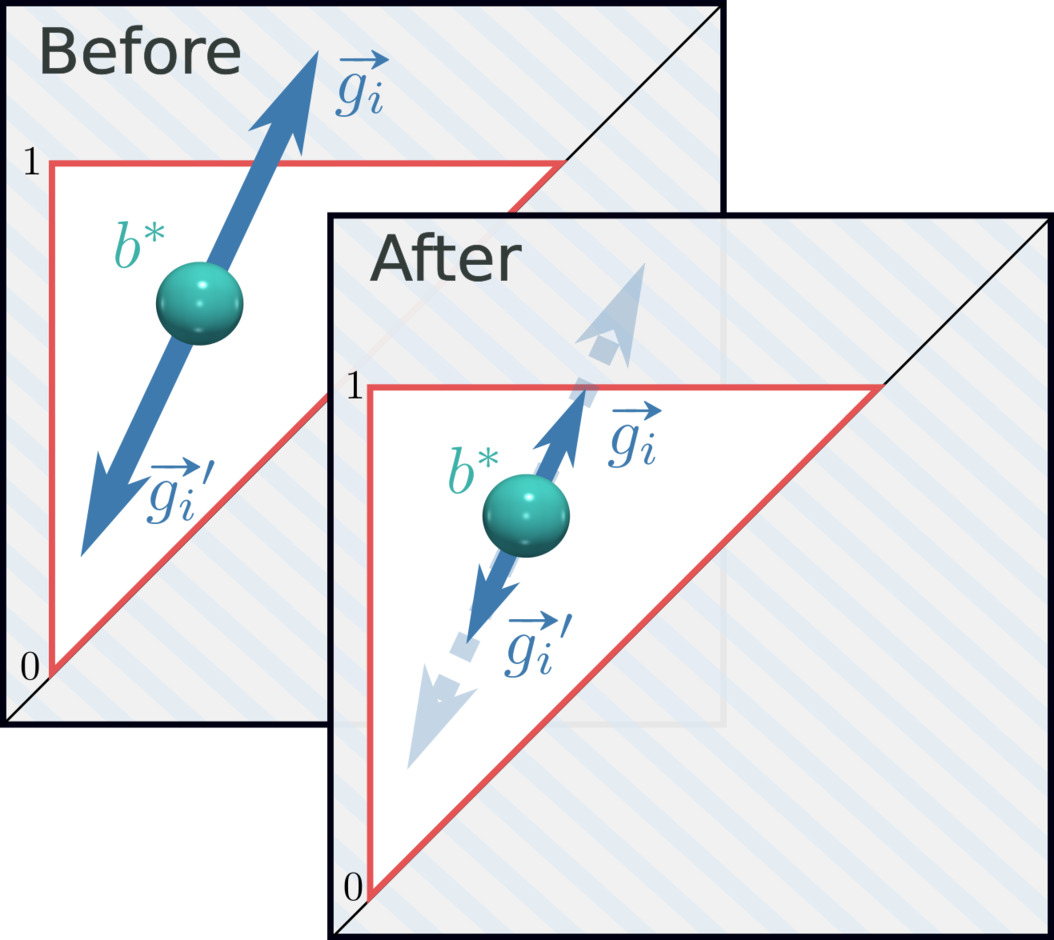}
\vspace{-1cm}
\end{center}
\end{wrapfigure}
vector $\vectorNotation{v_i}
= \vectorNotation{g_i} - \vectorNotation{g'_i}$
(with $\vectorNotation{g_i}$ and $\vectorNotation{g'_i}$ being the entries of
$\vectorNotation{\geodesictree_i}$ and $\vectorNotation{\geodesictree'_i}$
corresponding to the branch $\branch^*$) is temporarily scaled down  if
needed
(inset, right) by locally renormalizing $\alpha_i$, to guarantee
that the extremities $e_i$ ($\alpha_i = 1$) and $e'_i$ ($\alpha_i = 0$) describe
valid normalized birth/death locations
(i.e. $x_{e_i} \lt y_{e_i}$ and $[x_{e_i}, y_{e_i}] \subseteq [0, 1]$).

%

\begin{figure}[b]
\includegraphics[width=\linewidth]{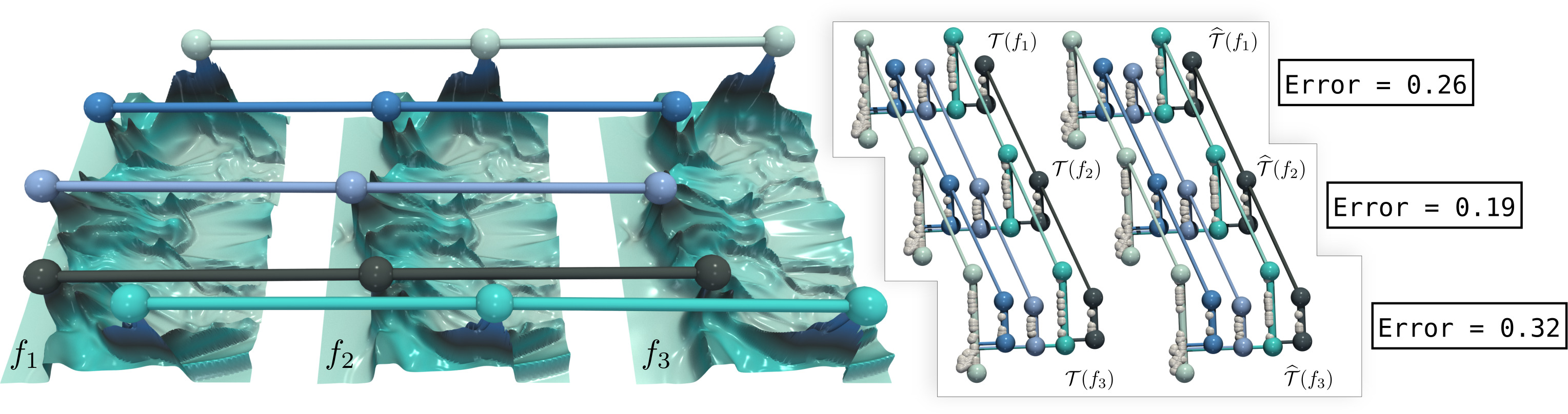}
\caption{Tracking features (the five most persistent maxima, spheres) in
time-varying (from left to right) 2D data (ion density during universe formation
\cite{scivisOverall}). The computed tracking \cite{pont_vis21} is identical,
when using the input original merge trees (inset, left) or their versions
compressed
by MT-PGA (inset, right), which are 
highly similar
visually.
Here, since the input
ensemble has few, small BDTs, the default reduction parameters ($\maxDimensions
= 3$ and
$\numberBranchinBarycenter = 20\%$) resulted in a modest compression factor
(\revision{$5.12$}).
\revision{The 
reported
relative reconstruction 
error (right) is given by the distance $\wassersteinTree$ 
between a tree and its reconstruction, divided by the maximum pairwise 
 distance $\wassersteinTree$ observed in the input ensemble.}
}
\label{fig_tracking}
\end{figure}

\subsection{Computational \revision{parameters}}
\label{sec_computationalAspects}
\revision{As described in previous work \cite{pont_vis21}, the}
Wasserstein metric $\wassersteinTree$ is subject to three parameters
($\epsilon_1$, $\epsilon_2$ and $\epsilon_3$, \autoref{sec_metricSpace}), for
which
we use the recommended default values
\cite{pont_vis21}. Otherwise, when switching $\epsilon_1$ to $1$,
$\wassersteinTree$
becomes equivalent to
$\wasserstein{2}$ (\autoref{sec_metricSpace}) and our framework
%
%
computes then a PGA basis of extremum persistence diagrams (PD-PGA for short).

Our
\revision{algorithm itself (\autoref{algo_mtPga})}
%
%
%
is subject to
two parameters.
\emph{(i)} $\numberBranchinBarycenter$ controls the
size of
$\branchtree^*$, and hence
the memory footprint of the MT-PGA basis.
In
practice, we set $\numberBranchinBarycenter$ to a 
conservative value:
20\% of the \revision{total} number of
branches in the ensemble ($\sum_{j = 1}^\ensembleSize |\branchtree(f_j)|$).
\emph{(ii)} $\numberGeodesicSamples$ controls the number of samples on each
geodesic axis,
to balance accuracy and speed. In practice, we set
$\numberGeodesicSamples$ to $16$.

\begin{figure}
\includegraphics[width=\linewidth]{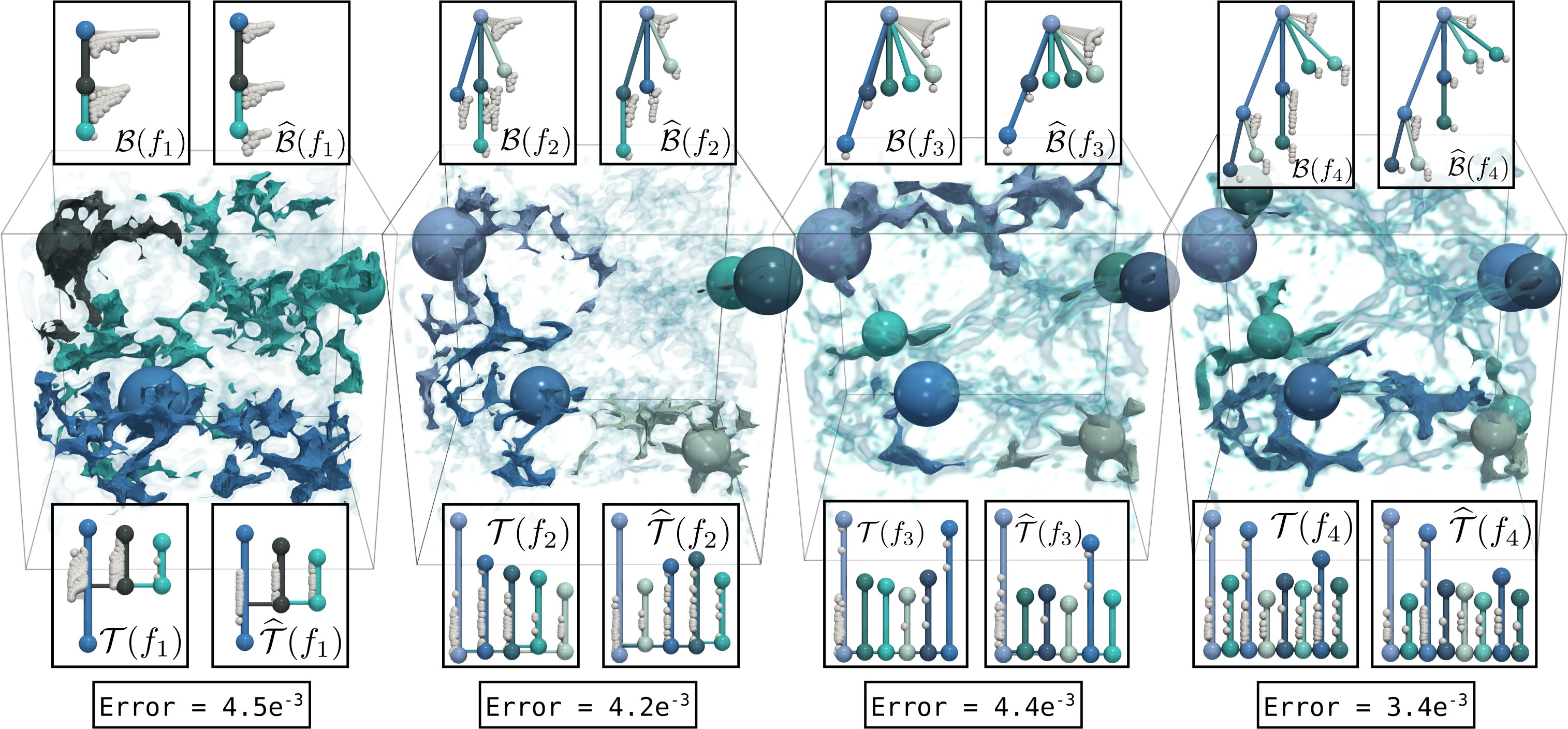}
\mycaption{Topological clustering of the $40$ members of a cosmology ensemble
\cite{scivisOverall} (one member is represented for each 
\revision{cluster}).
The computed clustering \cite{pont_vis21} is identical,
when using the input original merge trees (\revision{bottom left inset}) or their versions compressed
by MT-PGA (\revision{bottom right inset}).
\revision{The optimal assignment
between the left and right trees is visualized by the matching colors and the corresponding BDTs are reported at the top.}
Since this
ensemble has many members,
the default reduction parameters
($\maxDimensions = 3$ and $\numberBranchinBarycenter = 20\%$) resulted in an
important
compression factor
(\revision{$19.27$})\revision{. This compression still provided compressed trees (\revision{bottom right inset}) that are  visually similar to the input trees (\revision{bottom left inset}):
the number of prominent features (thick cylinders)
and
their persistence (heights of the cylinders  of matching color) are well preserved.
This is confirmed with the reconstructed BDTs (top right inset) which are mostly isomorphic to the input BDTs (thick cylinders of matching color).}
\revision{The 
reported
relative reconstruction
error (bottom) is given by the distance $\wassersteinTree$ 
between a tree and its reconstruction, divided by the maximum pairwise 
 distance $\wassersteinTree$ observed in the input ensemble.}
}
\label{fig_clustering}
\end{figure}









%
%

\begin{figure*}
\includegraphics[width=\linewidth]{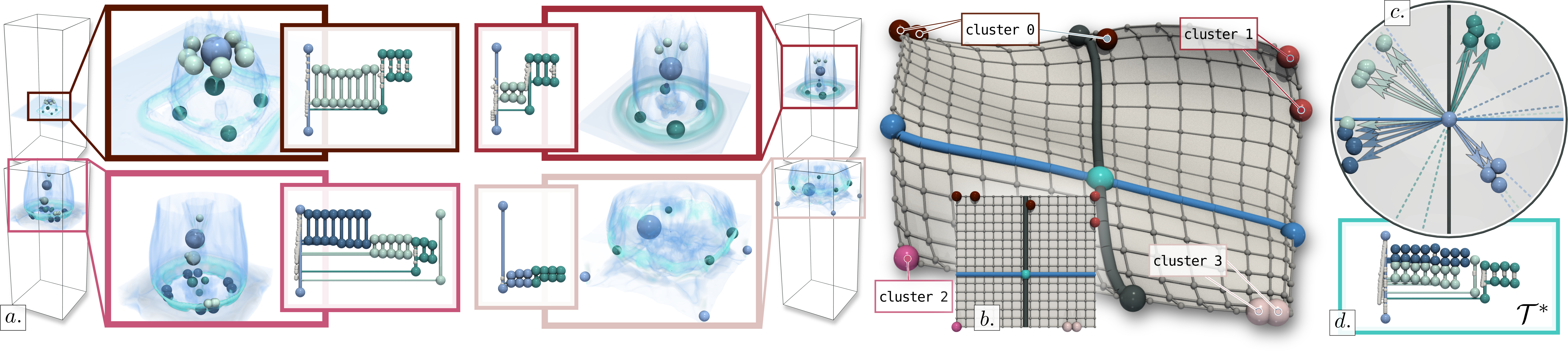}
\mycaption{Feature variability interpretation with \emph{Persistence Correlation
Views} \revision{(PCV)}
on the \emph{Ionization (3D)} ensemble.
\revision{\emph{(a)}}:
Four members of the ensemble
(one per ground-truth class).
\minorRevision{\emph{(b)}}:
\emph{Principal Geodesic Surface}
(PGS) and
its planar layout,
computed by MT-PGA (sphere color:
ground-truth classes). 
\minorRevision{\emph{(c)}:}
In the correlation view,
several
features (i.e. branches) are located near the disk boundary,
indicating high correlations.
The dark green features
are highly correlated
with the
direction $(0, \alpha_2)$, given their location (near the black axis,
upper half). Back in the data, this indicates that the
corresponding features (basis of the ionization front,
green spheres in the
data and merge tree
insets)
will
be the most persistent
for the ensemble members with high
$\alpha_2$ values
(clusters 0 and 1,
\minorRevision{\emph{(b)}}).
Light green features \revision{in the PCV} (upper left quadrant \minorRevision{, \emph{(c)}})
are 
highly
correlated with the direction $(-\alpha_1, \alpha_2)$:
these features (front extremities,
insets)
will be the most persistent
in the
top left corner
\revision{of the PGS}
(\minorRevision{\emph{(b)}}, cluster 0).
\revision{This is confirmed visually in the data (zoom insets), where the maxima are represented by spheres of matching color (the radius denotes persistence).}
}
\label{fig_ionization3D}
\end{figure*}

%

\section{Applications}

This section illustrates the utility of our
framework
in
concrete visualization tasks:
data reduction
and dimensionality reduction.

\subsection{Data reduction}
\label{sec_dataReduction}


\revision{Like any other data representation, merge trees can benefit from lossy compression, to facilitate their storage or transfer.
This can be particularly useful in scenarios where the
scalar fields
of
the
ensemble cannot all be stored permanently (because of their size or the induced IO bottleneck) but are represented instead individually by a topological signature (e.g. a persistence diagram or a merge tree), yielding an ensemble of signatures.
This can be the case for instance during large data acquisition campaigns, or large-scale simulations, where a topological signature
can be
typically computed in-situ \cite{AyachitBGOMFM15} to represent each time step
\cite{BrownNPGMZFVGBT21}. In this scenario, lossy compression is useful to facilitate the manipulation (i.e. storage and transfer) of the resulting ensemble of merge trees. 
Previous work \cite{pont_vis21} has investigated the 
lossy compression of a temporal sequence of merge trees. In this section, we extend this work to arbitrary ensembles of merge trees thanks to MT-PGA.}

Given its coordinates $\alpha^j \in [0,
1]^{\maxDimensions}$, 
each BDT $\branchtree(f_j)$ of 
$\branchtreeSet$ can be
reliably estimated with
\autoref{eq_reconstruction}.
Thus, we present now
an application to data reduction where the input ensemble of BDTs is compressed,
by only storing to disk: \emph{(i)} the origin $\branchtree^*$ of $\mtPgaBasis$,
\emph{(ii)} its axes $\{\axisNotation{\geodesicAxis_1},
\axisNotation{\geodesicAxis_2}, \dots,
\axisNotation{\geodesicAxis_{\maxDimensions}}\}$ and \emph{(iii)} the $N$ BDT
coordinates $\alpha^j \in [0,
1]^{\maxDimensions}$. The compression quality can be controlled with two
input parameters (\autoref{sec_overview}).
\emph{(i)} $\maxDimensions$ controls the
number of 
axes (and thus the ability of the basis to
capture mild variabilities).
\emph{(ii)} $\numberBranchinBarycenter$ controls the
size of
$\branchtree^*$ 
(and thus 
the ability of the
basis to capture small features). The reconstruction error
(\autoref{eq_mtPgaEnergy}) will be minimized for large values of both
parameters, while the compression factor will be maximized for low values.
In our reduction experiments,
we set $\maxDimensions$ to $3$ and
$\numberBranchinBarycenter$ to its
default
value
(\autoref{sec_computationalAspects}).
Figs. \ref{fig_tracking} and
\ref{fig_clustering}
show two examples of
visualization applications (feature tracking and ensemble clustering,
replicated from \cite{pont_vis21})
where the
BDTs compressed with the above strategy
($\reconstructed{\branchtree}(f_j)$) have been used as an input.
In both cases, the output is identical to the outcome obtained with the
\emph{original} BDTs.
This shows the
viability of
our
reconstructed BDTs (and MTs)
and
demonstrates the utility of our
reduction scheme.
\revision{A detailed analysis of the resulting compression factors and 
reconstruction errors for all our test ensembles is available in Appendix 
\minorRevision{D}.}



%

\subsection{Dimensionality reduction}
\label{sec_dimensionalityReduction}
The MT-PGA basis $\mtPgaBasis$ can also be used to
generate 
2D layouts of the ensemble, for its global visual
inspection.
This is achieved
by embedding each input BDT $\branchtree(f_j)$ 
as a point
in the 2D plane, given its first two coordinates in $\mtPgaBasis$.
To prevent an artificial anisotropic distortion 
($\alpha^j \in [0, 1]^2$), we 
scale the
coordinates of each BDT, to account for the variation in length of the 
axes, and we embed each BDT at coordinates
$\big(\alpha_1^j \times \wassersteinTree(\geodesicExtremity_1,
\geodesicExtremity_1'),
\alpha_2^j \times \wassersteinTree(\geodesicExtremity_2,
\geodesicExtremity_2')\big)$. This results in a summarization view of
the overall ensemble, which groups together 
BDTs which
are close (given the metric $\wassersteinTree$) and which
exhibit a similar
variability with regard to $\branchtree^*$.
Note that, 
given an arbitrary point
of the  2D
layout, its BDT (and MT) can be
efficiently reconstructed with \autoref{eq_reconstruction}.
This enables interactive navigations within the ensemble 
\revision{(\autoref{fig:teaser}\minorRevision{f})}.
We augment our planar layouts 
with the following two improvements, to further characterize the global and 
local variability in the ensemble.


\noindent
\textbf{\emph{(i)} Principal geodesic surface.}
To visually convey the curved (i.e. non-Euclidean) nature of the Wasserstein
metric space $\branchtreeSpace$, we introduce a 3D embedding of
our planar layouts, which we call Principal Geodesic Surface (PGS).
First, our 2D
layout is
sampled along a $\planarGridX\times\planarGridY$
regular grid $\regularGrid_\branchtreeSpace$ (in practice, $\planarGridX
= \planarGridY = \numberGeodesicSamples$). For each vertex of
$\regularGrid_\branchtreeSpace$,
a BDT can be
reconstructed
(\autoref{eq_reconstruction}), enabling the computation of a distance matrix
$\distanceMatrix_{\regularGrid_\branchtreeSpace}$, where
$\distanceMatrix_{\regularGrid_\branchtreeSpace}(i, j)$
is the
$\wassersteinTree$ distance between the
vertices $i$
and $j$ of $\regularGrid_\branchtreeSpace$. Then,
$\regularGrid_\branchtreeSpace$ is
embedded in 3D by multidimensional scaling \cite{kruskal78} of
$\distanceMatrix_{\regularGrid_\branchtreeSpace}$.
The resulting PGS
provides the same visual information as the planar
layout, but in the form of
a surface
in 3D, parameterized by
$\alpha_1$ and $\alpha_2$ (\autoref{fig_mtpgaOverview}), which
conveys
visually the curved nature of
$\branchtreeSpace$.

\begin{figure}
\includegraphics[width=\linewidth]{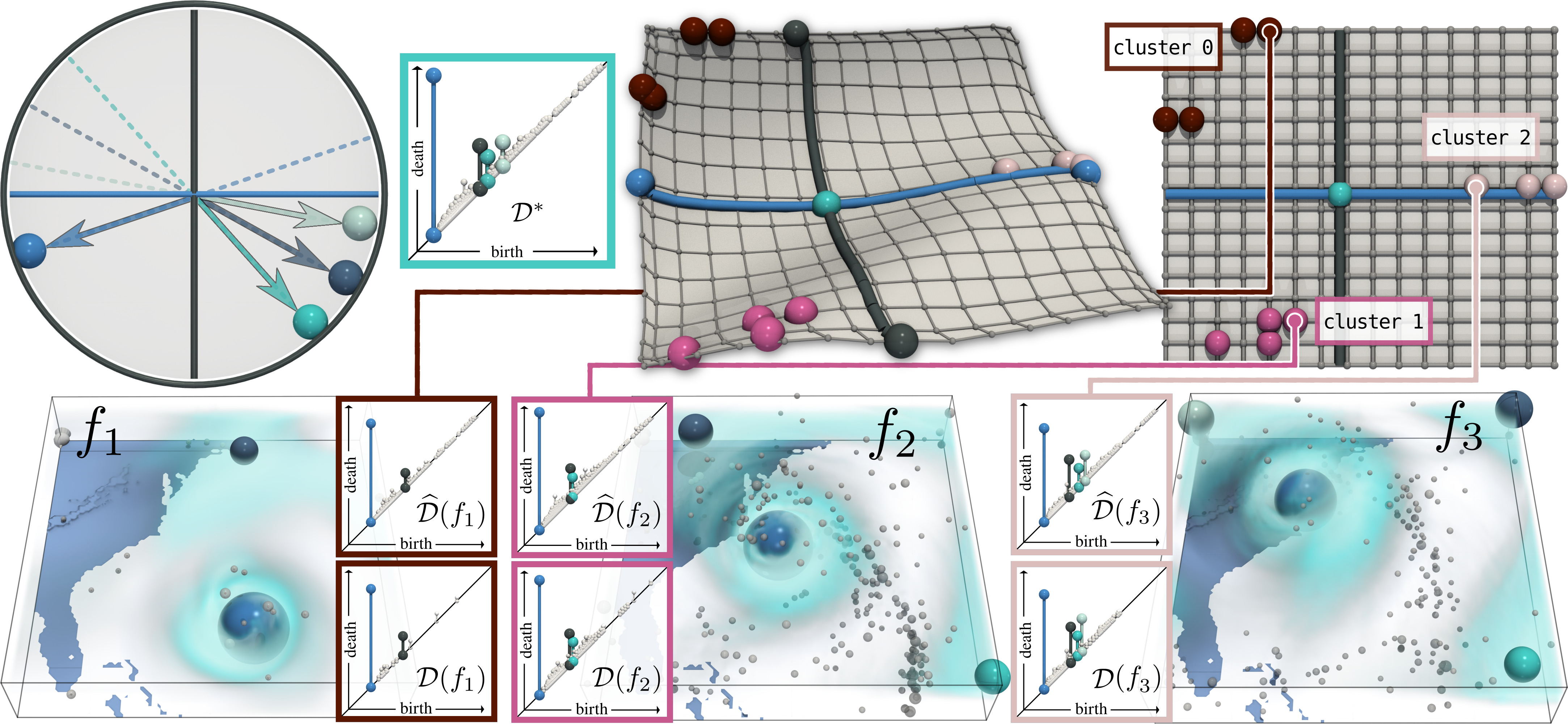}
\mycaption{Feature variability interpretation for the \emph{Isabel} ensemble,
with PD-PGA. In the
\revision{PCV}
(\minorRevision{top left}), the  blue feature
(eye of the
hurricane) is highly correlated with the direction  $(-\alpha_1, 0)$, which
indicates that its persistence will be stronger for the leftmost points in the
PGS
(clusters 0 and 1) \revision{and weaker for the rightmost points (cluster 2)}.
\revision{In other words, the winds in the eye of the hurricane are significantly weaker in the cluster 2 (landfall phase of the hurricane).}
On the contrary, the other highly correlated features
(right half \revision{of the PCV}) are correlated with the direction $(\alpha_1, 0)$:
their persistence will be stronger for the rightmost points
(cluster 2).
This is
confirmed visually when inspecting the corresponding
features in the data and the input diagrams (\minorRevision{bottom}). Similarly to MT-PGA, PD-PGA
enables a
data reduction
(\revision{compression factor: $5.49$}),
while guaranteeing
\revision{visually}
similar
reconstructed diagrams
($\widehat{\diagram}(f_j)$).
}
%
\label{fig_pdPga}
\end{figure}

\noindent
\textbf{\emph{(ii)} Persistence correlation view.}
As detailed in Appendix \minorRevision{E},
we compute
the correlation $\rho(p_i, \alpha_k)$  between
the coordinate $\alpha_k$ and
the persistence of the
$i^{th}$ branch (i.e. the branch in the input BDTs mapped to the $i^{th}$
branch of $\branchtree^*$ given the optimal assignment induced by
$\wassersteinTree$, \autoref{eq_wasserstein}).
Then, the
$i^{th}$ branch of
the barycenter BDT $\branchtree^*$ can be
embedded
in
\revision{a Persistence Correlation View (PCV),}
by
placing an arrow between the origin
and the point
$(\rho_{p_i, \alpha_1}, \rho_{p_i, \alpha_2})$.
This enables a
\emph{local} interpretation of the feature variability 
within the ensemble.
\revision{Specifically,
it
enables the visual identification of
the features
whose persistence is strongly correlated with a given direction in the MT-PGA basis.
Such features are located on the disk boundary of the correlation view (largest arrows in Figs. \ref{fig:teaser}, \ref{fig_ionization3D}, \ref{fig_pdPga}), and they are the most responsible for the variability in the ensemble. For each of these features, their matching to the
origin $\branchtree^*$ of the MT-PGA basis is encoded with the color map, which enables their direct inspection in the data. }

\revision{Together,
our Principal Geodesic Surface (PGS) coupled with our Persistence Correlation View (PCV) enable both a global and local inspection of the feature variability in the ensemble.}
\revision{In particular, in the example of \autoref{fig:teaser},
our visualization indicates overall that the global maximum of the seismic wave
(largest branch in the merge trees, purple, 
\autoref{fig:teaser}\minorRevision{c}) is mostly correlated with the direction 
$(-\alpha_1, 0)$ in the
PCV
(\autoref{fig:teaser}d), and therefore, mostly prominent in the cluster $1$ 
(located on the left of the PGS, \autoref{fig:teaser}\minorRevision{b}, for the 
lowest $\alpha_1$ values).
In contrast, the other features (blue, cyan and dark blue branches) are less correlated with this direction (shorter arrows in the PCV \autoref{fig:teaser}d), which indicates that they are slightly less variable through this direction:
their
persistence
decreases less quickly than the global maximum when transitioning from cluster $1$ to $2$ and $3$ 
(dark red, orange and pink spheres respectively in the PGS 
\autoref{fig:teaser}\minorRevision{g}). This
indicates that the initial energy of the seismic wave (represented by the global maximum) quickly spreads into multiple wavefronts (spheres of matching colors in the data, \autoref{fig:teaser}a), whose individual energy decreases through time more slowly than the global maximum (each cluster represents a different temporal phase in the simulation, from top to bottom in \autoref{fig:teaser}a).}
\revision{Figs. \ref{fig_ionization3D} and \ref{fig_pdPga} present a similar analysis and we refer the reader to the detailed captions for specific interpretations. Note that for
\autoref{fig_pdPga}, the features do not exhibit a clear global structure and the persistence diagram can be used instead of the merge tree.}

\section{Results}
\label{sec_results}

This section presents experimental results obtained on a computer with
two Xeon CPUs (3.2 GHz, 2x10 cores, 96GB of RAM). The input
merge trees were computed with FTM \cite{gueunet_tpds19} and pre-processed to
discard noisy features
(persistence simplification threshold:
$0.25\%$ of the data range).
%
We
implemented our approach in C++ (with the OpenMP task runtime),
as modules for TTK \cite{ttk17, ttk19}.
Experiments were performed on the benchmark of public ensembles
\cite{ensembleBenchmark} described in
\cite{pont_vis21}, which includes a variety of simulated and acquired 2D and 3D
ensembles extracted from previous work and past SciVis contests
\cite{scivisOverall}.

\subsection{Time performance}
Barycenters 
and geodesics 
are computed with the approach of Pont et al.
\cite{pont_vis21}, which implements fine-grain task-based shared-memory
parallelism.
Specifically,
during axis projection,
$\ensembleSize \times \numberGeodesicSamples$
geodesics need to be computed (\autoref{sec_optimization}),
each geodesic
requiring typically $\mathcal{O}(|\branchtree|^2)$
steps in practice,
where $|\branchtree|$ is the 
size of the
input BDTs.
In practice, this
is 
the most 
expensive part of 
our
algorithm. These geodesics are computed concurrently (by submitting each
geodesic to the task pool). In comparison, the evaluations of the numerical
expressions 
(\autoref{algo_mtPga},
lines \ref{alg_evaluateEnergy},
\ref{alg_line_optimize}, \ref{alg_line_secondConstraint} and
\ref{alg_line_lastConstraint}) have a nearly negligible 
cost.
\autoref{tab_timings} evaluates the 
time performance of our framework
for persistence diagrams (PD-PGA) and merge
trees (MT-PGA), for $\maxDimensions = 2$.
In sequential mode,
the running time is indeed a function of the size of the ensemble
($\ensembleSize$) and 
the size of 
trees ($|\branchtree|$). It is slightly slower for MT-PGA than for
PD-PGA, but timings remain comparable overall.
In parallel, speedups are the most important for the largest ensembles.
However, the iterative nature of 
our 
algorithm has an
impact on parallel efficiency
(the end of each 
loop
implies a
synchronization). Still, our parallelization significantly reduces
computation times,
with less than 6 minutes on average and at
most 30 minutes for the largest ensembles, which we believe 
\revision{is}
an
acceptable pre-processing time, prior to interactive exploration.








\begin{table} 
  \caption{Running times (in seconds)
  of our algorithm for
PD-PGA and MT-PGA computation (for $\maxDimensions = 2$, first
sequential, then with 20 cores).}
  \centering
  \scalebox{0.6}{
    \begin{tabular}{|l|r|r||r|r|r||r|r|r|}
      \hline
      \rule{0pt}{2.25ex} \textbf{Dataset} & $\ensembleSize$ & $|\branchtree|$ &
\multicolumn{3}{c||}{PD-PGA} & \multicolumn{3}{c|}{MT-PGA}\\
                       &      &                 & 1 c. & 20 c. & Speedup & 1 c. & 20 c. & Speedup\\
      \hline
      Asteroid Impact (3D) & 7 & 1,295 & 1,392.17 & 147.40 & 9.44 & 1,180.72 &
117.97 & 10.01 \\
      Cloud processes (2D) & 12 & 1,209 & 817.64 & 61.88 & 13.21 & 517.94 &
38.49 & 13.46 \\
      Viscous fingering (3D) & 15 & 118 & 86.69 & 9.17 & 9.45 & 42.89 & 4.71 &
9.11 \\
      Dark matter (3D) & 40 & 2,592  & 18,388.86 & 1,366.45 & 13.46 & 24,480.42
& 1,758.04 & 13.92 \\
      Volcanic eruptions (2D) & 12 & 811 & 460.17 & 37.99 & 12.11 & 1,004.37 &
81.75 & 12.29 \\
      Ionization front (2D) & 16 & 135 & 104.74 & 12.00 & 8.73 & 55.73 & 6.26 &
8.90 \\
      Ionization front (3D) & 16 & 763 & 3,750.00 & 300.96 & 12.46 & 4,029.71 &
294.29 & 13.69 \\
      Earthquake (3D) & 12 & 1,203 & 3,896.52 & 338.64 & 11.51 & 1,973.49 &
158.12 & 12.48 \\
      Isabel (3D) & 12 & 1,338 & 1,969.79 & 164.49 & 11.98 & 1,472.54 & 115.66 &
12.73 \\
      Starting Vortex (2D) & 12 & 124 & 17.71 & 2.72 & 6.51 & 11.51 & 1.65 &
6.98 \\
      Sea Surface Height (2D) & 48 & 1,787 & 12,420.98 & 670.00 & 18.54 &
27,791.00 & 1,669.52 & 16.65 \\
      Vortex Street (2D) & 45 & 23 & 18.75 & 2.69 & 6.97 & 35.79 & 3.93 & 9.11
\\
      \hline
    \end{tabular}
  }
  \label{tab_timings}
\end{table}

\begin{figure}
\includegraphics[width=\linewidth]{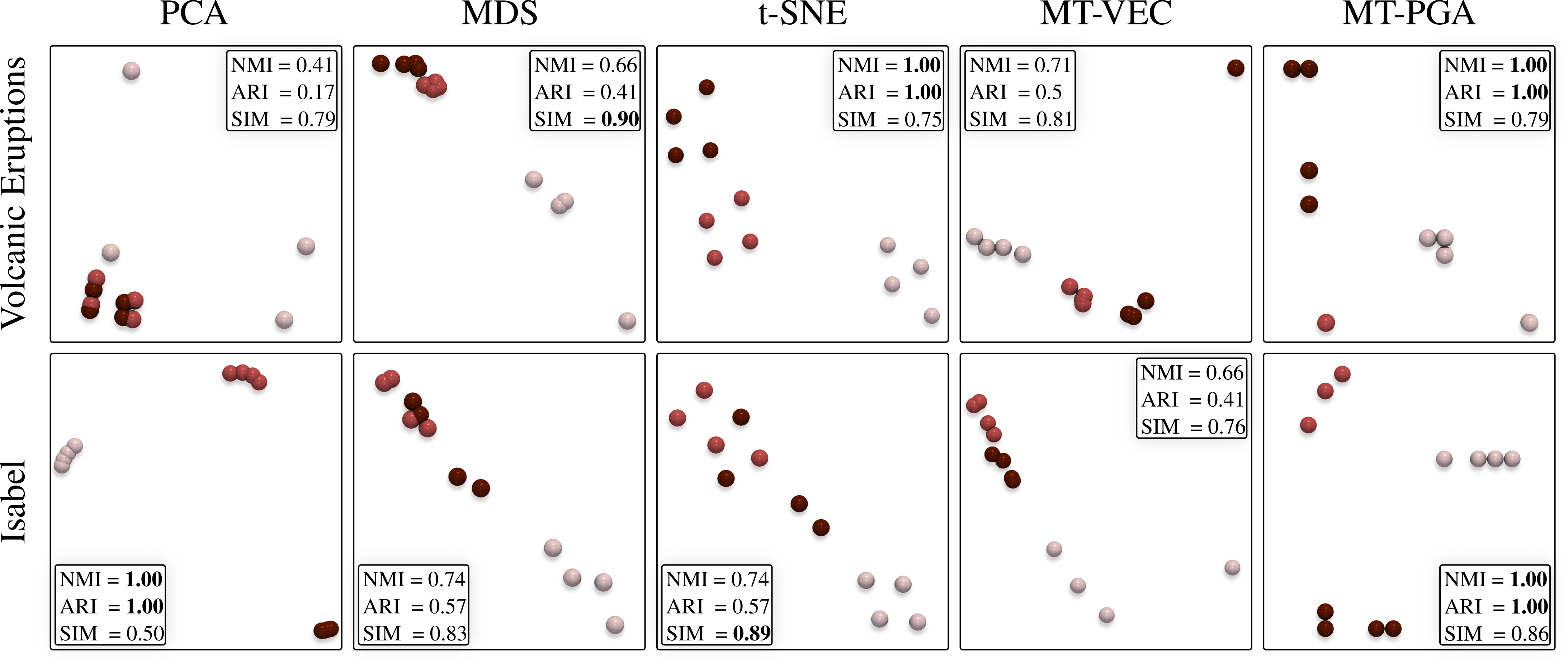}
\caption{Comparison of planar layouts for typical dimensionality reduction
techniques, on two input ensembles \revision{(bold: best value for each
quality score)}. The color encodes the classification
ground-truth
\cite{ensembleBenchmark}.
}
\label{fig_layoutComparison}
\end{figure}


\begin{table}
  \caption{Comparison of layout quality scores, averaged over all
ensembles (bold: best values).
  The layouts induced by MT-PGA better preserve
  the global structure of the ensemble and still preserve well
$\wassersteinTree$.}
  \centering
  \scalebox{0.75}{
    \begin{tabular}{|l||r|r|r|r||r|}
     \hline
     \textbf{Indicator} & PCA & MDS
     ($\wassersteinTree$) \cite{kruskal78}
     & t-SNE
     ($\wassersteinTree$) \cite{tSNE}
     & MT-VEC & \textbf{MT-PGA}\\
    \hline
    NMI & 0.79 & 0.82	&0.88	&0.78	&\textbf{0.94}\\
    ARI & 0.71 &	0.73	&  0.84	& 0.64	& \textbf{0.90}\\
    \hline
    SIM & 0.60 & \textbf{0.85}	& 0.75	& 0.76	&0.80\\
    \hline
    \end{tabular}
  }
  \label{tab_qualityIndicators}
\end{table}

\subsection{Framework quality}
\label{sec_quality}
Figs. \ref{fig_tracking} and
\ref{fig_clustering} report compression factors for our application to
data reduction (\autoref{sec_dataReduction}). These are 
ratios
between the storage size of the input 
BDTs and that
of the MT-PGA basis (barycenter, axes and coordinates).
This factor is
modest for a small example (\autoref{fig_tracking}), 
with few branches
($135$) and few members ($16$). Then, the overhead of the MT-PGA basis
is
non-negligible. In contrast, for a larger ensemble
(\autoref{fig_clustering}), this overhead 
is negligible and high
compression
factors ($30$)
can be
achieved, while providing reconstructed
merge trees which are highly similar visually
and which are still
viable for the applications.

\autoref{fig_layoutComparison} provides a visual comparison of the
planar
layouts generated by a selection of typical dimensionality
reduction techniques.
This 
includes the classical Euclidean
PCA (in $\mathbb{R}^{N_v}$, where $N_v$ is the number of vertices in 
$\domain$), MDS \cite{kruskal78} and t-SNE \cite{tSNE} 
(both
with
$\wassersteinTree$). 
Certain
approaches \cite{robins16, AnirudhVRT16, LI2021} applied PCA on top of 
\emph{vectorizations} of topological
descriptors (\autoref{sec_relatedWork}).
A similar
strategy can be
considered in our case, by embedding each input BDT $\branchtree(f_j)$ in
$\mathbb{R}^{2\times|\branchtree^*|}$, such that the $i^{th}$ entry of this
vector corresponds to the birth/death location of the $i^{th}$ branch of
$\branchtree(f_j)$ (i.e. the branch of $\branchtree(f_j)$ mapping to the
$i^{th}$ branch of $\branchtree^*$ by the optimal assignment
of $\wassersteinTree$, \autoref{eq_wasserstein}).
PCA is then computed for this vectorization (column
\emph{MT-VEC}). As shown in \autoref{fig_layoutComparison}, the layouts
generated with MT-PGA nicely separates the ground-truth classes, while
other techniques tend to artificially group them or even merge them.
To further
quantify this
structure preservation,
we run $k$-means in the
2D
layouts and evaluate the
quality of the resulting clustering (given the ground-truth 
\cite{ensembleBenchmark})
with
the normalized mutual information
($NMI$)
and adjusted rand index ($ARI$).
The MT-PGA layout
is the only one in
\autoref{fig_layoutComparison} which generates an exact clustering in both
cases ($NMI = ARI = 1$).
\revision{Appendix \minorRevision{F} extends this analysis to all our test 
ensembles.}
\autoref{tab_qualityIndicators} \revision{also} extends this
quantitative comparison to all our input ensembles
and
confirms the superiority, on average, of MT-PGA for the preservation of
the clusters.
\autoref{tab_qualityIndicators} also
includes a metric similarity indicator,  $SIM$, which evaluates the
preservation by a layout of the Wasserstein metric $\wassersteinTree$,
and whose expression is given in Appendix \minorRevision{G}. As expected, MDS
maximizes this score by design, while MT-PGA produces
the second best score. Overall, 
MT-PGA
preserves well $\wassersteinTree$ 
as well as the 
global ensemble structure. 
In comparison  to standard
techniques (e.g. MDS or t-SNE), it also supports additional
features, such as 
correlation views and interactive reconstructions
(\revision{Figs. \ref{fig:teaser}, \ref{fig_ionization3D} and \ref{fig_pdPga}}).
%



\begin{figure}
\makebox[\linewidth]{
\centering
\includegraphics[width=\linewidth]{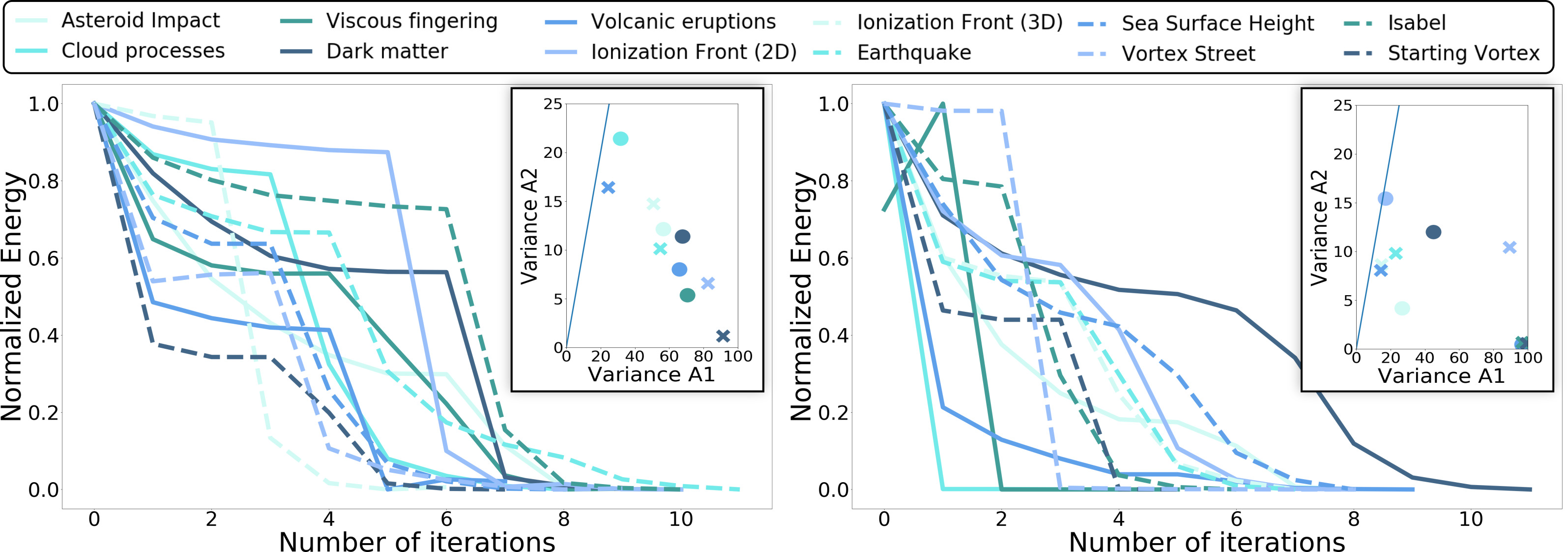}
}
\mycaption{Evolution of the normalized fitting energy, 
for
PD (left) and MT (right) PGA. 
The scatter plots show the projected
variance ($\axisNotation{\geodesicAxis_1}$ VS
$\axisNotation{\geodesicAxis_2}$). }
\label{fig_Energy}
\end{figure}

\autoref{fig_Energy} reports the evolution of the normalized fitting energy
for PD
and MT-PGA computations ($\maxDimensions = 2$). 
Sudden energy drops 
can be observed with clear kinks in the curves, indicating the
finalization of the first dimension (\autoref{sec_overview}), and the switch
to the second, which immediately improves
the overall fit.
The algorithm stops when the energy decreases by less than $1\%$.
The curve flat tails 
indicate that this
criterion
is reasonable.
\autoref{fig_Energy}
also reports
scatter plots of the \emph{projected variances}. For a given axis, it
is
the percentage of the variance of the projected trees (along the
axis) over the global variance of the input BDTs (i.e.
average of the
squared
$\wassersteinTree$
distances to the barycenter $\branchtree^*$). For all examples, the
corresponding point is located below the diagonal (blue line):
the projected variance
is indeed
larger
for the first axis
than
for the second.
This
confirms the ability of our algorithm, similarly to the classical PCA, to
identify in practice the most informative directions first. This is
confirmed
visually 
in our 2D layouts, where the first 
axis (blue) is always longer than the second (black).
\revision{This is also confirmed in Appendix \minorRevision{H}, which provides 
a detailed variance analysis, for all test ensembles, in up to 10 dimensions.}





\subsection{Limitations}
\label{sec_limitations}
Our overall strategy (alternation of fitting and constraint enforcement) is 
similar at a high-level to previous work 
on the optimal transport 
of histograms \cite{SeguyC15}.
Thus, it shares the same high-level limitations.
The constraint enforcement
(\autoref{sec_constraints})  induces, by construction,
an energy increase. Then, it is possible that, at the next 
iteration,
the fitting 
optimization (which is itself guaranteed to decrease the energy,
\autoref{sec_optimization}) does not manage to compensate the above
 increase.
This 
situation
occurs during the 
first 
iteration
for the \emph{Viscous Fingering} ensemble (dark green line,
\autoref{fig_Energy}, right), where the energy increased between two
iterations.
This temporary energy increase is the only one we observed
in
all our experiments. Moreover, it did not impact the
rest of the algorithm, as it
%
was immediately compensated at the 
next
iteration
(\autoref{fig_Energy}). Then, we believe that this theoretical limitation has a
very limited impact in practice 
and that
robust implementations can be obtained by stopping the algorithm in case of
multiple, consecutive energy increases (which we have not observed).
Another limitation involves the sampling of the geodesic axes
($\numberGeodesicSamples$, \autoref{sec_optimization}). When 
$\numberGeodesicSamples$ is too low,
BDTs which are close in $\branchtreeSpace$ may project to the
same points on the geodesic axes, possibly resulting overall in collocated
points in the MT-PGA basis. 
This can be easily resolved by
increasing
$\numberGeodesicSamples$, at the cost of increased computation times.
Finally, similarly to barycenter
optimization \cite{Turner2014, vidal_vis19,
pont_vis21}, the overall fitting energy (\autoref{eq_mtPgaEnergy}) is
non-convex and can in principle admit multiple local minimizers. However, our
experiments indicate that the axes returned by our approach are relevant, as
the projected variance does decrease for increasing dimensions
(\autoref{sec_quality}).





\section{Conclusion}

In this paper, we presented a computational framework for the Principal
Geodesic Analysis of merge trees (MT-PGA), with applications to data reduction
and dimensionality reduction. In particular, the visualizations  derived from
our core contribution
(Figs.
\ref{fig:teaser}, \ref{fig_ionization3D}, \ref{fig_pdPga})
enable the interactive, visual inspection of the variability in the
ensemble, both at a global level (with our two-dimensional layouts) and at a
feature level (with our persistence correlation views).
Our framework trivially extends to extremum persistence diagrams and
our algorithm enables in both cases PGA basis computations within minutes for
real-life ensembles.

A natural direction for future work is the extension of our framework to other
topological data representations, such as Reeb graphs or Morse-Smale complexes.
However, as detailed in \autoref{sec_basis}, this requires the definition of
several low-level geometrical tools, such as routines for geodesic or
barycenter computation, which is still an open research problem for the above
topological descriptors.
By adapting
Principal Component Analysis to merge trees, we
believe our work is an important practical step towards the definition of a
larger statistical framework on the space of merge trees. In the future,
we will continue our investigation of the adaptation of classical statistical
tools to ensembles of topological objects, as we believe it can become a
key solution in the long term for the advanced analysis of large-scale
ensembles.


\section*{Acknowledgments}
{\small We thank the reviewers for their careful feedback and suggestions.
This work is partially supported by the European Commission grant
ERC-2019-COG 
\emph{``TORI''} (ref. 863464, 
\url{https://erc-tori.github.io/}).}


%

%



\ifCLASSOPTIONcaptionsoff
  \newpage
\fi



%

\bibliographystyle{abbrv-doi}
\bibliography{template}




%

\begin{IEEEbiography}[{\includegraphics[width=1in,height=1.25in,clip,
keepaspectratio]{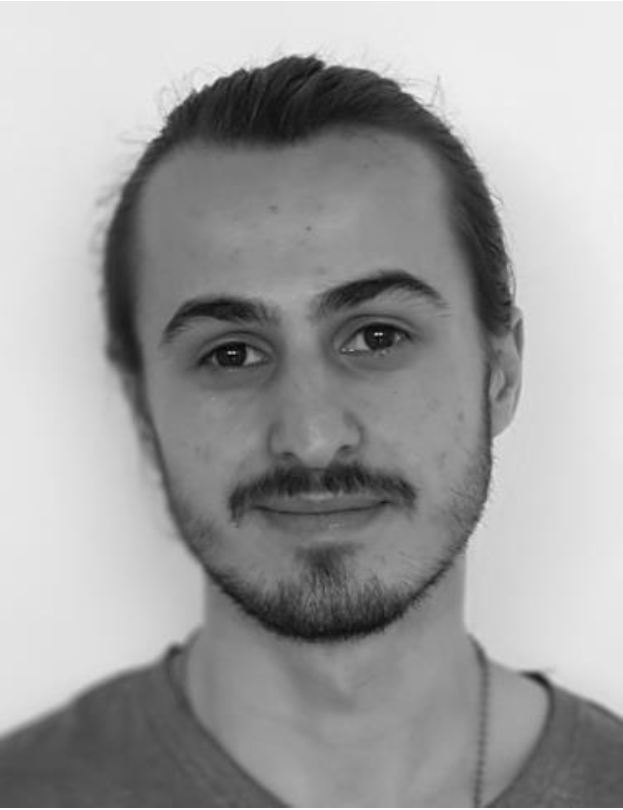}}]{Mathieu Pont}
is a Ph.D. student at Sorbonne Universite. He received a M.S. degree in Computer Science
from Paris Descartes University in 2020.
He is an active contributor to
the Topology ToolKit
(TTK), an open source library for
topological data analysis.
His notable contributions to TTK include distances, geodesics and barycenters
of merge trees, for feature tracking and ensemble clustering.
\end{IEEEbiography}

%
%

\begin{IEEEbiography}[{\includegraphics[width=1in,height=1.25in,clip,
keepaspectratio]{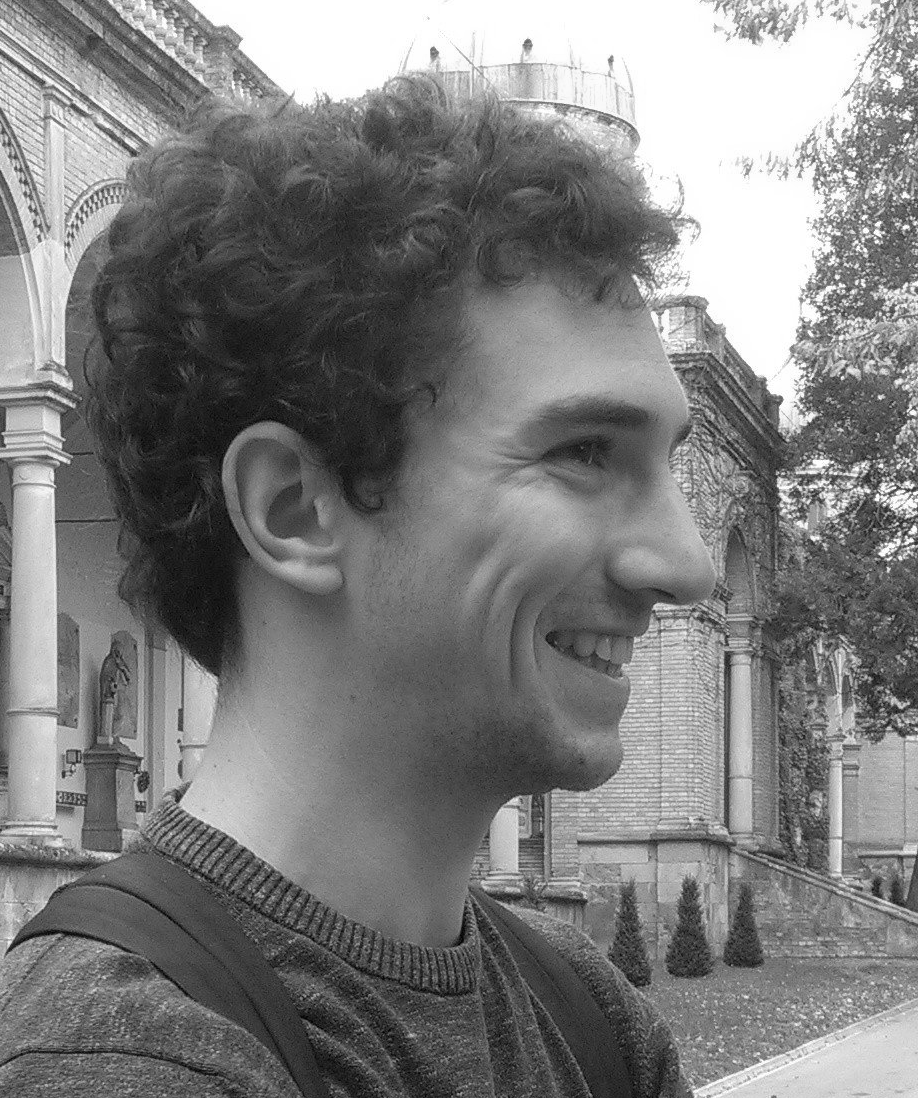}}]{Jules Vidal}
is a post-doctoral researcher, currently at Sorbonne Université, from where he received the Ph.D. degree in 2021.
He received the engineering degree
in 2018 from ENSTA Paris.
He is an active contributor to
the Topology ToolKit
(TTK), an open source library for
topological data analysis.
His notable contributions to TTK include the
efficient and progressive approximation of distances, barycenters and
clusterings of persistence diagrams.
\end{IEEEbiography}



\begin{IEEEbiography}[{\includegraphics[width=1in,height=1.25in,clip,
keepaspectratio]{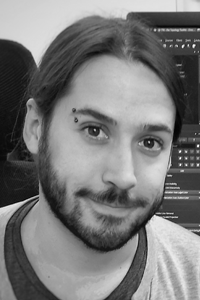}}]{Julien Tierny}
received the Ph.D. degree in Computer Science from the University
of Lille in 2008.
He is currently a CNRS research director, affiliated with
Sorbonne University. Prior to his CNRS tenure, he held a
Fulbright fellowship (U.S. Department of State) and was a post-doctoral
researcher at the Scientific Computing and Imaging Institute at the University
of Utah.
His research expertise lies in topological methods for data analysis
and visualization.
He is the founder and lead developer of the Topology ToolKit
(TTK), an open source library for topological data analysis.
\end{IEEEbiography}




\setcounter{section}{0}
\section*{Appendix}
\includegraphics{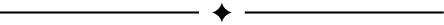}
\appendices
\section{\minorRevision{Glossary and notations}}
\minorRevision{This appendix includes a glossary and a notation table (Tabs \ref{tab_glossary} and \ref{tab_notations}), for the abbreviations and notations which are the most frequently used in the main manuscript.}

\begin{table}
  \caption{\minorRevision{Glossary of the abbreviations used in the main manuscript.}}
  \label{tab_glossary}
  \centering
  \scalebox{0.9}{
    \begin{tabular}{|l|l|}
      \hline
      \rule{0pt}{2.25ex} \textbf{Acronym} & ~\\
      \hline
      ARI & Adjusted Rand Index\\
      BDT & Branch Decomposition Tree\\
      CPU & Central Processing Unit\\
      FTM & Fibonacci Task-based Merge tree computation (TTK backend)\\
      GB & Giga Byte\\
      MDS & Multi-Dimensional Scaling\\
      MT & Merge Tree\\
      MT-PGA & Merge Tree Principal Geodesic Analysis\\
      MT-VEC & Merge Tree VECtorization\\
      NMI & Normalized Mutual Information\\
      PCA & Principal Component Analysis\\
      PCV & Persistence Correlation View\\
      PD & Persistence Diagram\\
      PD-PGA & Persistence Diagram Principal Geodesic Analysis\\
      PDF & Probability Density Function\\
      PGA & Principal Geodesic Analysis\\
      PGS & Principal Geodesic Surface\\
      PL & Piecewise Linear\\
      RAM & Random Access Memory\\
      SIM & SIMilarity indicator\\
      t-SNE & t-distributed Stochastic Neighbor Embedding\\
      TDA & Topological Data Analysis\\
      TTK & Topology Tool Kit\\
      \hline
    \end{tabular}
  }
  \label{tab_timings}
\end{table}

\begin{table}
  \caption{\minorRevision{Notations used in the main manuscript.}}
  \label{tab_notations}
  \centering
  \scalebox{0.7}{
    \begin{tabular}{|l|l|}
      \hline
      \rule{0pt}{2.25ex} \textbf{Symbol} & ~\\
      \hline
      $\ensembleSize$ & Size of the ensemble (number of members)\\
      $\domain$ & Input PL $d$-manifold\\
      $d$ & Dimension of $\domain$\\
      $f_i : \domain \rightarrow \range$& $i^{th}$ input PL scalar field ($i^{th}$ member of the input ensemble)\\
      $\sublevelset{{f_i}}(\isovalue)$ & Sub-level set of $f_i$ at the isovalue $\isovalue$\\
      $\superlevelset{{f_i}}(\isovalue)$ & Super-level set of $f_i$ at the isovalue $\isovalue$\\
      $\Index_i(c)$ & Index of the critical point $c$ in the field $f_i$\\
      $\diagram(f_i)$ & Persistence diagram of $f_i$\\
      $\jointree(f_i)$ & Join tree of $f_i$\\
      $\splittree(f_i)$ & Split tree of $f_i$\\
      $\mergetree(f_i)$ & Generic merge tree of $f_i$\\
      $\branchtree(f_i)$ & Branch decomposition tree (BDT) of $f_i$\\
      $|\branchtree(f_i)|$ & Number of nodes in $\branchtree(f_i)$\\
      $\pointMetric_{q}(p_i, p_j)$ & Ground distance between two persistence pairs $p_i$ and $p_j$\\
      $\wasserstein{q}\big(\diagram(f_i), \diagram(f_j)\big)$ & $L^q$-Wasserstein
distance between the diagrams $\diagram(f_i)$ and $\diagram(f_j)$\\
      $\phi$ & Bijective assignment between the \emph{augmented} diagrams $\diagram(f_i)$ and $\diagram(f_j)$\\
      $\Phi$ & Set of all possible $\phi$\\
      $\wassersteinTree\big(\branchtree(f_i), \branchtree(f_j)\big)$ & $L^2$-Wasserstein
distance between the BDTs $\branchtree(f_i)$ and $\branchtree(f_j)$\\
      $\branchtreeSpace$ & Metric space induced by the Wasserstein distance between BDTs\\
      $\phi'$ & Rooted partial isomorphism between $\branchtree(f_i)$ and $\branchtree(f_j)$\\
      $\Phi' \subseteq \Phi$ & Set of all rooted partial isomorphisms between $\branchtree(f_i)$ and $\branchtree(f_j)$\\
      $\epsilon_1, \epsilon_2, \epsilon_3$ & Parameters of the Wassertein distance between
      BTDs\\
      $\vectorNotation{\geodesictree}\big(\branchtree(f_i),
\branchtree(f_j)\big)$ & Geodesic between $\branchtree(f_i)$ and $\branchtree(f_j)$\\
      $\normalizedLocation\big(\branchtree(f_i)\big)$ & Normalization of $\branchtree(f_i)$ (with normalized persistence pairs)\\
      $\branchtreeSet = \{\branchtree(f_1),
\dots, \branchtree(f_\ensembleSize)\}$ & Ensemble of $\ensembleSize$ BDTs\\
      $\frechetEnergy(\branchtree)$ & Fr\'echet energy of $\branchtree$ with regard to $\branchtreeSet$\\
      $\branchtree^*  \in \branchtreeSpace$ & Wasserstein barycenter of $\branchtreeSet$\\
      $\vectorNotation{\geodesictree}\big(\geodesicExtremity,
\geodesicExtremity'\big)
\cdot
\vectorNotation{\geodesictree}\big(\geodesicExtremity,
\geodesicExtremity''\big)$ & Dot product between two geodesics on $\branchtreeSpace$\\
$\axisNotation{\geodesicAxis_i} =
\Big(\vectorNotation{\geodesictree}\big(\geodesicExtremity_O,
\geodesicExtremity_i\big),
\vectorNotation{\geodesictree}\big(\geodesicExtremity_O,
\geodesicExtremity_i'\big)\Big)$ & Geodesic axis $i$\\
$\vectorNotation{\directionVector_i} =
\vectorNotation{\geodesictree}\big(\geodesicExtremity_i',
\geodesicExtremity_i\big)$ & Direction vector of the axis $\axisNotation{\geodesicAxis_i}$\\
$\vectorNotation{\geodesicAxis_i}(\branchtree)$& Arc-length paramterization of $\branchtree \in \axisNotation{\geodesicAxis_i}$ along the geodesic axis $\axisNotation{\geodesicAxis_i}$\\
$\branchtree_{\axisNotation{\geodesicAxis_i}}$ & Projection of $\branchtree$ on the geodesic axis $\axisNotation{\geodesicAxis_i}$\\
$\axisNotation{\geodesicAxis_j}(\branchtree)$ & Translation of $\axisNotation{\geodesicAxis_j}$ along $\axisNotation{\geodesicAxis_i}$ with origin $\branchtree \in \axisNotation{\geodesicAxis_i}$ \\
$\mtPgaBasis = \{\axisNotation{\geodesicAxis_1},
\axisNotation{\geodesicAxis_2}, \dots, \axisNotation{\geodesicAxis_{d'}}\}$ &Orthogonal basis of $d'$ geodesic axes\\
$\alpha \in [0, 1]^{d'}$ & Coordinate vector of $\branchtree$ in $\mtPgaBasis$\\
$\widehat{\branchtree}$
&  Reconstruction of $\branchtree$ with $\mtPgaBasis$\\
$\mtPgaError(\mtPgaBasis)$ & Fitting energy of the basis $\mtPgaBasis$ with regard to $\branchtreeSet$\\
$d_{max}$& Maximum number of geodesics considered in the computation of $\mtPgaBasis$\\
$\beta' = ||\vectorNotation{\geodesictree_{d'}}|| /
(||\vectorNotation{\geodesictree_{d'}}||
+ ||\vectorNotation{\geodesictree'_{d'}}|| )$ & Ratio describing the relative norm of $\vectorNotation{\geodesictree_{d'}}$ with regard to $\axisNotation{\geodesicAxis_{d'}}$\\
$\projectionOperator_{\vectorNotation{\dummyVector_{d''}}}(
\vectorNotation{\dummyVector_{d'}})$
& Projection of
$\vectorNotation{\dummyVector_{d'}}$ onto the direction spanned by
$\vectorNotation{\dummyVector_{d''}}$\\
$N_1$ & Maximum size of $\branchtree^*$\\
$N_2$ & Number of samples along each geodesic axis\\
      \hline
    \end{tabular}
  }
  \label{tab_timings}
\end{table}

\begin{figure*}
\makebox[\linewidth]{
\centering
\includegraphics[width=\linewidth]{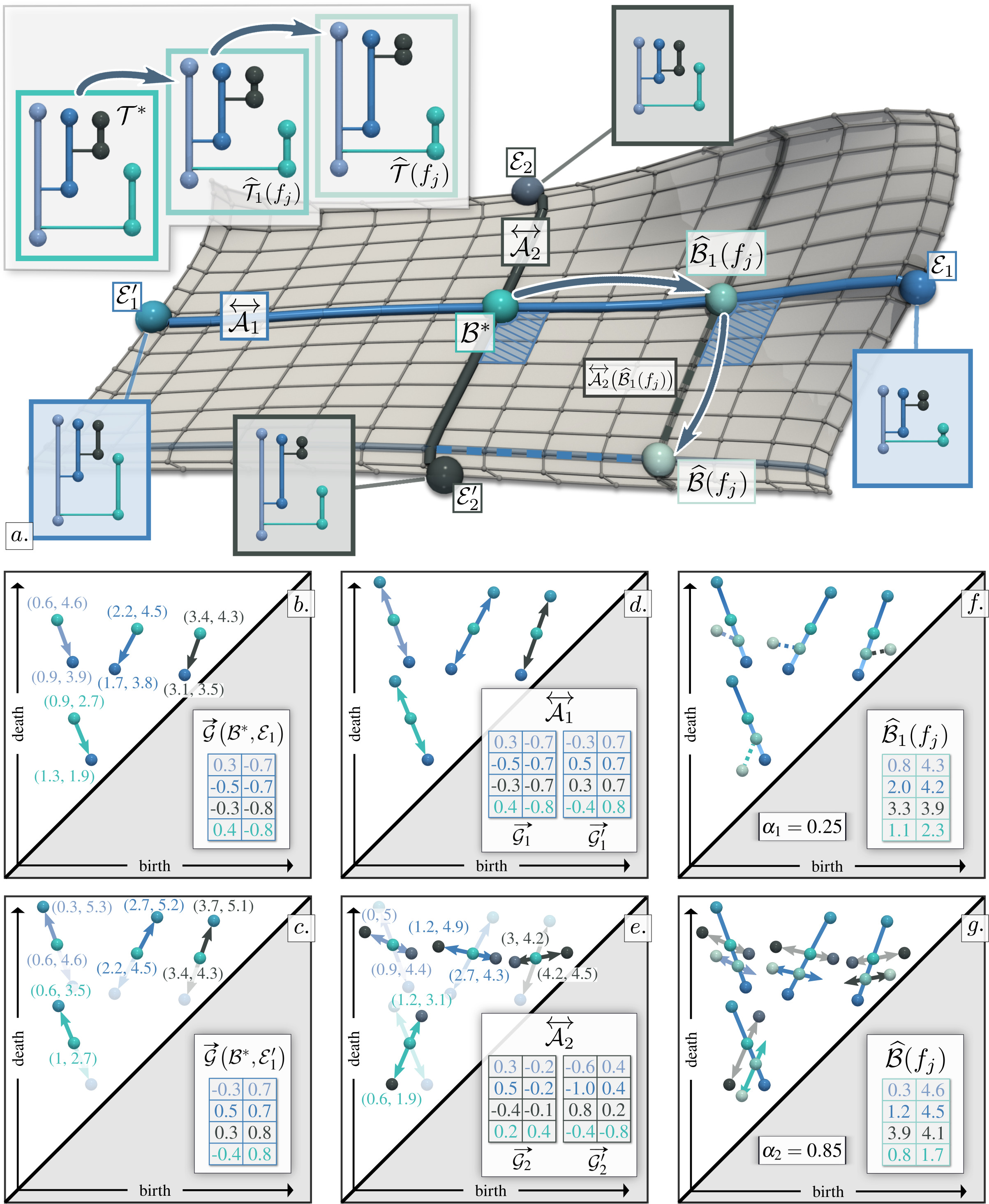}
}
\mycaption{\minorRevision{Illustration on a toy example of the geometrical notions 
formalized in our approach. We refer the reader to \autoref{sec_concepts} 
for a detailed description.}}
\label{fig_concepts}
\end{figure*}

\section{\minorRevision{Concept Illustrations}}
\label{sec_concepts}

\minorRevision{This appendix presents an illustration, on a toy 
example, of the low-level geometrical notions formalized in our approach.}

\noindent
\minorRevision{\textbf{\autoref{fig_concepts}\emph(a)}: a Principal Geodesic 
Surface  is show in gray, with its origin $\branchtree^*$ 
(cyan sphere), its BDT geodesic axis $\axisNotation{\geodesicAxis_1}$ (blue, 
with its extremities $\geodesicExtremity_1$ and $\geodesicExtremity'_1$ also in 
blue) and its axis $\axisNotation{\geodesicAxis_2}$ (black, with its 
extremities $\geodesicExtremity_2$ and $\geodesicExtremity'_2$ also in 
black). Given an input BDT $\branchtree(f_j)$, its \emph{translated projection} 
$\widehat{\branchtree}_1(f_j)$ along $\axisNotation{\geodesicAxis_1}$ is shown 
with a light blue sphere (on $\axisNotation{\geodesicAxis_1}$). The translated 
axis of $\axisNotation{\geodesicAxis_2}$ at $\widehat{\branchtree}_1(f_j)$, 
noted  $\axisNotation{\geodesicAxis_2}\big(\widehat{\branchtree}_1(f_j)\big)$, 
is shown with a black transparent curve. Finally, the reconstruction of  
$\branchtree(f_j)$ by this two-dimensional MT-PGA basis is given by the grey 
sphere $\widehat{\branchtree}(f_j)$.}

\minorRevision{All these concepts are illustrated in the birth-death space in 
the figures (b) 
to (g). In these figures, $\branchtree^*$ is represented by the cyan point cloud 
(origin of the basis, here: $|\branchtree^*| = 4$). 
Each $(2 \times |\branchtree^*|)$-dimensional geodesic vector is 
represented in the birth-death space by a set of $|\branchtree^*|$ 2D arrows,
modeling the individual branch-wise assignments (colored according to the branch 
in $\branchtree^*$ they start from, see the inset merge tree, top left corner 
of (a)). The coordinates of the arrows are reported, with matching colors, in 
the adjacent array (the numbers report the actual values, but rounded to one 
decimal after the point).}

\noindent
\minorRevision{\textbf{\autoref{fig_concepts}\emph(b)}:
The geodesic $\vectorNotation{\geodesictree}\big(\branchtree^*,
\geodesicExtremity_1\big)$ is reported with colored arrows as described above.}

\noindent
\minorRevision{\textbf{\autoref{fig_concepts}\emph(c)}:
The geodesic 
$\vectorNotation{\geodesictree}\big(\branchtree^*,
\geodesicExtremity'_1\big)$ is illustrated similarly. 
$\vectorNotation{\geodesictree}\big(\branchtree^*,
\geodesicExtremity_1\big)$ is reported in transparent, to illustrate the 
fact that the 
\emph{global} collinearity of two $(2\times|\branchtree^*|)$-dimensional  
geodesic vectors translates into a \emph{local} collinearity of the individual 
2D assignment vectors in the 
birth-death space.}

\noindent
\minorRevision{\textbf{\autoref{fig_concepts}\emph(d)}:
The geodesic axis $\axisNotation{\geodesicAxis_1}$ is 
reported with 
colored double arrows, along with their coordinates (arrays).}

\noindent
\minorRevision{\textbf{\autoref{fig_concepts}\emph(e)}:
The geodesic axis $\axisNotation{\geodesicAxis_2}$ is 
shown 
similarly.
$\axisNotation{\geodesicAxis_1}$ is reported in transparent, 
to illustrate the fact 
that the \emph{global} orthogonality between the 
$(2\times|\branchtree^*|)$-dimensional direction vectors 
$\vectorNotation{\directionVector_1} =
\vectorNotation{\geodesictree_1} - \vectorNotation{\geodesictree'_1}$ and 
$\vectorNotation{\directionVector_2} =
\vectorNotation{\geodesictree_2} - \vectorNotation{\geodesictree'_2}$ does not 
necessarily 
translate into a \emph{local} orthogonality between the individual 2D 
assignment vectors in the birth-death space. This can be explained by the fact 
that a 2D space can only 
capture orthogonality between $2$ directions, not 
$2\times|\branchtree^*|$.
Alternatively, the 
geodesic axes $\axisNotation{\geodesicAxis_1}$ and 
$\axisNotation{\geodesicAxis_2}$, when shown in the birth-death space, can be 
interpreted as 2-dimensional projections of orthogonal 
$(2\times|\branchtree^*|)$-dimensional direction vectors.}

\noindent
\minorRevision{\textbf{\autoref{fig_concepts}\emph(f)}:
The 
projection 
$\widehat{\branchtree}_1(f_j)$ (light blue sphere on the segments) is 
the BDT on $\axisNotation{\geodesicAxis_1}$ which minimizes its distance to 
$\branchtree(f_j)$.}

\noindent
\minorRevision{\textbf{\autoref{fig_concepts}\emph(g)}:
The translated axis 
$\axisNotation{\geodesicAxis_2}\big(\widehat{\branchtree}_1(f_j)\big)$ 
is shown with colored double arrows. The axis $\axisNotation{\geodesicAxis_2}$ 
is shown in transparent (black).}

\section{Gradient of the Fitting Energy}

Given the Wasserstein barycenter $\branchtree^*$ of the input set of BDTs
$\branchtreeSet = \{\branchtree(f_1), \dots, \branchtree(f_\ensembleSize)\}$ as
well as an
initialization of the $d'$-th geodesic axis of $\mtPgaBasis$, noted
$\axisNotation{\geodesicAxis_{d'}}$, we want to update
$\axisNotation{\geodesicAxis_{d'}}$ in order to
decrease the fitting energy associated to $\mtPgaBasis$ (Sec. 3, main
manuscript):
\begin{eqnarray}
\label{eq_projectionEnergy}
\mtPgaError(\mtPgaBasis) = \sum_{j = 1}^\ensembleSize
\wassersteinTree\Big(\branchtree(f_j),
\branchtree^* +
\sum_{i =
1}^{d'}\vectorNotation{\geodesicAxis_i}\big(\widehat{\branchtree}_i(f_j)
\big)
\Big)^2
.
\end{eqnarray}

Similarly to previous optimization strategies for the Fr\'echet
energy \cite{Turner2014, vidal_vis19, pont_vis21}, our approach consists in
alternating phases of \emph{(i)} \emph{Assignments}
(between the input BDTs and their projections along translations of
$\axisNotation{\geodesicAxis_{d'}}$, cf.
\autoref{eq_projectionEnergy})
and \emph{(ii)} \emph{Updates} (of the axis
$\axisNotation{\geodesicAxis_{d'}}$).

Thus, at the \emph{Update} phase \emph{(ii)},
the assignment
${\phi'_{d'}}^j$
between each input BDT $\branchtree(f_j)$ and its
projection
$\widehat{\branchtree}_{d'}(f_j)$
on the the translated axis
$\axisNotation{\geodesicAxis_{d'}}\big(\widehat{\branchtree}_{d'-1}(f_j)\big)$
is constant.
It follows that each
branch $\branch^* \in \branchtree^*$ can be considered independently for the
minimization of \autoref{eq_projectionEnergy}.
In particular, let
$\big(\branch^* + \sum_{i = 1}^{d'}(1 - \alpha_{i}^j) \vectorNotation{g_{i}} +
\alpha_{i}^j \vectorNotation{g'_{i}}\big)$ be the branch assigned to
$\branch^*$ in
$\widehat{\branchtree}_{d'}(f_j)$,
with
$\vectorNotation{g_{i}}$ and $\vectorNotation{g'_{i}}$ being two 2D vectors in
the
birth/death space (i.e. the entries of $\vectorNotation{\geodesictree_{i}}$ and
$\vectorNotation{\geodesictree_{i}'}$ corresponding to the branch $\branch^*$
of $\branchtree^*$). Then, the
individual fitting energy associated to
$\branch^*$, i.e. its contribution to \autoref{eq_projectionEnergy}, can be
expressed as:
\begin{eqnarray}
\label{eq_individualEnergy}
\individualEnergy_{\branch^*}(\vectorNotation{g_{d'}}, \vectorNotation{g'_{d'}})
 =
\sum_{j = 1}^\ensembleSize
d_2\big(b_j, \branch^* + \sum_{i = 1}^{d'} (1 - \alpha_{i}^j)
\vectorNotation{g_{i}} +
\alpha_{i}^j \vectorNotation{g'_{i}}\big)^2,
\end{eqnarray}
where $b_j$ stands for the branch in $\branchtree(f_j)$ assigned to $\branch^*$
and
$d_2$ stands for the $L_2$ norm in the 2D birth/death space (i.e. the
ground distance involved in $\wassersteinTree$, see Sec. 2.4, main manuscript).
Note that the usage of the $L_2$ norm implies that
$\individualEnergy_{\branch^*}$ is convex.
Our goal at this stage is to find the two vectors $\vectorNotation{g_{d'}}$
and $\vectorNotation{g'_{d'}}$ which minimize \autoref{eq_individualEnergy}.

\autoref{eq_individualEnergy} can be further detailed as follows, where
$\geodesictreeVec_{ix}$ and $\geodesictreeVec_{iy}$ stand for the X-Y
coordinates of the vector $\vectorNotation{g_{i}}$ in the birth/death plane:
\begin{equation}
\begin{aligned}
\label{eq_detailedIndividualEnergy}
 \individualEnergy_{\branch^*}(\vectorNotation{g_{d'}},
\vectorNotation{g'_{d'}}) = \sum_{j=1}^\ensembleSize  & \Big(\branch_{jx} -
\big(\branch^*_x + \sum_{i=1}^{d'} (1 - \alpha_i^j) \times \geodesictreeVec_{ix}
+ \alpha_i^j \times \geodesictreeVec_{ix}'\big)\Big)^2 \\ +
 & \Big(\branch_{jy} - \big(\branch^*_y + \sum_{i=1}^{d'} (1 - \alpha_i^j)
\times \geodesictreeVec_{iy} + \alpha_i^j \times
\geodesictreeVec_{iy}'\big)\Big)^2.
\end{aligned}
\end{equation}

In the following, we derive the gradient of the above convex energy. In
particular, we  detail the derivation for the $X$ coordinate only (the
derivation along the $Y$ coordinate being identical):
\begin{equation}
\begin{aligned}
\nonumber
 \frac{\partial \individualEnergy_{\branch^*}(\vectorNotation{g_{d'}},
\vectorNotation{g'_{d'}})}{\partial \geodesictreeVec_{d'x}} &= 2
\sum_{j=1}^\ensembleSize  (1 - \alpha_{d'}^j) \Big(\branch_{jx} -
\big(\branch^*_x \\
&\qquad\qquad\quad\qquad\qquad+
\sum_{i=1}^{d'} (1 - \alpha_i^j) \times \geodesictreeVec_{ix} + \alpha_i^j
\times \geodesictreeVec_{ix}'\big)\Big) \\
 \frac{\partial \individualEnergy_{\branch^*}(\vectorNotation{g_{d'}},
\vectorNotation{g'_{d'}})}{\partial \geodesictreeVec_{d'x}'} &= 2
\sum_{j=1}^\ensembleSize  \alpha_{d'}^j \Big(\branch_{jx} - \big(\branch^*_x +
\sum_{i=1}^{d'} (1 - \alpha_i^j) \times \geodesictreeVec_{ix} + \alpha_i^j
\times \geodesictreeVec_{ix}'\big)\Big).
\end{aligned}
\end{equation}

To minimize \autoref{eq_detailedIndividualEnergy}, we aim to find the values of
$\geodesictreeVec_{d'x}$ and $\geodesictreeVec'_{d'x}$ for which the above
partial derivatives equal zero. This yields the following linear system of two
equations (with two unknowns, $\geodesictreeVec_{d'x}$ and
$\geodesictreeVec'_{d'x}$):
\begin{equation}
\label{eq_system}
\begin{cases}
\begin{aligned}
 \sum_{j=1}^\ensembleSize  (1 - \alpha_{d'}^j) \Big(\branch_{jx} -
\big(\branch^*_x +
\sum_{i=1}^{d'} (1 - \alpha_i^j) \times \geodesictreeVec_{ix} + \alpha_i^j
\times \geodesictreeVec_{ix}'\big)\Big) & = 0 \\
 \sum_{j=1}^\ensembleSize  \alpha_{d'}^j \Big(\branch_{jx} - \big(\branch^*_x +
\sum_{i=1}^{d'} (1 - \alpha_i^j) \times \geodesictreeVec_{ix} + \alpha_i^j
\times \geodesictreeVec_{ix}'\big)\Big) & = 0.
\end{aligned}
\end{cases}
\end{equation}

Given the above system, we aim next at expressing
$\geodesictreeVec_{d'x}$ as a function of
$\geodesictreeVec_{d'x}'$.
To simplify
notations, we introduce the term $\branch^*_{d'x}$ as follows:
\begin{equation}
\begin{aligned}
\nonumber
 \branch^*_{d'x} := \branch^*_x + \sum_{i=1}^{d'-1} (1 - \alpha_{i}^j) \times
\geodesictreeVec_{ix} + \alpha_{i}^j \times \geodesictreeVec_{ix}'.
\end{aligned}
\end{equation}

Then, the first line of \autoref{eq_system} can be
re-written as:
\begin{equation}
\begin{aligned}
\nonumber
 & \sum_{j=1}^\ensembleSize  (1 - \alpha_{d'}^j) \Big(\branch_{jx} -
\big(\branch^*_{d'x} +
(1 - \alpha_{d'}^j) \times \geodesictreeVec_{d'x} + \alpha_{d'}^j \times
\geodesictreeVec_{d'x}'\big)\Big) = 0.
\end{aligned}
\end{equation}

Then, it follows that:
\begin{equation}
\label{eq_expression_g}
\geodesictreeVec_{d'x} = \cfrac{\sum_{j=1}^\ensembleSize (1 - \alpha_{d'}^j)
\Big(\branch_{jx} - \big(\branch^*_{d'x} + \alpha_{d'}^j \times
\geodesictreeVec_{d'x}'\big)\Big)}{\sum_{j=1}^\ensembleSize (1 -
\alpha_{d'}^j)^2}.
\end{equation}

Now, we apply the same reasoning with the second line of
\autoref{eq_system}, yielding the following expression of
$\geodesictreeVec'_{d'x}$:
\begin{equation}
\label{eq_expression_g'}
\geodesictreeVec_{d'x}' = \cfrac{\sum_{j=1}^\ensembleSize \alpha_{d'}^j
\Big(\branch_{jx} -
\big(\branch^*_{d'x} + (1 - \alpha_{d'}^j) \times
\geodesictreeVec_{d'x}\big)\Big)}{\sum_{j=1}^\ensembleSize {(\alpha_{d'}^j})^2}.
\end{equation}
%
%
%

At this stage, one can notice that the expression of $\geodesictreeVec'_{d'x}$
is itself a function of $\geodesictreeVec_{d'x}$. Thus, we insert the
expression of $\geodesictreeVec'_{d'x}$ (\autoref{eq_expression_g'}) into that
of $\geodesictreeVec_{d'x}$ (\autoref{eq_expression_g}), which results
eventually in the following expression (we omit the detailed, intermediate
steps):
\begin{equation}
\label{final_eq}
  \geodesictreeVec_{d'x} = \cfrac{\sum_{j=1}^\ensembleSize (1 -
\alpha_{d'}^j) \Bigg(\branch_{jx} - \branch^*_{d'x} - \alpha_{d'}^j
\cfrac{\sum_{k=1}^\ensembleSize \alpha_{d'}^k (\branch_{kx} -
\branch^*_{d'x})}{\sum_{k=1}^\ensembleSize
(\alpha_{d'}^k)^2} \Bigg)}{\sum_{j=1}^\ensembleSize (1 - \alpha_{d'}^j)^2 +
\sum_{j=1}^\ensembleSize (1
- \alpha_{d'}^j) \alpha_{d'}^j \cfrac{\sum_{k=1}^\ensembleSize \alpha_{d'}^k
\big(- (1 -
\alpha_{d'}^k)\big)}{\sum_{k=1}^\ensembleSize (\alpha_{d'}^k)^2}}.
\end{equation}

Finally, we can insert the expression \revision{of $\geodesictreeVec_{d'x}$ 
(\autoref{eq_expression_g})} into the original expression of
$\geodesictreeVec'_{d'x}$ (\autoref{eq_expression_g'}), resulting in the
following expression (again, we omit the detailed, intermediate steps):
\begin{equation}
\label{final_eq'}
\geodesictreeVec_{d'x}' = \cfrac{\sum_{j=1}^\ensembleSize \alpha_{d'}^j
\Bigg(\branch_{jx} - \branch^*_{d'x} - (1 - \alpha_{d'}^j)
\cfrac{\sum_{k=1}^\ensembleSize
(1 - \alpha_{d'}^k) (\branch_{kx} - \branch^*_{d'x})}{\sum_{k=1}^\ensembleSize
(1 -
\alpha_{d'}^k)^2} \Bigg)}{\sum_{j=1}^\ensembleSize (\alpha_{d'}^j)^2 +
\sum_{j=1}^\ensembleSize
\alpha_{d'}^j (1 - \alpha_{d'}^j) \cfrac{\sum_{k=1}^\ensembleSize (1 -
\alpha_{d'}^k) \big(-
\alpha_{d'}^k\big)}{\sum_{k=1}^\ensembleSize (1 - \alpha_{d'}^k)^2}}.
\end{equation}

Overall, \autoref{final_eq} and \autoref{final_eq'} provide the expression of
the $X$-coordinate of the vectors $\vectorNotation{g_{d'}}$ and
$\vectorNotation{g'_{d'}}$ which minimize the individual fitting energy
(\autoref{eq_detailedIndividualEnergy}). The same reasoning (not detailed
here) can be applied identically to retrieve the $Y$-coordinate of
$\vectorNotation{g_{d'}}$ and
$\vectorNotation{g'_{d'}}$.

Then, the entire
vectors $\vectorNotation{\geodesictree_{d'}}$ and
$\vectorNotation{\geodesictree_{d'}'}$ (defining
$\vectorNotation{\geodesicAxis_{d'}}\big(\widehat{\branchtree}_{d'}(f_j)
\big)$) which minimize
\autoref{eq_projectionEnergy} under the current assignments can be updated
similarly, by iterating the above computation for all the branches $\branch^*$
of $\branchtree^*$.

\begin{figure*}
\makebox[\linewidth]{
\centering
\includegraphics[width=1\linewidth]{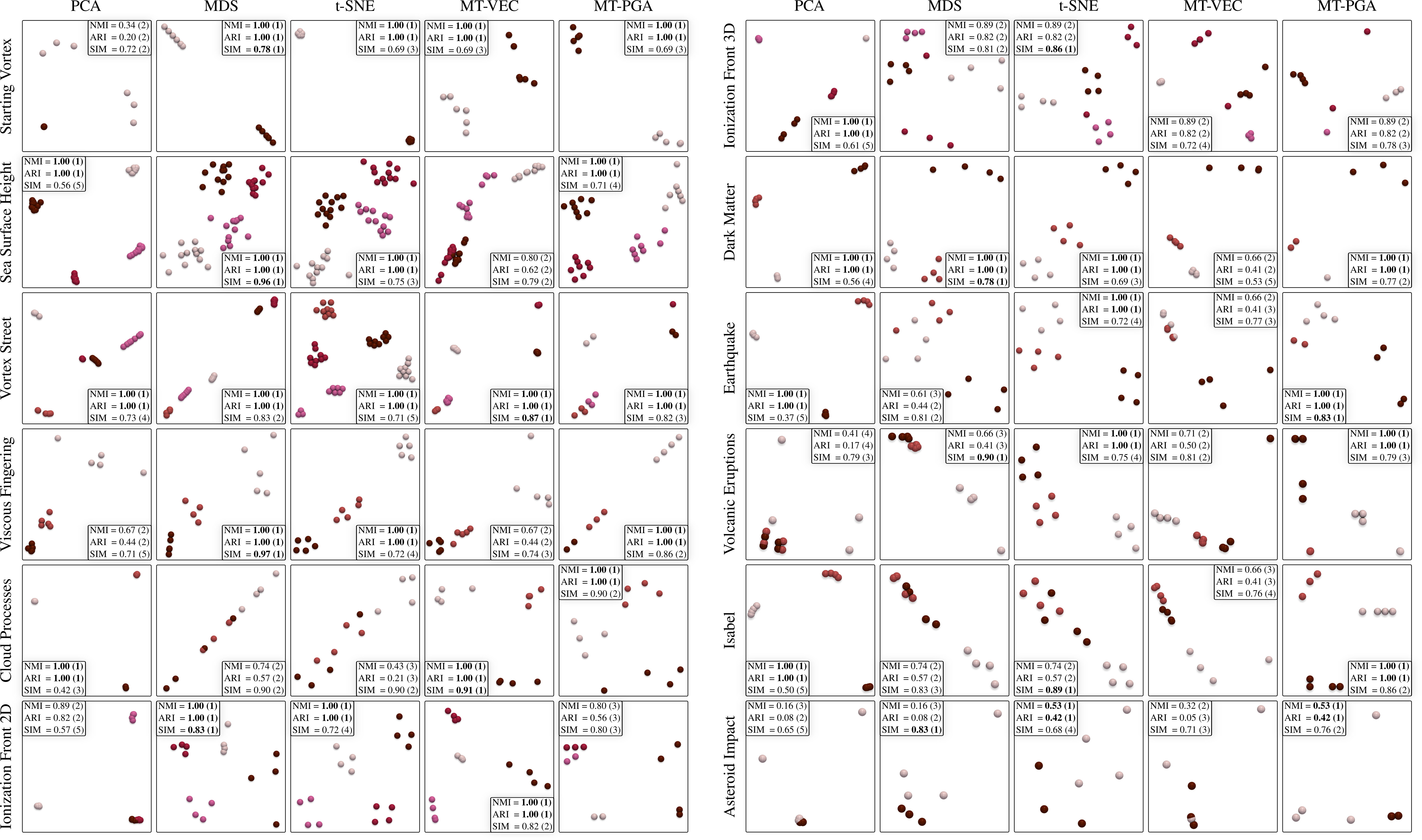}
}
\mycaption{\revision{Comparison of planar layouts for typical dimensionality
reduction techniques on all our test ensembles. The color encodes the
classification ground-truth \cite{pont_vis21}. For each quality score, the best
value appears bold and the rank of the score among all methods is in
parenthesis.}}
\label{fig_embeddingAll}
\end{figure*}

\section{Data Reduction}

\begin{table} 
  \caption{\revision{Compression factor and average relative reconstruction 
error of our 
  algorithm for PD-PGA and MT-PGA (for $d_{max}$ = 3 and $N_1$ = 20).}}
  \centering
  \scalebox{0.9}{
    \begin{tabular}{|l|r|r||r|r||r|r|}
      \hline
      \textbf{Dataset} & $N$ & $|\branchtree|$ & \multicolumn{2}{c||}{PD-PGA} & \multicolumn{2}{c|}{MT-PGA} \\
                       &      &                 & Factor & Error & Factor & Error \\      \hline
      Asteroid Impact (3D) & 7 & 1,295 & 2.97 & 0.07 & 4.84 & 0.22 \\
      Cloud processes (2D) & 12 & 1,209 & 5.94 & 0.19 & 7.39 & 0.01 \\
      Viscous fingering (3D) & 15 & 118 & 2.23 & 0.13 & 4.71 & 0.02 \\
      Dark matter (3D) & 40 & 2,592 & 10.00 & 0.04 & 19.27 & 0.04 \\
      Volcanic eruptions (2D) & 12 & 811 & 9.99 & 0.12 & 4.83 & 0.04 \\
      Ionization front (2D) & 16 & 135 & 2.56 & 0.14 & 5.12 & 0.40 \\
      Ionization front (3D) & 16 & 763 & 3.27 & 0.17 & 4.85 & 0.46 \\
      Earthquake (3D) & 12 & 1,203 & 1.42 & 0.18 & 2.19 & 0.33 \\
      Isabel (3D) & 12 & 1,338 & 5.49 & 0.27 & 9.25 & 0.05 \\
      Starting Vortex (2D) & 12 & 124 & 1.76 & 0.07 & 4.42 & 0.01 \\
      Sea Surface Height (2D) & 48 & 1,787 & 19.59 & 0.18 & 9.48 & 0.48 \\
      Vortex Street (2D) & 45 & 23 & 1.86 & 0.04 & 11.84 & 0.02 \\
      \hline
    \end{tabular}
  } 
  \label{tab_compresson}
\end{table}

\begin{figure*}
\makebox[\linewidth]{
\centering
\includegraphics[width=\linewidth]{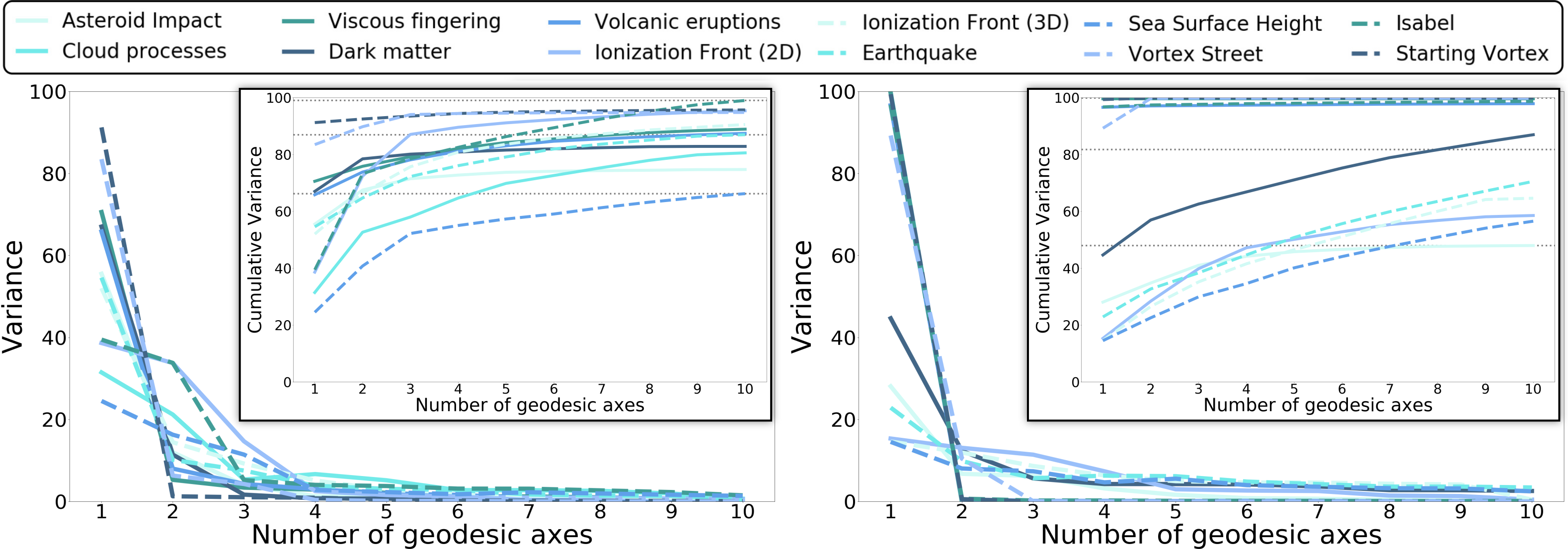}
}
\mycaption{\revision{Evolution of the projected variance (and cumulative
variance, inset) with the number of geodesic axes for PD-PGA (left)
and
MT-PGA (right).
}}
\label{fig_variance}
\end{figure*}

\revision{
\autoref{tab_compresson} reports the compression factors and average relative 
reconstruction error for all our test ensembles. In particular, for each input 
BDT $\branchtree(f_i)$, we compute its reconstruction error via the 
$\wassersteinTree$ distance to its reconstruction $\widehat{\branchtree}(f_i)$. 
In order to become comparable across ensembles, this distance is then divided 
by the maximum
%
$\wassersteinTree$ distance observed among two input BDTs in the 
ensemble. Finally, this relative reconstruction error is averaged over all the 
BDTs of the input ensemble.
}

\revision{
This table shows that our framework results in significant compression factors 
(9 or above) for the largest ensembles, i.e. the ensembles counting the most 
members and for which the BDTs are the largest (i.e. Dark matter, Sea Surface 
Height). On the contrary, modest compression factors tend to be  obtained for
the smallest ensembles, counting few members and few features.
The average relative reconstruction error seems acceptable overall:
$0.13$ and $0.17$ on average for PD-PGA and MT-PGA respectively.}

\revision{Finally, note that for each ensemble, the merge tree based clustering 
\cite{pont_vis21} computed from the input BDTs is strictly identical to the 
clustering computed from the reconstructed BDTs. This confirms the viability of 
our reconstructed BDTs, and their usability for typical visualization and 
analysis tasks.}


\section{Persistence correlation}
This section details the computation of the correlation between the persistence
of the $i^{th}$ feature of an input BDT $\branchtree(f_j)$ and its coordinate
$\alpha_k^j$ along the geodesic axis $\axisNotation{\geodesicAxis_k}$.
\minorRevision{It is derived from the seminal Pearson's correlation \protect\cite{pearsonCorrelation}, applied on the above terms (made available by our framework).}
Let $\mathbb{P}$ be an $(\numberBranchinBarycenter \times
\ensembleSize)$-matrix, such that the entry $\mathbb{P}(i, j)$ denotes the
topological persistence, in the $j^{th}$ input BDT,
of the
branch $\branch_j \in \branchtree(f_j)$ mapped to the $i^{th}$ most
persistent branch of $\branchtree^*$ given the optimal assignment induced by
$\wassersteinTree$ (Eq. 1 of the main manuscript).
Next, let $\mathbb{A}$ be a
 $(\maxDimensions \times \ensembleSize)$-matrix, such that then entry
$\mathbb{A}(k, j)$ denotes the coordinate $\alpha_k^j$
of the BDT $\branchtree(f_j)$ along the axis $\axisNotation{\geodesicAxis_k}$.

In the following, we aim at assessing how much the persistence of the
branches in $\branchtree(f_j)$ is correlated with the coordinate $\alpha_k^j$.
Let $\rho_{p_i, \alpha_k}$ be the correlation between the persistence $p_i$
($i^{th}$ line of $\mathbb{P}$) and the coordinate $\alpha_k$ ($k^{th}$ line of
$\mathbb{A}$). It is given by the following expression, where
$\overline{p_i}$ and $\overline{\alpha_k}$
are the average values for the the $i^{th}$ line of  $\mathbb{P}$ and the
$k^{th}$ line of
$\mathbb{A}$, and where $\sigma_{p_i}$ and $\sigma_{\alpha_k}$ stand for their
standard deviation:
\begin{eqnarray}
 \nonumber
\rho_{p_i, \alpha_k} = {{\sum_{j = 1}^\ensembleSize \big(
\mathbb{P}(i,j) - \overline{p_i}\big) \times \big(\mathbb{A}(k, j) -
\overline{\alpha_k}\big)}\over{\ensembleSize \times \sigma_{p_i} \times
\sigma_{\alpha_k}}}.
\end{eqnarray}

\section{Dimensionality reduction}

\revision{\autoref{fig_embeddingAll} extends Figure 12 (main 
manuscript) to all our test ensembles and it confirms visually the conclusions 
of the table of aggregated scores (Table 2 of the main manuscript).}

\revision{In particular, it confirms that MT-PGA provides a \emph{trade-off}
between 
the respective advantages of standard techniques such as MDS \cite{kruskal78} 
and t-SNE \cite{tSNE}. Specifically, MDS is known to preserve the input metric 
well, while t-SNE tends to better preserve the global structure of the data 
(i.e. the ground-truth classification), at the expense of metric violation.
}

\revision{
MT-PGA provides a \emph{balance} between these two behaviors: \emph{(i)} it 
improves structure 
preservation over MDS 
(it provides equivalent or better NMI/ARI scores for 11 out of 12 ensembles)
and \emph{(ii)} it improves 
metric preservation over t-SNE 
(it provides an equivalent or better SIM score for 9 out of 12 ensembles).
Visually, this means that MT-PGA groups together the members belonging to 
the same ground-truth class, while providing a layout which is more faithful 
than t-SNE's regarding the distances between the corresponding merge trees. 
}

\revision{As can be expected, the straightforward PCA preserves the 
$\wassersteinTree$ metric poorly (as it is based on the $L_2$ norm).
Since it relies on a rough approximation of $\branchtreeSpace$,
a
PCA derived from a vectorization of the BDTs (MT-VEC)
preserves poorly the
global structure of the ensemble (MT-PGA provides equivalent or better NMI/ARI scores for 11 out of 12 ensembles).
}
%
%
%
%

\section{Metric distortion}
The section details the computation of the metric distortion indicator $SIM$,
which evaluates the preservation of the Wasserstein metric in dimensionality
reduction tasks.

Specifically, given two points $x$ and $y$ in a planar layout (with BDTs
$\branchtree(f_x)$ and $\branchtree(f_y)$), we first measure their pairwise
distortion:
\begin{eqnarray}
\nonumber
\delta(x, y) = \Big(||x - y||_2 - \wassersteinTree\big(\branchtree(f_x),
\branchtree(f_y)\big)\Big)^2.
\end{eqnarray}
This measure is then normalized into:
\begin{eqnarray}
\nonumber
\delta'(x, y) = {{\delta(x, y)}\over{\max_{\forall x \neq y}\big(\delta(x,
y)\big)}}.
\end{eqnarray}
Finally, we evaluate the global indicator $SIM := 1 -
\overline{\delta'}$, where $\overline{\delta'}$ stands for the average
of $\delta'(x, y)$ for all pairs $(x, y)$ in the ensemble. $SIM$ values
lie within
the interval $[0, 1]$ and are optimal near $1$.

\section{Projected Variance}

\revision{\autoref{fig_variance} extends to 10 dimensions the scatter plots of 
projected variance reported in Figure 13 (main manuscript), which were 
computed for only 2 dimensions.
This figure confirms the conclusions of Figure 13 (main manuscript).
Except for a very mild oscillation for a specific dataset (\emph{``Cloud 
processes''}, PD-PGA, with a local maximum of low amplitude at 4 dimensions), 
overall,
the projected variance is indeed monotonically decreasing in practice
%
for 
an increasing  number of dimensions (i.e. the algorithm does tend to identify
the most informative directions first).
}

\section{Metric Parameter Analysis}

\revision{The Wasserstein metric between merge trees (introduced in previous 
work \cite{pont_vis21}) is subject to three parameters ($\epsilon_1$, 
$\epsilon_2$, and $\epsilon_3$), in order to 
adjust
the 
stability/discriminativeness balance of the metric.
Pont et al. \cite{pont_vis21} provided a detailed, experimental parameter 
analysis of their metric, in particular for the specific task of geodesic 
computation (Appendix 10.3 of \cite{pont_vis21}).}

\begin{figure*}
  \centering
  \includegraphics[width=.49\linewidth]{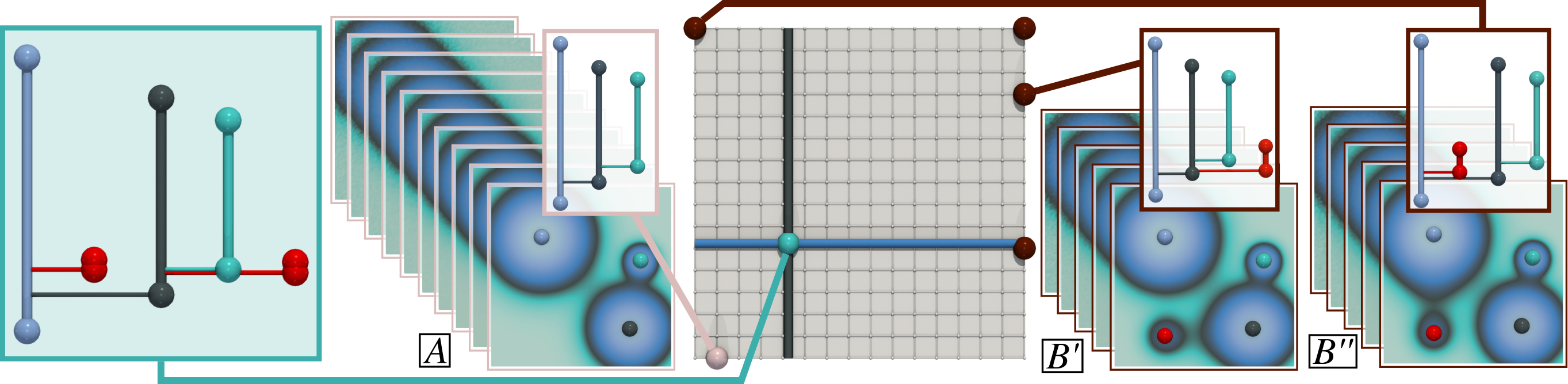}
  \hfill
  \includegraphics[width=.49\linewidth]{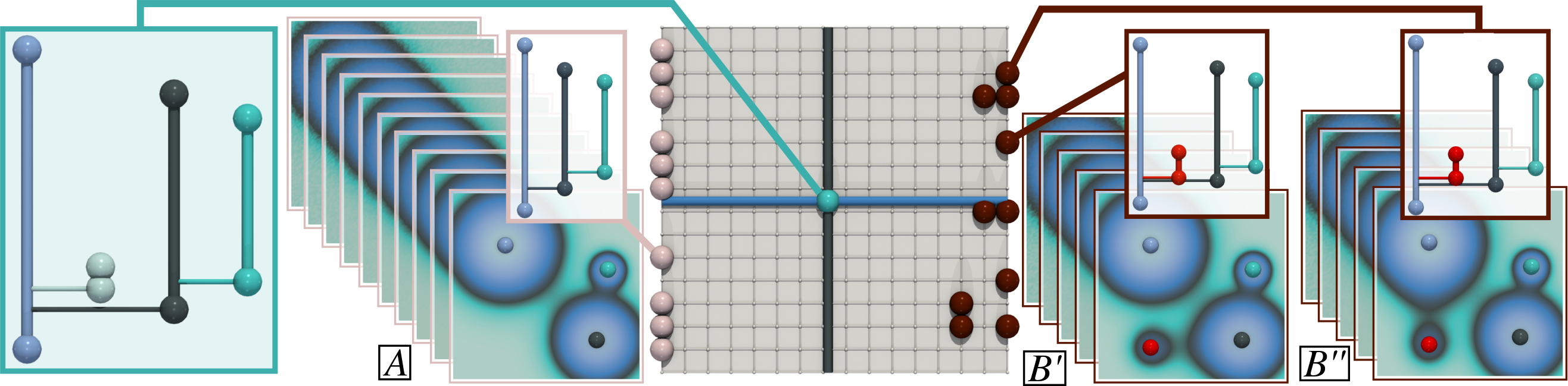}
  \caption{\revision{Impact of the parameter $\epsilon_1$ on MT-PGA computation 
(in this example, left: $\epsilon_1 = 0$, right: $\epsilon_1 = 0.3$).
Initially (left), the red 
branch in the cluster $B'$ is not
matched to the red branch in the cluster $B''$ as they have distinct depths in 
the corresponding BDTs (2 versus 1). However, these features are visually
similar in the data (Gaussians with the red maximum in $B'$ and $B''$, bottom 
left corner of the domain).
After the $\epsilon_1$ pre-processing
(right), the saddle of the
red
branch in $B''$ gets merged with its ancestor saddle (whose scalar value was 
less than $\epsilon_1$ away). Consequently, the red branch gets moved up the
BDT (i.e. the red branch is attached to the main light blue branch in $B''$, 
right). 
Since they now have identical depths in the corresponding BDTs, the
red branches of the sub-clusters $B'$ and $B''$ can now 
be matched together (right), which results in an overall matching 
between these two
trees 
which better conveys the resemblance between the two sub-clusters $B'$ 
and $B''$. 
Equivalently, one can interpret this procedure of 
saddle merge in
the input trees as a modification of the input scalar field, turning the 
Gaussian mixture $B'$ into $B''$. In particular, this field modification 
disconnects the Gaussian with the
red maximum from the Gaussian with the black maximum ($B'$) and 
reconnects it to the Gaussian with the light blue maximum ($B''$).
Overall, after the $\epsilon_1$ pre-processing (right), the MT-PGA better 
distinguishes the cluster $A$ (pink spheres, left of the planar layout) from the 
cluster $B$ (dark red spheres, right of the planar layout).
}}
  \label{fig_eps1}
\end{figure*}

\begin{figure*}
  \centering
\includegraphics[width=.49\linewidth]{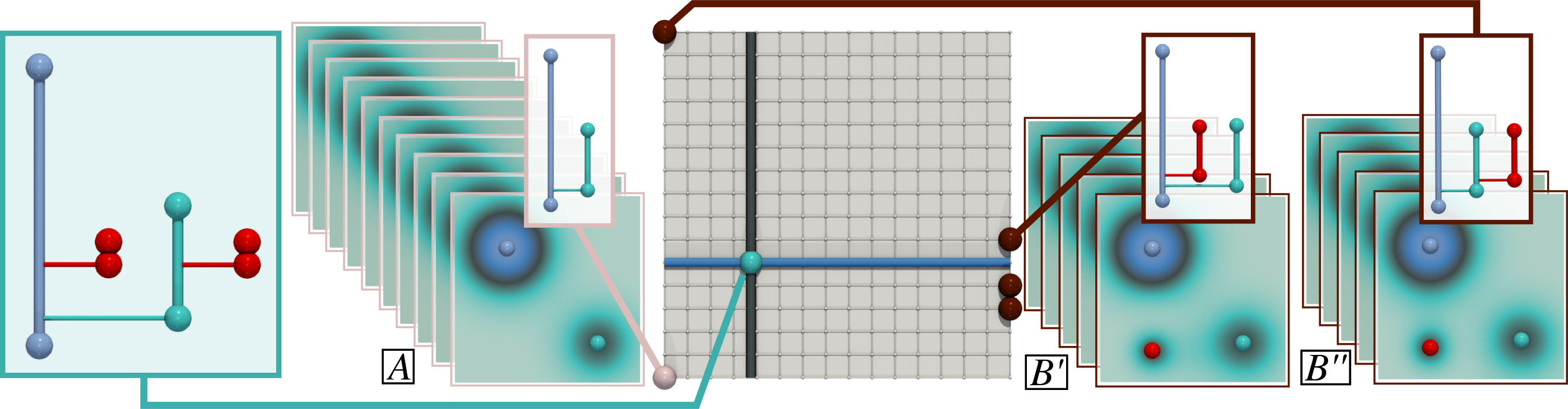}
  \hfill
  \includegraphics[width=.49\linewidth]{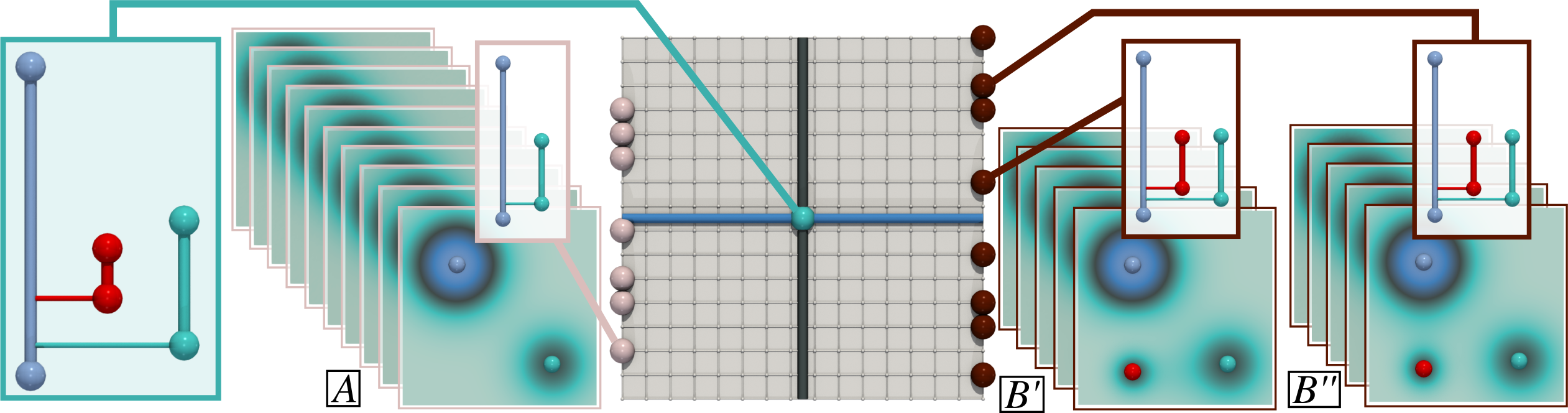}
  \caption{\revision{Impact of the parameter $\epsilon_2$ on MT-PGA  
computation (in this example, left: $\epsilon_2 = 1$, right: $\epsilon_2 = 
0.8$). 
Initially
(left), the red branch in $B'$ is not
matched to the red branch in $B''$
as they have distinct depths in the 
corresponding BDTs (1 versus 2). 
The
red branch in 
$B''$
has a persistence nearly identical to its parent 
(cyan). Thus, after local normalization (necessary to guarantee the topological
consistency of the interpolated trees), its normalized persistence would become 
artificially high, which can have an undesirable effect on the
metric. The BDT pre-processing addresses this issue and moves up the BDT 
branches with a relative persistence to their parent larger than $\epsilon_2$. 
After the $\epsilon_2$ pre-processing
(right), the red branch 
in 
$B''$
moves up the BDT and becomes adjacent to the main light blue
branch. Since they now have identical depths in the corresponding 
BDTs, the red branches of 
$B'$
and 
$B''$
can now be matched
together (right), which better conveys the resemblance between the two 
sub-clusters $B'$ and $B''$.
Equivalently, one can interpret this procedure of BDT
pre-processing as a modification of the input scalar field, turning 
the Gaussian mixture $B'$ into $B''$.
In particular, this field modification disconnects the Gaussian with the
red maximum from the Gaussian with the cyan maximum ($B''$) and reconnects it 
to the Gaussian with the light blue maximum ($B'$).
Overall, after the $\epsilon_2$ pre-processing (right), the MT-PGA better 
distinguishes the cluster $A$ (pink spheres, left of the planar layout) from 
the 
cluster $B$ (dark red spheres, right of the planar layout).
}}
  \label{fig_eps2}
\end{figure*}

\begin{figure*}
  \centering
\includegraphics[width=.49\linewidth]{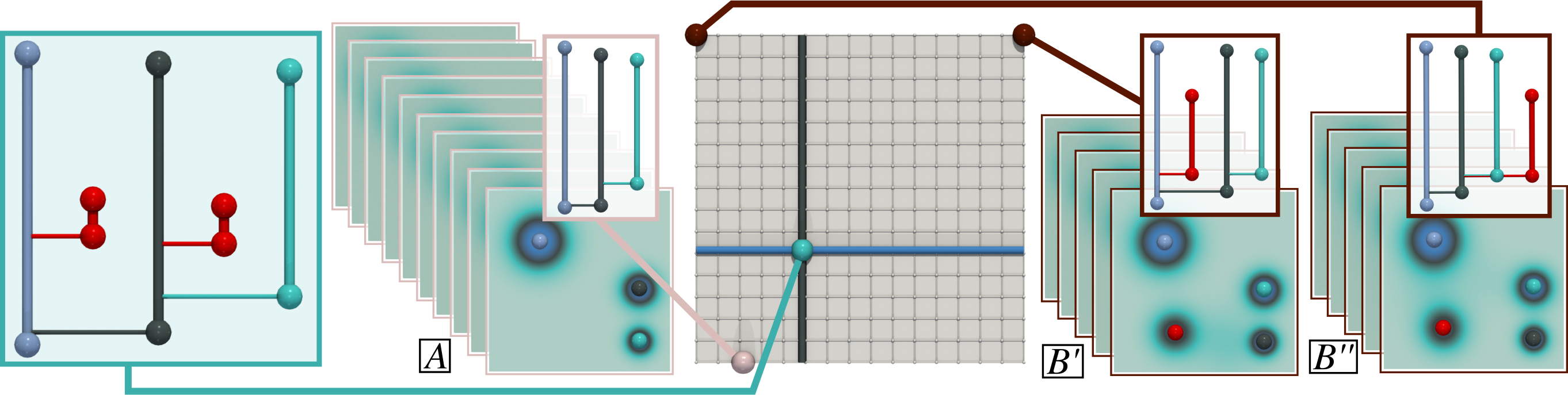}
  \hfill
  \includegraphics[width=.49\linewidth]{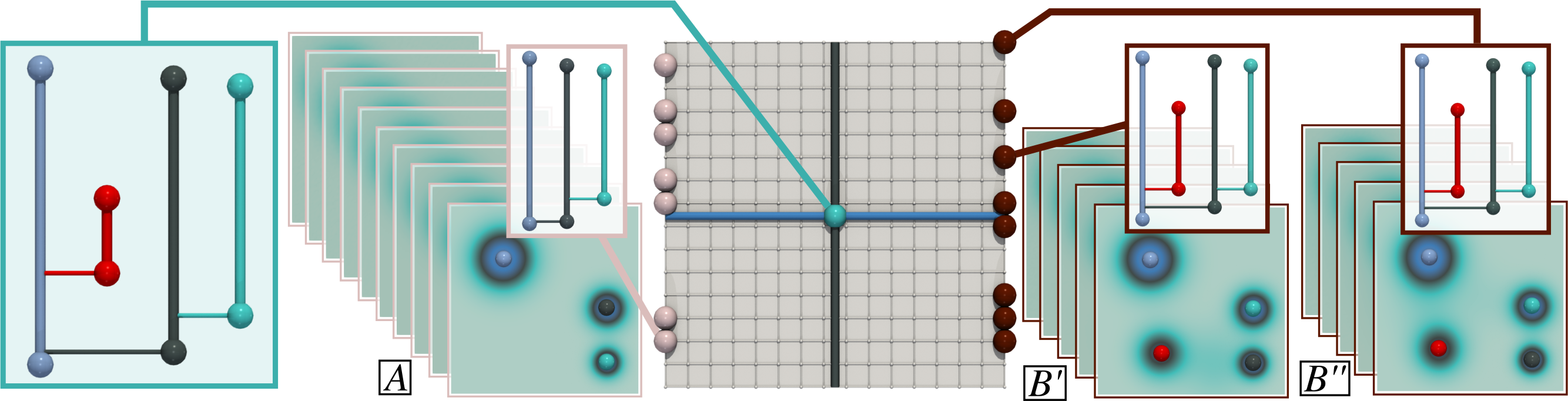}
  \caption{\revision{Impact of the parameter $\epsilon_3$ on MT-PGA  
computation (in this example, left: $\epsilon_3 = 0$, right: $\epsilon_3 = 
0.7$). 
Initially
(left), the red branch in $B''$ is not
matched to the red branch in $B'$ as they have distinct depths in the 
corresponding BDTs (1 versus 3). 
The parameter $\epsilon_3$
restricts the application of the above BDT pre-processing ($\epsilon_2$) and 
prevents the movement of the most persistent branches (relative persistence 
larger
than $\epsilon_3$). 
After the $\epsilon_3$ pre-processing
(right), the red branch
in $B''$ moves up the BDT and becomes adjacent to the main light blue branch. 
Since they now have identical depths in the corresponding BDTs, the red
branches of $B'$ and $B''$ can now be matched together
(right), which results in an overall matching 
between these 
two 
trees 
which better conveys the resemblance between the two 
sub-clusters $B'$ and $B''$.
Equivalently, one can interpret this procedure on the BDTs 
as a modification of the 
input scalar field, turning 
the Gaussian mixture $B'$ into $B''$.
In 
particular,
this field modification disconnects the Gaussian with the red maximum from the 
Gaussian with the cyan maximum ($B''$) and reconnects it to
the Gaussian with the light blue maximum ($B'$).
Overall, after the $\epsilon_3$ pre-processing (right), the MT-PGA better 
distinguishes the cluster $A$ (pink spheres, left of the planar layout) from 
the 
cluster $B$ (dark red spheres, right of the planar layout).
}}
  \label{fig_eps3}
\end{figure*}

\revision{In this section, we replicate the same experimental protocol, but this 
time for the specific task of MT-PGA computation. Specifically, 
Figures \ref{fig_eps1}, \ref{fig_eps2} and \ref{fig_eps3} 
respectively illustrate the effect of the parameters $\epsilon_1$, 
$\epsilon_2$, and $\epsilon_3$ on the MT-PGA basis.}

\revision{
As described by Pont et al. \cite{pont_vis21}, in the data, moving a 
branch up the BDT corresponds to only slight modifications, which consists in 
reconnecting maxima to distinct saddles. For each parameter ($\epsilon_1$, 
$\epsilon_2$, and $\epsilon_3$), the resulting 
pre-processing addresses cases where nearby saddles have close function values, 
which impacts the stability of the metric. 
As discussed in the main manuscript (section 2.3), 
similarly to Sridaharamurthy et al. 
\cite{SridharamurthyM20}, Pont et al. mitigate this effect with $\epsilon_1$, 
but they also introduce $\epsilon_2$ and $\epsilon_3$ to specifically limit the 
importance in the metric of branches with a persistence close to that of their 
parents \cite{pont_vis21}.
}

\revision{For each figure, we study an ensemble consisting of two main clusters: 
$A$ (pink spheres, left of Figures \ref{fig_eps1}, \ref{fig_eps2} and 
\ref{fig_eps3}) and $B$ (dark red spheres, right of Figures 
\ref{fig_eps1}, \ref{fig_eps2} and 
\ref{fig_eps3}). These clusters have been synthesized by considering 
Gaussian mixtures (such that $B$ has one more prominent feature than $A$)
and by generating 
the other members of the ensemble with
variants of these two  
patterns, by inserting a random additive noise. 
Specifically, the cluster $B$ is synthesized out of 2 slightly distinct 
Gaussian mixtures, yielding two \emph{artificial} sub-clusters $B'$ and $B''$, 
such that, in each case, the red branch directly connects to a different branch 
in the sub-clusters $B'$ and $B''$. Then each figure visualizes the impact 
of each parameter $\epsilon_1$, $\epsilon_2$, and $\epsilon_3$ on the 
displacement of the red branch, and hence on the resulting MT-PGA basis.}

\revision{Overall, as discussed in the detailed captions, these three 
parameters have the effect of moving branches up the input BDTs, hence 
reducing the structural impact of these branches on the metric, but also 
improving its stability. In Figures 
\ref{fig_eps1}, \ref{fig_eps2} and 
\ref{fig_eps3}, by moving the red branch up, a larger section of the 
BDTs of the sub-clusters $B'$ and $B''$ become isomorphic, and thus, the two 
sub-patterns of the cluster $B$ become closer to each other in 
$\branchtreeSpace$ and the clusters $A$ and $B$ become better differentiated in 
the MT-PGA basis (cluster $A$ on the left of the planar layout, pink 
spheres, cluster $B$ on the right of the planar layout, dark red 
spheres).}

\end{document}